\documentclass[12pt]{report}
\usepackage{graphicx,epsfig}
\newcommand{\be}{\begin{eqnarray}}
\newcommand{\ee}{\end{eqnarray}}
\newcommand{\QM }{quantum mechanics}
\newcommand{\CQM }{Quantum mechanics}
\newcommand{\LHV}{local hidden variables}
\newcommand{\ket}[1]{|#1\rangle}
\newcommand{\bra}[1]{\langle#1|}

\newcommand{\cyt}[1]{(\ref{#1})}

\newcommand{\dits}{qu$N$its}

\begin{document}
{
\thispagestyle{empty}
\vspace{1.5cm}
\begin{center}
{\large Dagomir Kaszlikowski}
\end{center}

\vspace{3cm} \begin{center}{{\Large Nonclassical Phenomena}}
\end{center}

\begin{center} {{\Large in Multi-photon Interferometry:}}
\end{center}

\begin{center}
{{\Large More Stringent Tests Against Local Realism}}
\end{center}

\vfill
\flushright{Praca doktorska napisana\\
w~Instytucie Fizyki Teoretycznej\\ i~Astrofizyki\\ Uniwersytetu
Gda\'nskiego\\ pod kierunkiem\\ dr hab. Marka \.Zukowskiego, prof. UG\\}

\vspace{2.cm}
\begin{center}
Gda\'nsk 2000
\end{center}
\newpage
\thispagestyle{empty}
\vspace*{2.5cm}
\begin{flushright}
{\bf So farewell elements of reality!\\
And farewell in a hurry.\\}
\vspace{.7cm}
David Mermin
\end{flushright}

\vspace*{9.5cm}

\begin{flushright}
\parbox{9cm}{I would like to thank
my thesis supervisor
dr hab. Marek
\.Zukowski for help, patience
and for
being 
my guide
through the maze of quantum world.\\[\baselineskip]
I also thank Maja for being with
me 
in difficult moments of my life.\\[\baselineskip]
This dissertation was supported by The Polish Committee for Scientific Research,
Grant No. 2 P03B 096 15 and by Fundacja Na Rzecz Nauki Polskiej. 
}
\end{flushright}}
\newpage
\setcounter{page}{1}
\chapter*{Abstract}
The present dissertation consists of three parts which are mainly based
on the following papers\footnote{The author's papers will be cited here and 
in the main text using the number
from the list in this page in square brackets, whereas other papers will be quoted
by the name of the first author and the year of publication.
}:

\begin{itemize}

\item[\cite{Moja1}] M. \.Zukowski, D. Kaszlikowski and E. Santos, 
Phys. Rev. A {\bf 60}, R2614 (1999) 

\item[\cite{Moja2}] M. \.Zukowski and D. Kaszlikowski, 
Acta Phys. Slov. {\bf 49}, 621 (1999) 

\item[\cite{Moja3}] M. \.Zukowski and D. Kaszlikowski, 
Phys. Rev. A {\bf 56}, R1682 (1997) 

\item[\cite{Moja4}] D. Kaszlikowski and M. \.Zukowski, 
Phys. Rev. A {\bf 61}, 022114 (2000)  

\item[\cite{Moja5}]  M. \.Zukowski and D. Kaszlikowski, 
Phys. Rev. A {\bf 59}, 3200 (1999) 

\item[\cite{Moja6}] M. \.Zukowski and D. Kaszlikowski, Vienna Circle Yearbook vol. 7, 
editors D. Greenberger, W. L. Reiter, 
A. Zeilinger, Kluwer Academic Publishers, Dortrecht (1999) 

\item[\cite{Moja7}] M. \.Zukowski, D. Kaszlikowski, A. Baturo and J. -\AA. Larsson, 
quant-ph/9910058 

\item[\cite{Moja8}] D. Kaszlikowski, P. Gnacinski, M. \.Zukowski, W. Miklaszewski
and A. Zeilinger, quant-ph/0005028 
\end{itemize}

The first two chapters have an introductory character.

In Chapter 3 it is shown 
\cite{Moja1} that the possibility of 
distinguishing between
single- and two- photon detection events, usually not met in the actual
experiments, is not a necessary requirement for the proof that the experiments
of Alley and Shih \cite{SHIH} and Ou and Mandel \cite{MANDEL88} are, modulo
a fair sampling assumption, valid tests of local realism. It is also shown 
that 
some other interesting phenomena (involving bosonic-type particle 
indistinguishability) can be observed during such tests.

Next in Chapter 4 it is shown again 
\cite{Moja2} that the possibility of 
distinguishing between single and two photon detection events
is not a necessary requirement for the proof that recent operational
realisation of entanglement swapping cannot find a local realistic
description. A simple modification of the experiment is proposed, which
gives a richer set of interesting phenomena.

In Chapter 5 a sequence of Bell inequalities
for $M$ particle systems, which involve three settings of each 
of the local measuring apparatuses, is derived \cite{Moja3}. 
For Greenberger-Horne-Zeilinger
states, quantum mechanics violates these inequalities by factors
exponentially growing with $M$. The threshold visibilities of the
multiparticle sinusoidal interference fringes, for which local realistic
theories are ruled out by these inequalities, 
decrease as $(2/3)^M$.

In Chapter 6 the Bell 
theorem for a pair of two
two-state systems (qubits) in a singlet state for the
entire range of measurement settings is presented \cite{Moja4}.

Chapter 7 is devoted to derivation of 
a series of Greenberger-Horne-Zeilinger
paradoxes for $M$ qu$N$its (particles described by
an $N$ dimensional Hilbert space) that are fed
into $M$ unbiased $2N$-port spatially separated beam splitters
\cite{Moja5,Moja6}.
 
In Part III a novel approach to the Bell theorem, via
numerical linear optimisation, is presented \cite{Moja7,Moja8}.

The two-qubit correlation obtained from the quantum state used 
in the Bell inequality is sinusoidal, but the standard Bell inequality
only uses two pairs of settings and not the whole 
sinusoidal curve. The highest to-date visibility of an 
explicit model reproducing
sinusoidal fringes is ${2\over\pi}$. We conjecture from a 
numerical approach \cite{Moja7} presented in Chapter 8 that the highest 
possible visibility for a local
hidden variable model reproducing 
the sinusoidal character of the quantum prediction for 
the two-qubit Bell-type interference
phenomena is ${1\over\sqrt 2}$. In addition, 
the approach can be applied directly to experimental data. 

In Chapter 9 the approach presented in Chapter 8 is applied
to three qubits in a maximally entangled Greenberger-Horne-Zeilinger 
state. For the first time the necessary and sufficient conditions
for violation of local realism for the case in which each
observer can choose from up to 5 settings of the measuring
apparatus are shown.

In Chapter 10 using
the modified approach developed in Chapter 8 
it is shown that violations of local realism are stronger
for two maximally entangled qu$N$its, than for two qubits \cite{Moja8}.
The magnitude of violation increases with $N$. It is objectively
defined by the required minimal admixture of pure noise
to the maximally entangled state such that a local realistic
description is still possible. Operational realisation of the
two qu$N$it measurement exhibiting strong violations of local
realism involves entangled photons and unbiased multiport
beamsplitters. The approach, extending at present to $N=9$,
neither involves any simplifications, or additional assumptions, nor
does it utilise any symmetries of the problem. 
\tableofcontents
\chapter{Introduction}
Bell's theorem \cite{Bell64}, formulated in 1964, initiated and
revitalised several branches of modern physics. The paper
was the first one to show that quantum entanglement
cannot be in any way simulated by classical correlations.
Within few years, a new branch of experimental physics
emerged: multi-particle quantum interferometry. Since then
it has evolved and extended its field of interest from two-photon
to multi-photon correlations. Recently Bell-EPR
correlations were observed for entangled atoms \cite{Hagley97}.
For as much as nearly 20-25 years the paper of Bell was 
studied mainly by people interested in the 
foundational-interpretational problems of quantum theory.
Suddenly with the discovery of the possibility of employing
Bell-EPR correlations \cite{ECKERT91} in "quantum cryptography" 
\cite{BB84}
and with the realisation of the importance of entanglement
in the hypothetical quantum computers \cite{FEYNMAN82,DEUTSCH89}, 
it turned out that
the paper of Bell can be thought of as the first one in 
the field of quantum information.

Studies of quantum information led to a proposal, employing
entanglement, of quantum teleportation \cite{TELEPORT95}. This phenomenon
was observed in 1997 and seems to be at the moment the crowning
achievement of quantum interferometry \cite{BOUW}. Interestingly,
the method to obtain quantum teleportation
of photon's polarisation was developed independently, as a
by product of theoretical and experimental research
towards obtaining Bell-EPR phenomena for particles originating
from independent sources \cite{YS,EVENT,Pan98}.

The same method was applied to obtain the first ever observations
of Greenberger-Horne-Zeilinger correlations (GHZ) \cite{INNSBRUCK}. 
The 1989 theoretical
discovery of GHZ correlations \cite{GHZ89} and the drastic amplification of
the Bell theorem, which is implied by them, was the event in the research
into foundations of quantum theory which amplified interest in 
entanglement (with the interesting sociological consequence:
the earlier terminology- correlated state- was replaced by the
Schr\"{o}dinger's term entanglement).

The technological progress of 1980's and 1990's has enabled all that 
experimental and theoretical activity to flourish. The phenomenon of
parametric down conversion (PDC) turned out to be a versatile source
of entangled photons. The simplicity of the phenomenon of
PDC has lead to an explosion of the number of experiments studying
various aspects of entanglement or the basic phenomena
linked with quantum information and quantum communication (dense coding 
\cite{MATTLE96},
Bell-state measurement \cite{MICHLER96}, etc.). 
Pulsed down conversion enabled to observe
two-photon interferometric effect for independently emitted
photons \cite{Pan98}. The future of experimental quantum information
most probably will be associated with trapped atoms and 
microcavity-atom interactions \cite{Hagley97}- 
fields in which extensive studies of 
entanglement are currently carried out.

Interestingly, all these developments led to studies concerning entanglement
properties of mixed states. In the case of the Bell theorem the pioneering
works were by Werner (1989) \cite{WERNER89} and the Horodecki Family (1995) 
\cite{HOROD95}. In recent
years one observes an avalanche of works on the problem of separability
of density matrices initialised by Peres \cite{PERESSEP96} and again
the Horodecki Family (see e.g. \cite{HOROD96}).
The research into the separability has shown once more (recall 
Bell-Kochen-Specker theorem \cite{KOCHEN67}) 
the qualitative difference between systems described 
by 2-dimensional Hilbert space (qubits) and those described by 
Hilbert spaces of higher dimension (qu$N$its).

The present work tries to answer some questions on the relation
of the Bell theorem with various performed or proposed multiparticle
(essentially, multiphoton) quantum interference experiments.

The first part deals with some of performed experiments. Proposals
of improvements and re-interpretations are given \cite{Moja1,Moja2}.

Next, in part two, we study new methods of deriving Bell inequalities both
for the standard two qubit experiments as well as to multi-qubit
GHZ experiments \cite{Moja3,Moja4}. Derivation of GHZ paradoxes
for gedanken experiments involving $M$ entangled qu$N$its observed beyond
multiport beamsplitters \cite{ZZH} is also shown \cite{Moja5,Moja6}.

Finally the last part is devoted to the Bell theorem without
inequalities via a numerical approach utilising linear optimisation 
\cite{Moja7,Moja8}.

\chapter{Some history and basic notions}
\section{Preliminaries}
According to a prevailing common opinion \QM\ is
a fundamental theory which applies to all physical systems. Its
predictive power is astonishing. Up to the present day there has not
been a single experiment which invalidates it.
However, the conclusions that can be drawn from \QM\ force us to
entirely abandon the picture of nature implied by classical
physics and the common sense. One of the main sources of difference 
between the quantum
world and the classical one, in which we are particularly interested
in this work, is {\it entanglement}.

The no\-tion of en\-tan\-gle\-ment was in\-tro\-duced for the first time  
by Schr\"{o}dinger to describe a situation
in which  
\begin{quote}
Maximal knowledge of a total system does not necessarily
include total knowledge of all its parts, not even when these are
fully separated from each other and at the moment are not
influencing each other at all...
\end{quote}
The work of Schr\"{o}edinger \cite{Erwin35} was partially
motivated by the seminal paper of Einstein, Podolsky and Rosen (EPR)
\cite{EPR} in which the authors used an entangled state 
(an EPR state) of
two qubits to show that \QM\ could not be considered as a complete
physical theory aiming to describe the phenomena occurring
in micro world.  Although they did not state it clearly, EPR effectively
postulated the existence of {\it local hidden variables}, 
in the form of "elements of reality",
which were to play
the same role in \QM\ as the positions and velocities of particles
in statistical classical mechanics and that were to solve the
interpretational problems of \QM . That paper directly influenced the 
formulation of the Bell theorem\footnote{According to Stapp 
\cite{Stapp77}, the Bell theorem is one 
of the most
important discoveries in modern physics.} 
\cite{Bell64,Bell66,Bell87}.

In his 1964 paper Bell for the very first time showed\footnote{A theorem by 
von Neumann \cite{vonNeumann} excluding 
the possibility of the existence of hidden variables was formulated
in 1930's but, as pointed 
out by Bell, the assumptions used by von Neumann were 
much too restrictive.} that the idea of local hidden variables was in
contradiction with \QM\ and, what is even more important, that
it could be tested experimentally! That way the subject mainly
discussed by physicists at the parties was brought to the realm of
experimentally verifiable physics.

\section{Entanglement}
To discuss basic features of entanglement let us consider the following
state of two two-state systems (qubits)\footnote{If we take as
the qubits two spin ${1\over 2}$ particles 
and put $\ket{0}=\ket{-\frac{1}{2}}, \ket{1}=\ket{+\frac{1}{2}}$
the corresponding state
is a rotationally invariant state with the total spin equal zero-
the so called singlet state.}
\begin{eqnarray}
&&|\psi\rangle ={1\over\sqrt2}(|0\rangle_{1}\otimes|1\rangle_{2}-
|1\rangle_{1}\otimes|0\rangle_{2}),
\label{singlet}
\end{eqnarray}
where $\otimes$ denotes the tensor product\footnote{This symbol
will be only used when it makes the notation easier to read.} and kets 
$|0\rangle_{i},|1\rangle_{i}$ describe two orthogonal states
of the $i$-th qubit.
The above pure state describes a coherent superposition of two
product states that occur with equal probability. 

According to \QM\ 
$|\psi\rangle$ contains all available information about the state of
the qubits. If we write (\ref{singlet}) in the form of a density matrix 
$\rho_{12}$ 
\begin{eqnarray}
&&\rho_{12}=|\psi\rangle\langle\psi|={1\over2}\left[|0\rangle_{1}{}_{1}\langle 0|
\otimes
|1\rangle_{2}{}_{2}\langle 1|-
|0\rangle_{1}{}_{1}\langle
1|\otimes|1\rangle_{2}{}_{2}\langle 0|\right.\nonumber\\
&&\left.-|1\rangle_{1}{}_{1}\langle 0|\otimes
|0\rangle_{2}{}_{2}\langle 1|
+|1\rangle_{1}{}_{1}\langle 1|\otimes|0\rangle_{2}{}_{2}\langle 0|\right],
\label{sdens}
\end{eqnarray}
and trace out one qubit
we obtain a density matrix describing the other qubit, which reads
\begin{eqnarray}
&&\rho_{k}={1\over2}\left(|1\rangle_{k}{}_{k}\langle 1|+
|0\rangle_{k}{}_{k}\langle 0|\right),
\label{sdens1}
\end{eqnarray}
$k=1,2$.
Such a density matrix describes a situation in which we have a chaotic
mixture of two orthogonal pure states. This is a  
situation typical for
the entanglement. We have the full possible information about the 
state of two qubits as a
whole but we do not have any information about the state of each 
qubit separately, which in fact is not even defined! In addition, 
the properties of 
the qubits are tightly correlated. 

To see this more clearly
let us consider the measurement of two {\it dichotomic} observables
with spectrum consisting of $\pm 1$, which spectral decomposition 
has the following
form
\begin{eqnarray}
&&\hat{O}_{k}(\phi_{k})=|+,\phi_{k}\rangle_{k}{}_{k}\langle +,\phi_{k}|-
|-,\phi_{k}\rangle_{k}{}_{k}\langle -,\phi_{k}|
\label{operators}
\end{eqnarray}
with $k=1,2$ and
\be
&&|+,\phi_{k}\rangle_{k}={1\over\sqrt2}\left(|0\rangle_{k}+
e^{i\phi_{k}}|1\rangle_{k}\right)\nonumber\\
&&|-,\phi_{k}\rangle_{k}={1\over\sqrt2}\left(|0\rangle_{k}-e^{i\phi_{k}}
|1\rangle_{k}\right),
\ee
where $\phi_{k}\in [0,2\pi]$. The
mean value $E_{QM}(\phi_{1},\phi_{2})$ of the joint measurement of the 
observable 
$\hat{O}_{1}(\phi_{1})$ on
the first qubit and the observable $\hat{O}_{2}(\phi_{2})$ on the second one- 
the so called correlation function-
reads
\begin{eqnarray}
&&E_{QM}(\phi_{1},\phi_{2})=\langle\psi|\hat{O}_{1}(\phi_{1})
\hat{O}_{2}(\phi_{2})|\psi\rangle=
-\cos(\phi_{1}+
\phi_{2}).
\label{corrfunc}
\end{eqnarray}

It is easy to see that, for example, whenever the sum of the 
phases $\phi_{1}$ and 
$\phi_{2}$ is $0$ modulo $2\pi$ we observe the so called {\it 
perfect correlations}
between the results of the measurements performed by the
observers; if the first observer obtains $+1$ as the result
of his measurement the second one obtains $-1$ and vice versa.
However, each observer alone measures
$+1$ and $-1$ with equal probability, which can be easily seen
using the density matrix (\ref{sdens1}) 
\be
&&\langle O_{k}(\phi_{k})\rangle_{\rho_{k}}=Tr(O_{k}(\phi_{k})\rho_{k})=0.
\ee

The above formulas clearly demonstrate that
all information about the state (\ref{singlet}) is contained in the joint
properties of the qubits.

\section{Elements of reality}
In their famous paper \cite{EPR} EPR used the perfect correlations
observed when measuring local observables on an entangled system of spatially
separated qubits
(\ref{singlet}) to demonstrate that \QM\ is incomplete, i.e., that
one needs some additional parameters to fully describe phenomena occurring
in micro world. 
We briefly present their reasoning in the version of Bohm
\cite{Bohm-EPR1,Bohm-EPR2} to introduce the notion of 
{\it local realism}, which plays a central part in 
the Bell theorem.

EPR reasoning goes as follows. According to \QM\ all we know about
the system of two entangled qubits is encoded in the 
state (\ref{singlet}). \CQM\ also tells us that we cannot consider
simultaneous measurements
of two non commuting observables and therefore does not even
define predictions for such cases. 
Let us imagine that the observers
one and two are spatially separated and
that they simultaneously measure the observables $\hat{O}_{1}(\phi_{1}=0)$ and
$\hat{O}_{2}(\phi_{2}=0)$ (see the picture \cyt{EPR-fig}) on 
subsystems described by the singlet 
state. 
\begin{figure}[htbp]
\begin{center}
\includegraphics[angle=0, width=13.5cm, height=3cm]{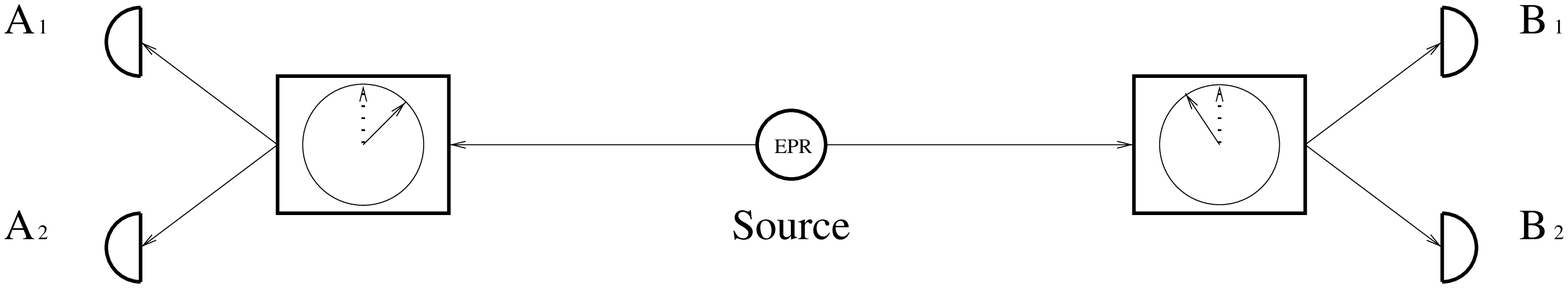}
\caption{The scheme of the Bell type experiment. 
Solid arrows represent actual positions
of the measuring apparatus, dashed arrow the future ones. Detectors
$A_1, A_2, B_1, B_2$ register incoming qubits.\label{EPR-fig}}
\end{center}

\end{figure}

For such observables 
the
perfect correlations occur, which means that if the first observer has
obtained $+1$ the second one, because of (\ref{corrfunc}), must
have obtained
$-1$ and vice versa. Moreover, due to the spatial separation of
the observers and the non superluminal velocity of propagation
of any interaction (information) in nature (locality), 
the outcome of the first (second) observer cannot be influenced
by the choice of observable measured
by
the second (first) observer and neither by its outcome. All correlations
between the results of measurements must have been established in
the source.

However, the second observer could have measured
the observable $\hat{O}_{2}(\phi_{2}'={\pi\over 2})$ instead of the
previous one characterised by $\phi_{2}=0$. Then, the
result of this would-be measurement, would have enabled him to infer
{\it with probability equal to one
and without disturbing} the first qubit 
the result of the measurement of 
the observable $\hat{O}_{1}(\phi_{1}'={\pi\over 2})$ by the first
observer. At this stage of reasoning EPR introduce the notion of
{\it physical reality} \cite{EPR}:
\begin{quote}
If, without in any way disturbing a system, we can predict with
certainty (i.e. with probability equal to unity) the value of a
physical quantity, then there exists an element of physical
reality corresponding to this physical quantity.
\end{quote}  
According to this definition the inference made by the
second observer\footnote{The reasoning can be reversed and the
inference can be made by the first observer about the second one.} 
about the result of the measurement that could have been made by the 
first observer has a well defined physical meaning. This would mean 
that it is possible to ascribe the definite values to the results of 
the measurement 
of two non commuting observables $\hat{O}_i(\phi_i=0)$
and $\hat{O}_i(\phi_{i}'={\pi\over 2})$ ($i=1,2$), the statement which
makes quantum purist's hair stand on ends.
Therefore, according to EPR, \QM\ is incomplete
and should be completed at least by introducing elements
of reality into the description of quantum state.

\section{The Bell theorem}
The hypothesis of local hidden variables, which effectively
stems from the notion of elements of reality, was proved 
to be unacceptable for systems described by quantum 
mechanics
by Bell, 
\cite{Bell64} who found an inequality which should be
obeyed by any local hidden variable theory, and which
is violated by predictions of \QM. 
Here we present the derivation of the Clauser-Horne-Shimony-Holt (CHSH) 
inequality 
\cite{CHSH69}. 

To this end, let us again
consider the experiment in which two spatially separated observers 
perform the 
measurements
of the observables defined in (\ref{operators}) on 
the state (\ref{singlet}). Each of them as a result of their measurement
obtains only one of the two possible values $\pm 1$ (they measure bivalued 
observables). Although the different results of the measurement at 
each observer's
side occur with equal probability 
the results of joint measurements are correlated, which is
expressed by the formula (\ref{corrfunc}).
If deterministic \LHV\ exist they predetermine
the results of each single measurement for each observer at the time
of the emission of each pair of qubits. In other words, 
the value of the local hidden variable for a particular emission
of a pair of qubits would allow us to predict the result of the measurement
made by each observer with certainty. However this value is hidden.
To model the probabilistic 
nature of quantum experiments
we assume that there exists some probability distribution of \LHV\ 
associated with a given quantum mechanical state, 
which represents our lack of knowledge about them.

To express the above idea of \LHV\ mathematically we assume that
there is a set of hidden variables $\Lambda$ on which we can define
a probabilistic measure $d\rho(\lambda)$. We also assume the existence
of two bivalued functions $O_{1}, O_{2}$ defined on the space
$\Lambda$ which take only the values $\pm 1$. Each of these two functions
must 
depend on the local parameters 
$\phi_{1}$ and $\phi_{2}$ characterising the experiment performed by the
first and the second observer respectively. Because of the assumption
of locality the function $O_1$ can solely depend on the $\phi_{1}$ and the
other one on $\phi_{2}$.
Thus, the correlation function based
on the idea of \LHV\ must have the following form
\be
&&E_{LHV}(\phi_{1},\phi_{2})=\int_{\Lambda}d\rho(\lambda)O_{1}(\lambda,\phi_{1})
O_{2}(\lambda,\phi_{2}),
\label{hidcorfunc}
\ee
where $\int_{\Lambda}d\rho(\lambda)=1$.

Now, let us imagine that each observer performs two mutually exclusive
experiments characterised by $\phi_{i}, \phi_{i}'$ and let us consider
the following expression made out of the four local hidden variables 
correlation
functions (\ref{hidcorfunc})
\be
&&C_{LHV}=E_{LHV}(\phi_{1},\phi_{2})+E_{LHV}(\phi_{1}',\phi_{2})
+E_{LHV}(\phi_{1},\phi_{2}')-E_{LHV}(\phi_{1}',\phi_{2}')\nonumber\\
&&=\int_{\Lambda}d\rho(\lambda)\left[O_{1}\left(\lambda,\phi_{1})
(O_{2}(\lambda,\phi_{2})+O_{2}(\lambda,\phi_{2}')\right)
+O_{1}(\lambda,\phi_{1}')
\left(O_{2}(\lambda,\phi_{2})-O_{2}(\lambda,\phi_{2}')\right)\right].
\label{CHSH(LHV)}
\ee
It is easy to see that the modulus of the expression in the square 
brackets is either $-2$ or $+2$. Therefore, the hypothesis of \LHV\ implies
that the following inequality (Bell-CHSH inequality) must be valid 
\be
&&-2\leq C_{LHV} \leq 2.
\label{CHSH(LHV)ineq}
\ee 

Is the CHSH inequality always satisfied by quantum predictions for \cyt{singlet}? 
To answer this question
let us put $\phi_{1}=0,\phi_{1}'={\pi\over 2}$ and $\phi_{2}=-\pi/4,
\phi_{2}'=\pi/4$. For these values of local parameters one has
\be
&&C_{QM}=E_{QM}(\phi_{1},\phi_{2})+E_{QM}(\phi_{1}',\phi_{2})
+E_{QM}(\phi_{1},\phi_{2}')-E_{QM}(\phi_{1}',\phi_{2}')\nonumber\\
&&=-\cos(\phi_{1}+\phi_{2})-\cos(\phi_{1}'+\phi_{2})
-\cos(\phi_{1}+\phi_{2}')+\cos(\phi_{1}'+\phi_{2}')=-2\sqrt2.
\label{CHSH(QM)ineq}
\ee
Because $-2\sqrt 2 < -2$, we have a contradiction.

The above result, known as the Bell theorem\footnote{The inequality
which must be obeyed by any local and realistic theory is usually called the Bell
inequality whereas the violation of such an inequality by \QM\ is called the Bell theorem.}, 
needs some further explanation. In our
reasoning we have made two crucial assumptions without which the theorem would not be valid.
These assumptions are: {\it locality} and {\it realism}. 
The Bell theorem tells us that either notion of locality, or realism, or both are false 
in quantum theory \cite{Redhead87}.
  
Another remark is that the Bell-CHSH inequality can be directly applied to any experimental
data. Also, even if \QM\ is not valid and we will find another better theory we
can still, using the Bell inequality, verify whether this new theory 
fulfils the necessary condition for the local realistic description or not.

The final remark is that one may consider the existence of the so called stochastic
\LHV\ \cite{CH}, which do not predict with certainty the results of local measurements but
give merely the probabilities of their occurrence. In such a case instead of functions $O_{n}$ 
($n=1,2$) appearing in (\ref{hidcorfunc}) we have the probabilities 
$P_{1}(m|\lambda,\phi_1),
P_{2}(m'|\lambda,\phi_2)$  
giving the ratio of occurrence of the results $m,m'$ when 
measuring observables characterised by parameters $\phi_1,\phi_2$ respectively
(obviously they sum up to
one, i.e., $\sum_{m=\pm 1}P_{i}(m|\lambda,\phi_i)=1$, $i=1,2$). The
relation between $P_{i}(m|\lambda,\phi_i)$ and $O_i(\lambda,\phi_i)$ ($i=1,2$)
is such that within this description $O_i=\pm 1$ have to be replaced by
\be
&&O_i(\lambda,\phi_i)=\sum_{m=\pm 1}mP_{i}(m|\lambda,\phi_i).
\ee
with the values of modulus bounded by 1.
With such 
probabilities the idea of stochastic \LHV\ is to 
reproduce the quantum probabilities 
$P_{QM}(m,m'|\phi_1,\phi_2)$, i.e., the probabilities of
obtaining the result $m$ and $m'$ by the first and the second 
observer when measuring
the observable characterised by $\phi_1$ and $\phi_2$ respectively,
by local hidden variables probabilities of the form
\be
&P_{HV}(m,m'|\phi_1,\phi_2)=\int d\rho(\lambda)
P_{1}(m|\lambda,\phi_1)P_{2}(m'|\lambda,\phi_2).&
\label{stochLHV}
\ee
It is clear that any deterministic local hidden variables theory
can be always treated as a stochastic one. 
Fine \cite{Fine} proved that a stochastic 
local hidden variable theory implies the existence
of  an underlying deterministic one\footnote{We do not take into account
non Kolmogorovian probability calculus.}.

In general, all Bell-type inequalities found since the famous Bell
paper \cite{Bell64} 
constitute only necessary conditions for the existence of local hidden variables. 
The exceptions are 
the full set of four Clauser-Horne inequalities (CH) \cite{CH}
and the Bell-CHSH inequality, 
which were proved 
by Fine \cite{Fine} to be
also sufficient ones for dichotomic observables\footnote{The CHSH inequality 
is sufficient if one assumes certain
symmetries of the probabilities.} (see also \cite{PERESBELL}).       

In 1989 Greenberger, Horne and Zeilinger (GHZ) \cite{GHZ89} used
a maximally entangled state of four qubits to show
that the discrepancy between local realism and quantum mechanics
is much stronger than that observed in two qubit 
correlations\footnote{Later Mermin simplified the proof
and derived GHZ paradox for three qubits \cite{MERMIN-GHZ}.}. 
By a clever trick they showed that the idea
of local realism breaks already at the stage of defining elements of 
reality. 

The Bell theorem
does not have to be restricted to two or three
qubit correlations. The discrepancy
between local realism and quantum mechanics
can be also proved for entangled particles each 
described by an $N$ dimensional Hilbert space- so called qu$N$its 
(see, for instance,
\cite{MERMIN80,MERMIN82,GARG82,GISIN92,PERES92,WODKIEWICZ94})-
as well as for $M$ entangled qubits (see, for instance, 
\cite{MERMIN90}).
\section{Experimental tests of the Bell theorem}
\subsection{First experiments}
The first experimental test of Bell inequality was performed
by Freedman and Clauser \cite{FREEDMAN72} with
photons from atomic cascade decays. 
They observed violation of Bell inequality and confirmation
of quantum mechanical predictions. However, in the experiment
the detection efficiency and the angular correlation of the photon pairs
was low, and no care was taken to make
the two polarising settings detection stations to be set
independently \cite{CH,SANTOS}.

Such care was taken in the most quoted ex\-per\-i\-ment
by As\-pect, Grang\-ier and Dal\-ibard \cite{ASPECT82}.
In this experiment fast switchings of the analyser
position to prevent "communication" between the source and
the analyser was used. However, as it was pointed out in
\cite{Zeilinger86} the periodic switching was not truly 
random and
was predictable after a few periods of the switch\footnote{In 1998
Weihs {\it et. al.} \cite{WEIHS98} for the first time performed
an experiment in which true locality condition was 
enforced. In the experiment two observers were spatially
separated by the distance of 400m, which means that the time
of the measurement had to be shorter than 1.3$\mu$s to prevent
communication with the speed of light  between observers. 
They succeeded to achieve
the time of measurement within 100 ns and 
the violation of the CHSH inequality by 30 standard deviations
was observed.
}.

In all experiments
(except one, with systematic errors) 
the violation of the Bell inequalities (with certain additional 
assumptions) 
was
observed. An excellent review concerning these
pioneering experiments can be found 
in \cite{CLAUSER78}.

\subsection{New experiments}
Recently the experiments, in which the entangled pairs of
photons are generated in the process of parametric down 
conversion (PDC),
dominated the field of laboratory tests of local
realism. In the PDC process the pairs of photons
are spontaneously created. The propagation directions and
the frequencies of created photons (the photons are called
idlers and signals) are highly correlated, which
is used to generate an entangled state. In the type-II
down conversion one has also correlated polarisations.

Among the Bell-type experiments with PDC process
the following ones will be mainly addressed to in this work 
\begin{itemize}
\item Alley-Shih and Ou-Mandel experiments \cite{SHIH,MANDEL88}
\item entanglement swapping \cite{Pan98}
\item GHZ experiment \cite{INNSBRUCK}.
\end{itemize}

First Bell-type experiments with a PDC source of
correlated photons were experiments by
Alley, Shih and Ou, Mandel. In both of them
the violation of the Bell inequality (by 3 standard deviations
in \cite{SHIH} and 6 standard deviations in \cite{MANDEL88}) 
and confirmation
of quantum mechanics was reported. However, in both
experiments only coincident counts were measured (half
of the events were discarded). That raised some doubts about
the validity of the experiments as tests of local
realism \cite{SANTOS,GARUCCIO}. The situation was 
clarified in \cite{SHZ} where it was shown that one
does not need to discard "wrong" events to test local realism
in the experiments. 
In this dissertation we perform
further theoretical analysis of these experiments (see Part I). 

In 1993 \.Zukowski {\it et. al} \cite{EVENT} showed
experimental conditions to
entangle particles (photons) originating from
independent sources\footnote{The first proposal was given by Yurke and Stoler 
\cite{YS}.}. Five years
later, in 1998, the first entanglement
swapping experiment was performed \cite{Pan98}. 
The visibility
of around $65\%$ was observed (the notion of visibility
is explained in the next subsection).

In 1999 Bouwmeester {\it et. al} \cite{INNSBRUCK} 
reported the first experimental
observation of the GHZ correlations. The experiment was based
on the techniques developed in the teleportation experiment \cite{BOUW}
and entanglement swapping experiment \cite{Pan98}. In the experiment 
pairs of entangled photons (entangled polarisations) 
produced in a nonlinear crystal pumped by a short pulse
of ultraviolet light from the laser were used. The applied technique to
obtain GHZ correlations rests upon an observation that when a single
particle from two independent entangled pairs is detected in a manner
such that it is impossible to determine from which pair the single
came, the remaining three particles become entangled.
The high visibility
of around $60\%$ was observed.

The experimental realisation of entanglement does not have to
be restricted to massless particles (photons).
Hagley {\it et. al.} reported the experiment in which
entangled atoms were produced \cite{Hagley97}. They demonstrated
the entanglement of pairs of atoms at centimetric distances and measured
their correlations. 
The visibility of only around $25\%$ was observed.
\subsection{Problems encountered in the Bell type experiments}
However, in all experiments performed thus far there
has been the problem
with a low quantum efficiency $\eta$ of detectors used to register
incoming particles\footnote{More precisely the quantum efficiency
describes the full detection stations, including all devices that
collect the incoming radiation (lenses, etc.).}. The quantum efficiency is 
defined as the ratio
of the number of detected particles to the number of the emitted
ones\footnote{In the operational terms it is defined for a detection
station A as the ratio of coincident correlated counts at the pair of 
detection stations A and B to the ratio of singles at the station B.}.
If it lies
below the threshold value $\eta^{tr}_2=2\sqrt2 -2$
there 
is no violation of the Bell-CHSH and the CH inequality for maximally 
entangled two qubits.

Since the collection efficiency in all experiments done so far was much lower
than $\eta^{tr}_2$, all claims about the violations of Bell inequalities in
performed experiments are based on the assumption 
that the observed
sub ensemble of particles is a representative one 
for the emitted ensemble
(so called
"fair sampling assumption"). 

In the real experiment one usually cannot obtain a pure maximally entangled
state of two qubits due to
some 
imperfections
in the source producing the state and other difficulties. As a simple
generic model of such experimental imperfections 
one can take
\be
&&\hat{\rho}(F_2)=F_2\hat{\rho}_{noise}+(1-F_2)\hat{\rho}_{pure},
\label{noise0}
\ee
where for qubits $\hat{\rho}_{noise}={1\over4}I\otimes I$ ($I$ is a unit $2\times 2$ 
matrix),
$\hat{\rho}_{pure}=\ket{\psi}\bra{\psi}$ (for the definition
of $\ket{\psi}$ see \cyt{singlet}) and the real parameter 
$F_2$ (the index 2 stands for qubit) 
lies between zero and one ($0 \leq F_2\leq 1$).
The $\hat{\rho}_{noise}$ describes a totally chaotic mixture of two qubits.
Results of any measurements carried out on the 
$\hat{\rho}_{noise}$ are completely uncorrelated therefore
the parameter $F_2$ can be interpreted as the number telling us 
how much noise is contained 
in the system.

It is easy to
check that for $\hat{\rho}(F_2)$ the correlation function 
defined in \cyt{corrfunc} 
reads
\be
&&E_{QM}^{F_2}(\phi_{1},\phi_{2})=Tr[\hat{\rho}(F_2)\hat{O}_{1}(\phi_{1})
\hat{O}_{2}(\phi_{2})]
=-(1-F_2)\cos(\phi_{1}+\phi_{2}).
\label{corrfuncV}
\ee
We see that if $F_2 > 0$ the amplitude of the correlation function is
reduced. 

In quantum interferometry the number $1-F_2$ 
is directly linked with
the visibility (contrast)
of interferometric
two-qubit fringes\footnote{The traditional definition of
interference visibility (contrast) is given by
\be
&&V=\frac{I_{max}-I_{min}}{I_{max}+I_{min}}
\ee 
where $I$ is the intensity (of light) in an interference
pattern. $I_{max}$ refers in the case
of spatial pattern  to the maximum intensity and $I_{min}$
to its neighbouring minimum. The same formula can be used
for an output of a Mach-Zehnder interferometer into one
of its exit arms. In this case $I_{max}$ is the maximum intensity
and $I_{min}$ its neighbouring minimum which occurs after suitable
change of the phase shift. In the case of a quantum process for
a single particle the visibility is defined, in analogy, as
\be
&&V=\frac{P_{max}-P_{min}}{P_{max}+P_{min}},
\label{X}
\ee
where $P_{max}$ and $P_{min}$ are the maximal and minimal
probabilities for detection of the particle in a specified
output of an interferometer (also obtainable for certain
different phase settings).

The visibility of two particle interference is again defined
using the same general rule \cyt{X}. However, in this case
$P$'s refer to the probability of coincident counts at a pair
of detectors. In the case when both single particle and two
particle interference occurs in the experiment, the relation
between the two visibilities is quite subtle, and the two-particle
visibility has to be redefined \cite{YAGER}. However, throughout the present
work we shall discuss only the cases for which no single particle
interference occurs.}. 

Therefore, sometimes it is convenient
to consider the parameter $V_2=1-F_2$ instead of $F_2$, which is usually
done in the description of real experiments. Throughout the dissertation
both parameters will be used depending on the context. 

In the entanglement swapping experiment \cite{Pan98} 
and in the
experiments with atoms \cite{Hagley97} the observed visibility 
was quite low: around $65\%$ in the entanglement swapping and
$25\%$ in the experiment with entangled atoms. 

It is obvious that if there is too much noise in the system, i.e.,
the visibility is low, then one cannot observe violations
of Bell inequalities.

To summarise, if in a Bell-type experiment with
two maximally entangled qubits $V_2\leq {1\over\sqrt 2}$
($F_2\geq\frac{2-\sqrt 2}{2}$)
the Bell-CHSH inequality (and also the CH inequality)
cannot be violated. Furthermore, in the case of 
the Bell-CHSH inequality, if $\eta_2\leq 2\sqrt 2 -2$, then even if
we have the perfect case, i.e., $V_2=1$ ($F_2=0$), the violation of
the inequality has only a {\it bona fide} status- one has to use
the fair sampling assumption. In all experiments thus far
performed one has $\eta_2\ll 2\sqrt 2 -2$. Similar situation (to
be described in more details later) occurs for the GHZ
correlations (with new specific threshold $\eta$'s and $V$'s).
\part{New theoretical analysis of Alley-Shih, Ou-Mandel
and entanglement swapping experiments}
\newpage
~~~
\thispagestyle{empty}
\newpage
\chapter*{}
\begin{center}
\Large{*}
\end{center}
\phantom{~~~}

Some of the performed Bell-type experiments
have a distinguishing trait. Not all event observed follow 
the standard pattern assumed in the usual derivations
of Bell inequalities.
To such experiments belong the Alley-Shih-Ou-Mandel
experiment \cite{SHIH,MANDEL88} and the entanglement swapping 
experiment \cite{Pan98}.

In all these experiments even in the ideal, gedanken case,
half or more of the emissions do not lead to correlated counts at
spatially separated detector stations. Therefore in order to
prove that such experiments can be indeed considered as tests 
of local realism one has to perform an analysis which takes
into account this characteristic trait. Such an analysis
will be presented below for the Alley-Shih, Ou-Mandel
experiments and the entanglement swapping experiment\footnote{The 
experiments proposed by Franson (1989) share properties which were
thought to be of a similar nature. By a closer inspection it turns out
that they are different. For explanation see \cite{AERTS}.}. Special
care will be taken to discuss the question of whether two versus
single photon counts distinguishability is the necessary requirement
for the studied experiments.

\chapter{Alley-Shih and Ou-Mandel experiments: resolution of the problem
of distinguishability of single and two photon events [1]}
\section{Introduction}
The first Bell-type experiments which employed parametric 
down conversion process as the source of entangled photons
were those reported in refs \cite{SHIH} and  \cite{MANDEL88}.
However, the specific traits of those experiments have led to a 
protracted dispute
on their validity as tests of local realism.
In this case the issue was not the standard problem of detection efficiency
(which up till now permits a local realistic interpretation of
all performed experiments).
The  trait that distinguishes the experiments is that even in the  
perfect {\it gedanken} situation (which assumes perfect detection) 
only in $50\%$ of the detection events 
each observer receives a
photon, in the other $50\%$ of events one observer receives both
photons of a pair while the other observer receives none.
The early ``pragmatic" approach was  to discuss only the events of the 
first type (as only such ones lead to spatially
separated coincidences). 
Only those were used as the data input to the
Bell inequalities in \cite{SHIH} and \cite{MANDEL88}. 
This procedure was soon 
challenged (see e.g. \cite{KWIAT1,KWIAT2}, and 
especially in the theoretical analysis of ref. \cite{GARUCCIO}), as it 
raises justified doubts whether such
experiments could be ever genuine tests of local realism (as the effective
overall collection efficiency of the photon pairs, $50\%$ in the gedanken
case, is much below what is usually required for tests of local realism). 
Ten years after the first experiments  of this type were made, finally
the dispute was resolved \cite{SHZ}.
It was proposed, to take into account also those
``unfavourable" cases and to
analyse the entire
pattern of events. 
In this way one can indeed show that  the experiments 
are true test of local realism (namely, 
that the CHSH inequalities are violated by quantum predictions
for the idealised case). The idea was based upon
a specific value assignment for the ``wrong events". 
However, the scheme presented by 
Popescu et al \cite{SHZ} has one drawback. The authors assumed in 
their analysis that the detecting scheme employed in the 
experiment should be able to 
distinguish between two and one photon detections. 
This was not the case in the actual experiments.
The aim of this chapter is to show that even this is unnecessary, all one 
needs 
is the use of the specific value assignment procedure of \cite{SHZ}.

Finally, we shall also give 
prediction of all effects occurring in the experiment.
It is quite often overlooked that a kind of Hong-Ou-Mandel dip phenomenon
\cite{HOM} can be observed in the experiment. 

\section{Description of the experiment}
In the class of experiments we consider
(see \cyt{figura1}) \cite{SHZ} 
a type I parametric down-conversion source  \cite{MANDEL2} 
is used to generate pairs of
photons which are degenerated in frequency and  polarisation
(say $\hat x$) but propagate in two different directions. 
One of the photons
passes through a wave plate ($WP$) which rotates its polarisation by 
$90^o$.
The two photons are then directed 
onto the two input ports of a (nonpolarising) ``$50-50$''
beamsplitter ($BS$). The observation stations
 are located in the exit beams of the beamsplitter. Each
local observer 
is equipped with a polarising beamsplitter\footnote{Following 
references \cite{GARUCCIO} and \cite{SHZ}
we assume that both local detection stations 
are equipped with polarising beamsplitters, and 
each of the output ports is observed by a detector. 
In the actual experiments \cite{SHIH} and \cite{MANDEL88} at each station 
only one of the 
outputs was monitored.}, orientated along an
arbitrary axis (which, in principle can be 
randomly chosen, in the delayed-choice manner, 
just before the photons are supposed to
arrive). Behind each polarising beamsplitter are two detectors, $D_1^+$,
$D_1^-$ and $D_2^+$, $D_2^-$ respectively, where the lower index indicates
the corresponding observer and the upper index the two exit ports of the
polarised beamsplitter ($``+"$ meaning parallel with the polarisation axis of
the beamsplitter and $``-"$ meaning orthogonal to this axis). All
optical paths are assumed to be equal.
\begin{figure}[htbp]
\begin{center}
\includegraphics[angle=270,width=14.5cm]{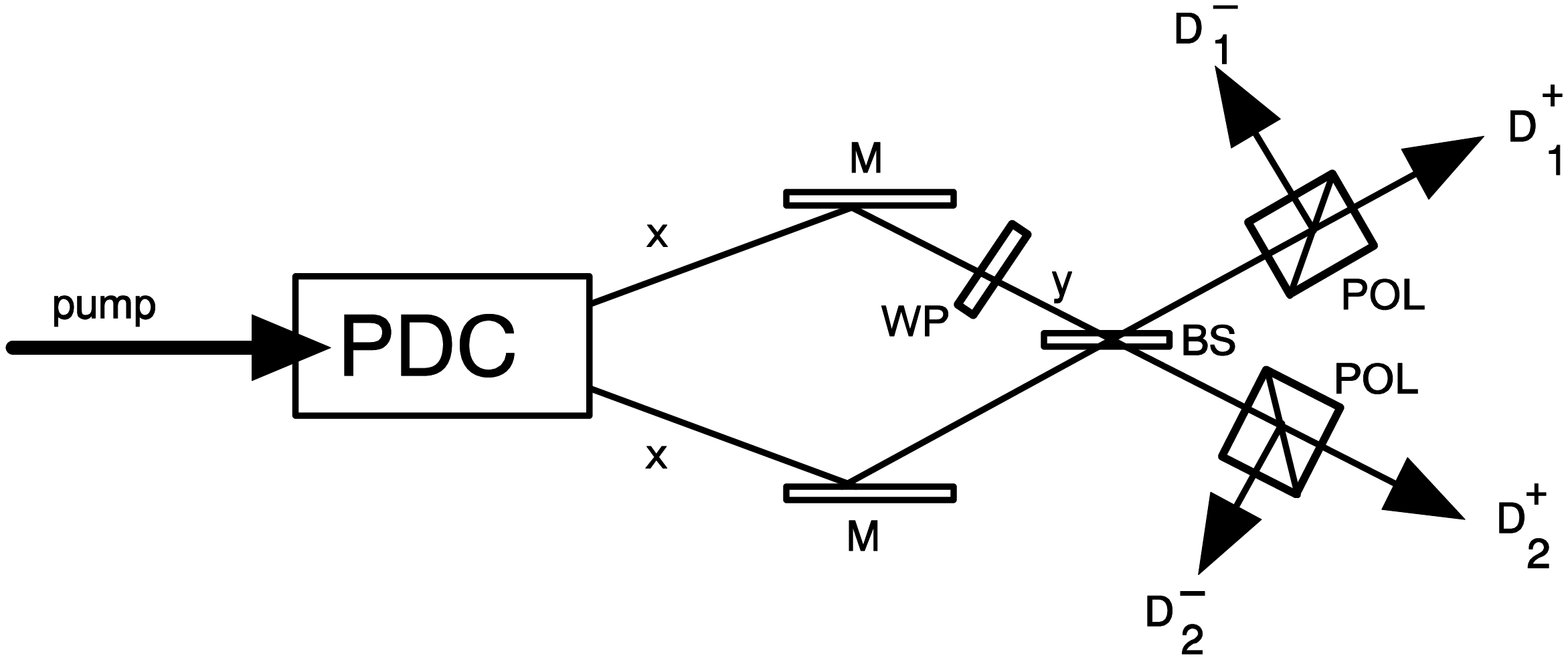}
\caption{Schematic of the setup. For explanations see the 
main text.\label{figura1}}  
\end{center} 
\end{figure}

\section{Quantum predictions}
Let us calculate the quantum predictions for the experiment.
We will use the second quantisation 
formalism, which is very convenient here, since the whole phenomenon observed
in the experiment rests upon the indistinguishability 
of photons. 

After the action of the wave-plate one can approximate quantum mechanical 
state describing two 
photons emerging from a non - linear crystal along the "signal" and the 
"idler"
beam by

\begin{equation}
|\Psi_{0}\rangle =a_{1\vec{x}}^{\dagger}a_{2\vec{y}}^{\dagger}|0\rangle,
\end{equation}
where $a_{1\vec{x}}^{\dagger}$ and $a_{2\vec{y}}^{\dagger}$ are creation
operators and $|0\rangle$ denotes the vacuum state. Subscripts 
$\vec{x},\vec{y}$
decode the  polarisation of the photon (either along $\vec{x}$ 
or $\vec{y}$ axis). The beamsplitter action can be described by

\begin{eqnarray}
&a_{1\vec{x}}^{\dagger}={1\over\sqrt{2}}(ic_{\vec{x}}^{\dagger}
+d_{\vec{x}}^{\dagger})
&\nonumber\\
&a_{2\vec{y}}^{\dagger}={1\over\sqrt{2}}(c_{\vec{y}}^{\dagger}+
id_{\vec{y}}^{\dagger}),&\\
\nonumber
\end{eqnarray}
where $c_{\vec{x}}^{\dagger},d_{\vec{x}}^{\dagger}
,c_{\vec{y}}^{\dagger},d_{\vec{y}}^{\dagger}$
are operators describing output modes of the beamsplitter ($c$ stands for
the first observer and $d$ for the second one). Thus
our state $|\Psi_{0}\rangle$ changes to :

\begin{equation}
|\Psi\rangle=
\frac{1}{2}(ic_{\vec{x}}^{\dagger}c_{\vec{y}}^{\dagger}-
c_{\vec{x}}^{\dagger}
d_{\vec{y}}^{\dagger}+c_{\vec{y}}^{\dagger}d_{\vec{x}}^{\dagger}+
id_{\vec{x}}^{\dagger}d_{\vec{y}}^{\dagger})|0\rangle
\end{equation}

Next comes the action of the polarisers in both beams, 
which can be described 
as

\begin{eqnarray}
&n_{\vec{x}}^{\dagger}=\cos(\theta_{1})n_{\parallel}^{\dagger}+
\sin(\theta_{1})n_{\perp}^{\dagger}& \nonumber\\
&
n_{\vec{y}}^{\dagger}=\sin(\theta_{1})n_{\parallel}^{\dagger}-
\cos(\theta_{1})n_{\perp}^{\dagger},& \nonumber\\
\end{eqnarray}
where $n^{\dagger}=c^{\dagger}$ or $d^{\dagger}$, and $n_{\parallel}^{\dagger}$ describes 
the mode parallel 
to
polarizer's axis  and $n_{\perp}^{\dagger}$ describes the mode perpendicular
to polarizer's axis; $\theta$ is the angle between the $\vec{x}$
axis and polarizer's axis. 
Thus the final state reaching the detector reads
\begin{eqnarray}
&|\psi_{final}\rangle=\frac{1}{2}\big[\sin(\theta_{1}-
\theta_{2})|c_{\parallel},
d_{\parallel}\rangle&\nonumber\\
&+\cos(\theta_{1}-\theta_{2})|c_{\parallel}
d_{\perp}\rangle&\nonumber\\
&
-\cos(\theta_{1}-\theta_{2})|c_{\perp},d_{\parallel}\rangle
+\sin(\theta_{1}-\theta_{2})|c_{\perp},d_{\perp}\rangle
&\nonumber\\
&+i{1\over{\sqrt{2}}}\sin(2\theta_{1})|2c_{\parallel}\rangle
+i{1\over{\sqrt{2}}}\sin(2\theta_{1})|2c_{\perp}\rangle
&\nonumber\\
&-i\cos(2\theta_{1})|c_{\perp},c_{\parallel}\rangle
+i{1\over{\sqrt{2}}}\sin(2\theta_{2})|2d_{\parallel}\rangle
&\nonumber\\
&+i{1\over{\sqrt{2}}}\sin(2\theta_{2})|2d_{\perp}\rangle
-i\cos(2\theta_{2})|d_{\parallel},d_{\perp}\rangle\big],&\nonumber\\
\label{final1}
\end{eqnarray}
where e.g. $|c_{\parallel},
d_{\parallel}\rangle$ denotes one photon in the mode $c_{\parallel}$, and
one in
$d_{\parallel}$, whereas $|2c_{\parallel}\rangle={1\over{\sqrt{2}}}
{c_{\parallel}^{\dagger}}^2|0\rangle$ denotes two photons
in the mode $c_{\parallel}$.

Let us denote by $P(i,\theta_{1};j,\theta_{2})$ the joint
probability for the outcome $i$ to be registered by observer 1 when
her 
polariser is oriented along the direction that makes an angle $\theta_{1}$
with the $\vec{x}$ direction and the outcome $j$ to be registered by
observer 2 when her polariser is oriented along the direction that
makes an angle $\theta_{2}$ with the $\vec{x}$ direction. Here
$i,j=1-6$ and have the following meaning \cite{SHZ}:

1=one photon in $D^{-}$, no photon in $D^{+}$

2=one photon in $D^{+}$, no photon in $D^{-}$

3=no photons

4=one photon in $D^{+}$ and one photon in $D^{-}$

5=two photons in $D^{+}$, no photon in $D^{-}$

6=two photons in $D^{-}$, no photons in $D^{+}$.

The quantum predictions for joint probabilities of those events
are given by:

\begin{eqnarray}
&P(1,\theta_{1};1,\theta_{2})=P(2,\theta_{1};2,\theta_{2})={1\over 8}
[1-\cos2(\theta_{1}-\theta_{2})],&\\
&P(2,\theta_{1};1,\theta_{2})=P(1,\theta_{1};2,\theta_{2})={1\over 8}
[1+\cos2(\theta_{1}-\theta_{2})],&\\
&P(5,\theta_{1};3,\theta_{2})=P(6,\theta_{1};3,\theta_{2})={1\over{8}}
\sin^2(2\theta_{1}),&\\
&P(3,\theta_{1};5,\theta_{2})=P(3,\theta_{1};6,\theta_{2})={1\over{8}}
\sin^2(2\theta_{2}),&\\
&P(4,\theta_{1};3,\theta_{2})={1\over{4}}\cos^2(2\theta_{1}),&\\
&P(3,\theta_{1};4,\theta_{2})={1\over{4}}\cos^2(2\theta_{2}).&
\end{eqnarray}
Following \cite{SHZ} we associate with each outcome registered 
by the observer 
1 and 2
a corresponding value $a_{i}$ and
$b_{j}$ respectively, where $a_{1}=b_{1}=-1$
while all the other values are equal to 1. Let us denote by
$E(\theta_{1},\theta_{2})$ the expectation value of their product

\begin{equation}
E(\theta_{1},\theta_{2})
=\sum_{i,j}a_{i}b_{j}P(i,\theta_{1};j,\theta_{2}).
\end{equation}
After simple calculations one has:

\begin{eqnarray}
&E(\psi_{1},\psi_{2})&\nonumber\\
&=-{1\over{2}}\cos(\psi_{1}+\psi_{2})+\frac{1}{2},&
\label{perfect}
\end{eqnarray}
where we have put $2\theta_{k}=(-1)^{k-1}\psi_{k}$.

The above formula for the correlation function is valid if one 
assumes that it is possible to
distinguish between single and double photon detection. This is 
usually not
the case. Thus it is convenient to have a parameter $\alpha$ that
measures the distinguishability of the double and single detection at
one detector ( $0\leq\alpha\leq1$, and gives
the probability of distinguishing by the employed detecting scheme 
of the double counts). The partial distinguishability blurs the 
distinction
between events 1 and 6 (2 and 5) and thus part of the events of
the type 6 are interpreted as of type 1 and are ascribed by the local
observer a wrong value, e.g. an event of type 6, if both photons
go to the $`` - "$ exit of the polariser, can be interpreted as a firing due 
to a single photon and is ascribed a $-1$ value. Please note that
such events like 1 or 2 in station 1 accompanied by 3 (no photon) at station 2
do not contribute to the correlation function because for any
$\alpha$ $P(1,\theta_{1};3,\theta_{2})=P(2,\theta_{1};3,\theta_{2})$.

If the parameter $\alpha$ is taken into account the correlation function 
acquires
the following form:

\begin{eqnarray}
&E(\psi_{1},\psi_{2}; \alpha)&\nonumber\\
&=-{1\over{2}}\cos(\psi_{1}+\psi_{2})
+{1\over{2}}\alpha&\nonumber\\
&+{1\over{4}}(1-\alpha)(\cos^2\psi_{1}+\cos^2\psi_{2})&
\label{quant}
\end{eqnarray}

\section{Conditions to violate local realism}
After the insertion of the  quantum correlation
function (\ref{quant}) into the CHSH
inequality, 

\begin{eqnarray}
&-2\leq E(\psi_{1},\psi_{2};\alpha)+E(\psi_{1}',\psi_{2};\alpha)& \nonumber\\
&+E(\psi_{1},\psi_{2}';\alpha)-E(\psi_{1}',\psi_{2}';\alpha))\leq2,& \nonumber
\end{eqnarray}
one obtains:
\begin{eqnarray}
&-2\leq -{1\over{2}}[\cos(\psi_{1}+\psi_{2})+\cos(\psi_{1}'+\psi_{2})
&\nonumber\\
&+\cos(\psi_{1}+\psi_{2}')-\cos(\psi_{1}'+\psi_{2}')]+
\alpha &\nonumber\\
&+{1\over{2}}(1-\alpha)(\cos^2\psi_{1}+
\cos^2\psi_{2})\leq 2.&
\end{eqnarray}

The interesting feature of this inequality is that it can be violated for all
values of $\alpha$. What is perhaps even more important, it can be 
robustly violated even when one is not able to distinguish between
single and double clicks at all ($\alpha=0$). The actual value of
the CHSH expression can reach in this 
case $2.33712$ (a numerical result), 
which is only slightly less than the maximal value
for $\alpha=1$, which is $\sqrt{2}+1\approx2.41421$. Therefore we
conclude that in the experiment {\it one can observe violations of
local realism even if one is not able to distinguish between
the double and single counts at one detector}. That is, the essential
feature of the method of \cite{SHZ} to reveal
violations of local realism in the experiment of this type is the
specific value assignment scheme and not the double-single photon
counts distinguishability. 

The specific angles at which the maximum violation of 
the CHSH inequality is achieved for $\alpha=0$ differ very much from those 
for $\alpha=1$ (for which the standard result is reproduced), 
and they read (in radians) $\psi_1=2.93798$,
$\psi_1'=4.25513$, $\psi_2=-0.20241$ and $\psi_2'=1.11708$.

Let us notice that with the setup of 
\cyt{figura1} one is able to observe effects of similar
nature to the famous Hong-Ou-Mandel dip \cite{HOM}. These are 
revealed by the probabilities pertaining to the wrong events (3.8-3.11).
Simply for certain orientations of the polarisers, if the two photons emerge 
on 
one side of the experiment only, then they must exit the 
polarising beamsplitter via a single output port
(this effect is due to the bosonic-type indistinguishability of 
photons, see \cite{HOM}). 

Finally let us discuss what is the critical efficiency of the detection of the 
experiments of this type.
To this end, in our calculations we will use a very simple model of imperfect 
detections: 
we insert a beamsplitter with reflectivity $\sqrt{1-\eta}$,
in front of an ideal detector, which observes only the transmitted light. This 
results in the
system behaving just like a detector of efficiency $\eta$. If we assume that 
the incoming light
is described by a creation operator $a^ \dagger$ then transmitted
mode is denoted as $t_{a}^ \dagger$ whereas reflected
mode is denoted as $r_{a}^ \dagger$ and one has
\begin{equation}
a^ \dagger=\sqrt{1-\eta}r_{a^ \dagger}^ \dagger + \sqrt{\eta}t_{a^ \dagger}^ 
\dagger.
\end{equation}
For instance, if one takes the following part of the state vector 
(\ref{final1}):
\begin{equation}
c_{||}^ \dagger d_{||}^ \dagger|0\rangle.
\end{equation}
the beamsplitter model of an imperfect
 detector transforms this term into:

\begin{eqnarray}
&[(1-\eta)r_{c_{||}}^ \dagger r_{d_{||}}^ \dagger +
\sqrt{\eta (1-\eta)}r_{c_{||}}^ \dagger t_{d_{||}}^ \dagger&\nonumber\\
& +
\sqrt{\eta (1-\eta)}t_{c_{||}}^ \dagger r_{d_{||}}^ \dagger +
\eta t_{c_{||}}^ \dagger t_{d_{||}}^ \dagger]|0\rangle.&\nonumber\\
\end{eqnarray}
The probabilities now read:

\begin{eqnarray}
&&P(3,\theta_{1};2,\theta_{2})=P(2,\theta_{1};3,\theta_{2}) \nonumber\\
&&P(1,\theta_{1};3,\theta_{2})=P(3,\theta_{1};1,\theta_{2})=
\eta (1-\eta)\\
&&P(1,\theta_{1};1,\theta_{2})=P(2,\theta_{1};2,\theta_{2})=
{1\over 4}\eta^2[\sin(\theta_{1}-\theta_{2})]^2\\
&&P(2,\theta_{1};1,\theta_{2})=P(1,\theta;2,\theta_{2})=
{1\over 4}\eta^2[\cos(\theta_{1}-\theta_{2})]^2\\
&&P(5,\theta_{1};3,\theta{2})=P(6,\theta_{1};3,\theta_{2})=
{1\over 8}\eta^2[\sin(2\theta_{1})]^2\\
&&P(3,\theta_{1};5,\theta{2})=P(3,\theta_{1};6,\theta_{2})=
{1\over 8}\eta^2[\sin(2\theta_{2})]^2\\
&&P(4,\theta_{1};3,\theta_{2})={1\over 4}\eta^2[\cos(2\theta_{1})]^2\\
&&P(3,\theta_{1};4,\theta_{2})={1\over 4}\eta^2[\cos(2\theta_{2})]^2
\end{eqnarray}

The correlation function, which includes the inefficiency of the detection 
reads

\begin{equation}
E(\psi_{1},\psi_{2};\eta,\alpha)=\eta^2E(\psi_{1},\psi_{2};\alpha)
+ (1-\eta)^2,
\end{equation}
where
$E(\psi_{1},\psi_{2};\alpha)$ is given by (\ref{quant}). 
We have tacitly assumed here that the parameters $\alpha$ and $\eta$
are independent of each other (this assumption may not hold for 
specific technical arrangements).
Putting this prediction into CHSH inequality, assuming that $\alpha=1$
(full distinguishability) we obtain a minimum
quantum efficiency needed for violation of local realism equal to $0.91$, 
whereas for other values of $\alpha$ we have: for $\alpha=0$ $\eta=0.926$;
for $\alpha=0.5$ $\eta=0.92$;
for $\alpha=0.75$ $\eta=0.92$;
for $\alpha=0.875$ $\eta=0.91$. One should note here that the method of value
assignment of \cite{SHZ} is in accordance with the method given by Garg and 
Mermin \cite{GM}
for the optimal estimation of required detector quantum efficiency to violate 
local realism
in a Bell-test. Thus the obtained efficiencies are indeed the lowest possible, 
and show that
experiments of this type are not good candidates for a "loophole-free" 
Bell-test \cite{SANTOS}, nevertheless due to the fact that the whole 
observable effect is a consequence of quantum principle of 
particle indistinguishability
such test are very interesting by themselves - they reveal the
entanglement inherently associated with this principle.
\section{Conclusions}
To conclude, we state that the possibility of distinguishing
between single and two photon detection events, usually not met in
the actual experiments, is not a necessary requirement for the proof
that the experiments of Shih-Alley and Ou-Mandel are, modulo fair
sampling assumption, valid tests of local realism. We also show that
some other interesting phenomena (involving bosonic type particle
indistinguishability) can be observed during such tests.
\chapter{Better entanglement swapping [2]}
\section{Introduction}
Until recent years it was commonly believed that
particles producing EPR-Bell phenomena have to originate from a single
source, or at least have to interact with each other.  However,
under very special conditions, by a suitable monitoring procedure of the
emissions of the independent sources one can {\it pre-select} an ensemble of
pairs of particles, which either reveal EPR-Bell correlations, or are in an
entangled state.  
The first explicit proposal to use two independent sources of particles in a
Bell test was given by Yurke and Stoler \cite{YS}. 
However, they did not discuss the importance of very specific
operational requirements necessary to implement such schemes in real
experiments. Such conditions were studied in \cite{EVENT} and 
\cite{ZZW}.

The method of entangling independently radiated photons, which share no
common past, \cite{EVENT} is essentially a pre-selection procedure. The
selected registration acts of the idler photons define the ensemble which
contains entangled signal photons (see next sections).  Surprisingly, such a
procedure enables one to realize the Bell's idea of "event-ready" detection.
This approach for many years was thought to be completely infeasible and thus
no research was being done in that direction \cite{CLAUSER78}.   This so-called 
{\it entanglement swapping} technique \cite{EVENT}, was
also  adopted to observe
experimental quantum states teleportation \cite{BOUW}.

The first entanglement swapping experiment was 
performed in 1998 \cite{Pan98}. High visibility  (around $65\%$) of two 
particle
interference fringes were observed on a pre-selected subset of photons
that never interacted.
This is very close to the usual threshold visibility of 
two particle fringes to violate some Bell inequalities, which 
is $70.7\%$. Therefore there exists a strong temptation for
breaking this limit, and in this way showing that the two particle fringes 
due to entanglement swapping have no local and realistic model. 

However, due to the spontaneous nature of the sources involved,
the initial condition for entanglement swapping cannot
be prepared. Simply the probability that the two sources 
would produce a pair of entangled states each is of the same order as the 
probability that
one of them produces two entangled pairs.
In the latter case no entanglement swapping results. Nevertheless,
such events can excite the trigger detectors (which in the 
case of the right initial condition select the antisymmetric 
Bell state of the two independent idlers). Therefore they are an unavoidable 
feature of the experiment, and have to be taken into account in 
any analysis of the possibility of finding a local realistic
description for the experiment.

The aim of this chapter is to perform such an analysis.
We shall show that if all firings of the trigger detectors are
accepted as pre-selecting the events for a Bell-type test\footnote{As was the
case in the actual experiment.},
one must necessarily, at least partially, be able to 
distinguish between two and single photon events
at the detectors observing the signals to enable demonstrations
of violations of local realism. Whereas, if one accepts additional 
selection at the trigger detectors, based on the polarisation 
of the idlers,  detectors possessing this ability are unnecessary.
We shall present our argumentation assuming that the reader
knows the methods and results of \cite{EVENT,ZZW,Pan98}.
The analysis will be confined to the gedanken situation
of perfect detection efficiency (the results can be easily generalised
to the non-ideal case).
\begin{figure}
\centerline{\psfig{width=5.0cm,angle=-90,file=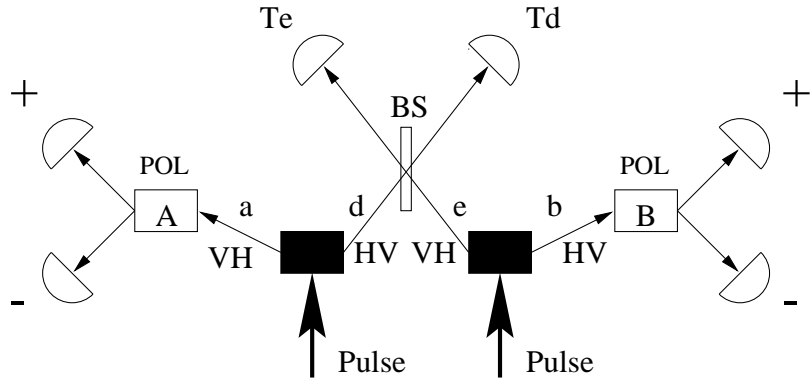}}
\vspace{2cm} 
\caption{
Entanglement swapping. Two type-II 
down conversion crystals are pumped
by a pulsed laser. The radiation from each of the crystals   
is entangled in polarisations (e.g. if one has an H polarised photon in 
mode $a$, then in mode $b$ is a V polarised photon). The idler photons
(in modes $d$
and $e$) are fed into a non-polarising beam splitter BS.
Simultaneous firing of the trigger detectors $T_e$ and $T_d$
pre-selects, in the event-ready way, a sub-ensemble of detection events
behind the polarising beam splitters $A$ and $B$.
The orientation of the 
polarising beam splitters can be set at will by the 
local observers. The output 
signal photons 
are registered by the two local detection stations, consisting of 
detectors denoted by $+$ and $-$.\label{forbidden} }

\end{figure}

\section{Description of the experiment}
Consider the set-up of \cyt{forbidden}, which is in 
principle the scheme used in the Innsbruck experiment
\cite{Pan98}.
Two pulsed type-II 
down conversion sources are 
emitting their radiation into the 
spatial propagation modes $a$ and $b$ (signals), $c$ and $d$ (idlers).
Due to
the statistical properties of the PDC radiation,
the initial state that is fed to the 
interferometric set-up has the following form:
\begin{eqnarray}
&|\psi\rangle=\sum_{n=0}^{\infty}({\gamma\over \sqrt2}
(a_{H}^{\dagger}d_{V}^{\dagger}+a_{V}^{\dagger}
d_{H}^{\dagger}))^{n}&\nonumber\\
&\times\sum_{m=0}^{\infty}({\gamma\over \sqrt2}
(e_{H}^{\dagger}b_{V}^{\dagger}+e_{V}^{\dagger}
b_{H}^{\dagger}))^{m}|0\rangle,&
\label{e1}
\end{eqnarray}
where, for instance, $a_{H}^{\dagger}$ denotes the creation operator of
the photon in beam $a$ having ``horizontal" polarisation.
As for the entanglement swapping to work one cannot have too excessive 
pump powers \cite{ZZHW},
the $\gamma$ coefficient can be assumed small.
Therefore we select only those  terms that are proportional 
to $\gamma^{2}$, as these are the lowest order terms
terms that can induce simultaneous firing of both trigger detectors.
They read
\begin{eqnarray}
&|\psi'\rangle={1\over 2}{\gamma}^{2}((a_{H}^{\dagger}d_{V}^{\dagger}+
d_{V}^{\dagger}d_{H}^{\dagger})(e_{H}^{\dagger}b_{V}^{\dagger}+
e_{V}^{\dagger}b_{H}^{\dagger})&\nonumber\\
&+(a_{H}^{\dagger}d_{V}^{\dagger}+a_{V}^{\dagger}d_{H}^{\dagger})^{2}+
(e_{H}^{\dagger}b_{V}^{\dagger}+
e_{V}^{\dagger}b_{H}^{\dagger})^{2})|0\rangle.&
\label{e2}
\end{eqnarray}
The factor 
$\frac{1}{2}\gamma^2$ simply gives the order of magnitude of 
the probability of the two trigger detectors to fire, and therefore we
drop it from further considerations. The action of the 
non-polarising beam splitter (BS) is described by
$d_{x}^{\dagger}={1\over \sqrt2}({\tilde d}_{x}^{\dagger}+
i{\tilde e}_{x}^{\dagger})$ and
$e_{x}^{\dagger}={1\over\sqrt2}({\tilde e}_{x}^{\dagger}+
i{\tilde d}_{x}^{\dagger})$
where $x=H$ or $x=V$, and $\tilde{e}$ and $\tilde{d}$
represent the modes monitored by the trigger detectors behind
the beam splitter.  
Taking into account only the terms in (\ref{e2}) that lead to clicks
at two trigger detectors we arrive at
\begin{eqnarray}
&|\psi''\rangle=(i(a_{H}^{\dagger 2}+b_{H}^{\dagger 2})
{\tilde e_{V}^{\dagger}}{\tilde d_{V}^{\dagger}}+
i(a_{V}^{\dagger 2}+b_{V}^{\dagger 2}){\tilde e_{H}^{\dagger}}
{\tilde d_{H}^{\dagger}}&\nonumber\\
&+i(a_{H}^{\dagger}a_{V}^{\dagger}+b_{H}^{\dagger}b_{V}^{\dagger})
({\tilde e_{V}}^{\dagger}{\tilde d_{H}}^{\dagger}+
{\tilde e_{H}}^{\dagger}{\tilde d_{V}}^{\dagger})&\nonumber\\
&{1\over2}(a_{H}^{\dagger}b_{V}^{\dagger}-a_{V}^{\dagger}
b_{H}^{\dagger})({\tilde e_{V}}^{\dagger}{\tilde d_{H}}^{\dagger}-
{\tilde e_{H}}^{\dagger}{\tilde d_{V}}^{\dagger}))|0\rangle.&
\label{e4}
\end{eqnarray}
It is convenient to normalise and rewrite the above state into the form:
\begin{eqnarray}
&|\psi_{N}\rangle={1\over\sqrt13}
\left[i\sqrt2({1\over\sqrt2}a_{H}^{\dagger 2}+
{1\over\sqrt2}b_{H}^{\dagger 2})
|VV\rangle\right.&\nonumber\\
&\left.+i\sqrt2({1\over\sqrt2}a_{V}^{\dagger 2}+
{1\over\sqrt2}b_{V}^{\dagger 2})|HH\rangle\right.&\nonumber\\
&\left.+(i(a_{H}^{\dagger}a_{V}^{\dagger}+
b_{H}^{\dagger}b_{V}^{\dagger})+
{1\over2}(a_{H}^{\dagger}b_{V}^{\dagger}-a_{V}^{\dagger}
b_{H}^{\dagger}))|VH\rangle\right.&\nonumber\\
&\left.+(i(a_{H}^{\dagger}a_{V}^{\dagger}+
b_{H}^{\dagger}b_{V}^{\dagger})-
{1\over2}(a_{H}^{\dagger}b_{V}^{\dagger}-a_{V}^{\dagger}
b_{H}^{\dagger})\right]|HV\rangle,&
\label{e4a}
\end{eqnarray}
where $|VV\rangle={\tilde 
e_{V}^{\dagger}}{\tilde d_{V}^{\dagger}}|0\rangle$, 
$|HH\rangle=
{\tilde e_{H}^{\dagger}}
{\tilde d_{H}^{\dagger}}|0\rangle$, 
$|VH\rangle=
{\tilde e_{V}}^{\dagger}{\tilde d_{H}}^{\dagger}|0\rangle$ and 
$|HV\rangle=
{\tilde e_{H}}^{\dagger}{\tilde d_{V}}^{\dagger}|0\rangle$.
We see  clearly that several processes may lead to 
the simultaneous firing of the trigger detectors (which observe 
the spatial modes) 
$\tilde{e}$ and $\tilde{d}$. 
The signal photons enter the polarising 
beam splitters.  Their action can be
described by the following relations
\begin{eqnarray}
&x_{V}^{\dagger}=\cos(\theta_{i})x_{+}^{\dagger}+
\sin(\theta_{i})x_{-}^{\dagger}&\nonumber\\
&x_{H}^{\dagger}=-\sin(\theta_{i})x_{+}^{\dagger}+
\cos(\theta_{i})x_{-}^{\dagger},&
\end{eqnarray}
with $x=a,b; i=1,2$ respectively and $+$, $-$ 
denoting the output spatial modes.

\section{Quantum predictions}
The probabilities of various
two-particle processes
that may occur at the spatially separated observation stations, under the
condition of both trigger detectors firing simultaneously, are given by:
\begin{eqnarray}
&P(1a_{+},1a_{-};0b_{+},0b_{-})=P(2a_{+},0a_{-};0b_{+},0b_{-})&\nonumber\\
&=P(0a_{+},2a_{-};0b_{+},0b_{-})=P(0a_{+},0a_{-};1b_{+},1b_{-})&\nonumber\\
&=P(0a_{+},0a_{-};2b_{+},0b_{-})=P(0a_{+},0a_{-};0b_{+},2b_{-})=
{2\over13},&\nonumber\\
&P(1a_{+},0a_{-};1b_{+},0b_{-})=P(0a_{+},1a_{-};0b_{+},1b_{-})
={1\over26}\left[\sin(\theta_{1}-\theta_{2})\right]^2,&\nonumber\\
&P(1a_{+},0a_{-};0b_{+},1b_{-})=P(0a_{+},1a_{-};1b_{+},0b_{-})
={1\over26}\left[\cos(\theta_{1}-\theta_{2})\right]^2,&
\label{e5}
\end{eqnarray}
where, for example, $P(0a_{+},0a_{-};2b_{+},0b_{-})$  denotes
the probability of observing two photons at the output $b_{+}$, and no 
photons in the other outputs.

The Bell correlation function 
for the product of the measurement results
on the signals at the two sides of the experiment
can be redefined in the way proposed in the previous chapter, i.e., 
all standard Bell-type events are assigned their usual values 
whereas all non-standard events are assigned the value of one.
I.e., if no photons are registered at one side, the local value of the 
measurement is one, if two photons are registered at one side again
the local measurement value
is one. The latter case includes both the event in which the two photons 
end-up at a single detector, as well as those when two detectors 
at the local station fire. 
Please note, that the experiment considered 
is a realization of Bell's idea of 
``event ready detectors'' (see e.g. \cite{CLAUSER78}). Therefore,
non-detection events are operationally well defined (as the simultaneous 
firing of the trigger detectors pre-selects the sub-ensemble
of time intervals in which one can expect the signal detectors to fire).

The above value assignment method, as it has been in the previous chapter,
works perfectly if one 
assumes that it is possible to
distinguish between single and double photon detection at a single 
detector. Therefore, again it is convenient to introduce the parameter
$\alpha$.

The partial distinguishability blurs the 
distinction
between events (at one side) in which 
there was one photon detected at say the output $\pm$, and 
events in which two photons entered a the detector observing output
$\pm$, but the detector failed to distinguish this event
from a single photon count.
In such a case the local event is sometimes ascribed by the local
observer a wrong value
namely $-1$ instead of $1$ (if both photons
go to the $`` - "$ exit of the polariser
and the devices fail to inform the experimenter 
that it is a two photon event, this is interpreted as a firing due 
to a single photon and is ascribed a $-1$ value). 
Please note, that if one includes less than perfect detection efficiency
of the detectors this problem is more frequent and more involved
(we shall not study this aspect here).

\section{Conditions to violate local realism}
Under such a value assignment the correlation function reads:
\begin{eqnarray}
&E_{\alpha}(\theta_{1},\theta_{2})=-{1\over 13}\cos(2\theta_{1}-2\theta_{2})+
{4\over 13}(1+2\alpha),&
\end{eqnarray}
where $\alpha$ is the numerical value of the distinguishability.
When we put into the standard CHSH inequality this correlation
function it violates the standard bound of 2, only if the distinguishability
satisfies $\alpha\geq{9-\sqrt2\over8}\approx0.948$. 
Such values are definitely
beyond the current technological limits. As the efficiency
of real detectors makes this problem even more acute, one has to
propose a modification of the experiment that gets rid of this problem.

\section{Proposal of modification of the experiment}
Therefore,
in front of the idler detector $Te$ we propose to put 
polarising beam splitter that transmit only vertical polarisation
whereas in front of the idler detector $Td$ one that transmits only
horizontal polarisation. This further reduces the 
relevant terms in our state, i.e. those that can
induce firing of the trigger detectors, to the following 
ones:

\begin{eqnarray}
&|\psi\rangle=\sqrt{2\over5}\left(i(a_{H}^{\dagger}a_{V}^{\dagger}+
b_{V}^{\dagger}b_{H}^{\dagger})+
{1\over2}(a_{H}^{\dagger}b_{V}^{\dagger}-
a_{V}^{\dagger}b_{H}^{\dagger})\right){\tilde e}_{V}^{\dagger}
{\tilde d}_{H}^{\dagger}|0\rangle.&
\label{e5a}
\end{eqnarray}
Again we have normalised the above state.

Using the above formula we can calculate the probabilities of all possible
processes in this interferometric set-up, conditional on
firings of the two trigger detectors:

\begin{eqnarray}
&P(1a_{+},1a_{-};0b_{+},0b_{-})={2\over5}\left[\cos(2\theta_{1})\right]^2,&\nonumber\\
&P(2a_{+},0a_{-};0b_{+},0b_{-})=P(0a_{+},2a_{-};0b_{+},0b_{-})
={1\over5}\left[\sin(2\theta_{1})\right]^2,&\nonumber\\
&P(0a_{+},0a_{-};1b_{+},1b_{-})={2\over5}\cos(2\theta_{2})^2,&\nonumber\\
&P(0a_{+},0a_{-};2b_{+},0b_{-})=P(0a_{+},0a_{-};0b_{+},2b_{-})
={1\over5}\left[\sin(2\theta_{2})\right]^2,&\nonumber\\
&P(1a_{+},0a_{-};1b_{+},0b_{-})
=P(0a_{+},1a_{-};0b_{+},1b_{-})={1\over10}
\left[\sin(\theta_{1}-\theta_{2})\right]^2,&
\nonumber\\
&P(1a_{+},0a_{-};0b_{+},1b_{-})=P(0a_{+},1a_{-};1b_{+},0b_{-})
={1\over10}\left[\cos(\theta_{1}-\theta_{2})\right]^2.&
\label{e7}
\end{eqnarray}

Under the earlier defined value assignment the correlation function
for the current version of the experiment reads:
\begin{eqnarray}
&E_{\alpha}(\theta_{1},\theta_{2})=-{1\over5}\cos(2\theta_{1}-2\theta_{2})&
\nonumber\\
&+{2\over5}(1-\alpha)\left[(\cos2\theta_{1})^{2}+(\cos2\theta_{2})^2\right]
+{4\over5}\alpha.&
\end{eqnarray}
When such a correlation functions are inserted into the CHSH inequality
one has:
\begin{eqnarray}
&-2\leq-{1\over5}[\cos2(\theta_{1}-\theta_{2})+\cos2(\theta_{1}-\theta_{2}')&
\nonumber\\
&+\cos2(\theta_{1}'-\theta_{2})-\cos2(\theta_{1}'-\theta_{2}')]&\nonumber\\
&+{4\over5}(1-\alpha)\left[(\cos2\theta_{1})^2+(\cos2\theta_{2})^2\right]+
{8\over5}\alpha\leq 2.&
\label{CHSH}
\end{eqnarray}
Please note that some of the terms of the correlation function 
which depend only on one local angle cancel upon insertion into CHSH
inequality.

For $\alpha=1$ (perfect distinguishability) the 
middle expression in (\ref{CHSH}) reaches
2.16569, i.e. we have a clear violation 
of the local realistic bound.
This maximal violation occurs at angles (in radians)
$2\theta_{1}=-1.30278,$ $2\theta_{1}'=-2.87435,$ $2\theta_{2}=1.05326,$
$2\theta_{2}'=2.62386$.
What is more interesting, for $\alpha=0$, i.e. for a complete lack of
distinguishability
between two and single photon events at one detector, the
expression in (\ref{CHSH})
reaches a value which is not much lower, namely 
2.11453. This can be reached for the orientation angles 
$2\theta_{1}=0.0837317,$ $2\theta_{1}'=-1.0749,$ 
$2\theta_{2}=3.05769,$ $2\theta_{2}'=4.21568$). 

\section{Conclusions}
Therefore we conclude
that the proposed modification of the entanglement swapping experiment, 
despite the unwanted additional events
due to the impossibility of controlling the spontaneous emissions
at the two separate sources, makes it possible to consider it
as test of Bell inequalities. The standard configuration can serve
as a test of local realism only under the condition of
extremely high distinguishability between two and single photon counts.

Finally let us mention that the proposed
modification in the configuration
enables one to observe,  in the event ready mode,
a bosonic interference effect similar to the 
Hong-Ou-Mandel dip \cite{HOM}. It is described by the first four formulas of
(\ref{e7}). E.g. if $\theta_1=\pi/4$, 
no coincidences between firings of the two detectors
of the station $a$
are allowed. All two photon events at this station
are, under this setting, double counts at a single detector.
Thus, we have two very interesting non-classical 
phenomena in one experiment.



\part{New generalised Bell inequalities and GHZ paradoxes for qu$N$its}
\newpage
~~~
\thispagestyle{empty}
\newpage

\chapter*{}
\begin{center}
\Large{**}
\end{center}
\phantom{~~~}

In the first chapter of this part the approach to the Bell theorem
employing Bell inequalities is generalised to GHZ correlations
for which each local observer is allowed to use more then
two settings of his or her measuring apparatus.

In the second chapter the functional Bell inequality is derived.
Although the functional inequalities for two qubits are of less practical
importance they seem to be the first step towards
an analytic search for the critical visibility of two-qubit
sinusoidal interference fringes which violate local realism.

In the third chapter we derive the series of GHZ paradoxes for
$N$ and $N+1$ maximally entangled 
qu$N$its observed via unbiased multiport
beamsplitters.

\chapter{Wringing out better Bell inequalities for GHZ experiment [3]}
\section{Introduction}
Greenberger-Horne-Zeilinger correlations \cite{GHZ89} lead to a
strikingly more direct refutation of the argument of Einstein Podolsky and
Rosen (EPR), on the possibility of introducing elements of reality to
complete quantum mechanics \cite{EPR}, than considerations involving only
pairs of qubits.  The EPR ideas are based on the observation that for some
systems quantum mechanics predicts perfect correlations of their properties.
However, in the case of three or more qubits, in the entangled GHZ state,
such correlations cannot be consistently used to infer at a distance hidden
properties of the qubits. In contradistinction to the original two
qubit Bell theorem, the idea of EPR, to turn the exact predictions of
quantum mechanics against the claim of its completeness, breaks down already
at the stage of defining the elements of reality.

The reasoning of GHZ involved perfectly correlated qubit systems.
However, the actual data collected in a real laboratory would reveal less
than perfect correlations, and the imperfections of the qubit collection
systems would leave many of the potential events undetected.  Therefore the
original GHZ reasoning cannot be ever tested in the laboratory, and one is
forced to make some modifications (already, e.g., in
\cite{GHSZ}).

To face these difficulties several $M$ qubit Bell inequalities appeared in
the literature \cite{MERMIN90,ROY91,ARDEHALI92,BIELINSKI93,ZUK93}.
All these works show that quantum predictions for GHZ states violate these
inequalities by an amount that grows exponentially with $M$.  The increasing
number of qubits, in this case, does not bring us closer to the classical
realm, but rather makes the discrepancies between the quantum and the
classical more profound.

The study of three or more qubit 
interference effects does not seem to be a good route towards a loophole free
test of the hypothesis of local hidden variables. However, 
to interpret the results of such experiments\footnote{First observation
of GHZ correlations has been already reported \cite{INNSBRUCK}.} one
should know the
borderline between the quantum and the classical (i.e., local realism).
According to current literature (with the exception of \cite{ZUK93}) we enter
the non-classical territory when the fringes in a $M$ qubit interference
experiment have visibilities higher than ${2}^{\frac{1}{2}(1-M)}$. The
principal aim of this chapter is to show that, if one allows each of the local
observers to have {\it three} measurements to choose from (instead of the
usual {\it two}), the actual threshold is lower (for $M>3$). 

\section{Geometrical method of finding Bell inequalities}
Let us first explain the method that will be used in the next 
two sections 
(it is called a geometrical method)
\cite{ZUK93} on an example of two maximally
entangled qubits. The method will be presented in the case
of stochastic local hidden variables. Application to deterministic 
case is straightforward \cite{ZUK93}.

We consider the state \cyt{singlet} on which
two spatially separated observers $a,b$
measure dichotomic observables $O_{n}(\phi_{n})$ with eigenvalues $\pm 1$
referring to eigenstates 
\be
&&|\pm,\phi_{n}\rangle_{n}=\frac{1}{\sqrt{2}}\left(|0\rangle_{n}
\pm e^{(i\phi_{n})}|1\rangle_{n}\right)
\ee
controlled by knobs (local settings) $\phi_{n}$ ($n=a,b$).

Let us further assume that in the experiment observer $a$ chooses between,
say, $N_a$ settings of the local apparatus denoted by $\phi^{i}_{a}$, 
$i=1,2,\dots,N_a$, and the observer $b$ chooses between
$N_b$ settings denoted by $\phi_{b}^{j}$ $j=1,2,\dots,N_b$. The quantum 
prediction for 
observer $a$ to obtain the result $m=\pm 1$, and observer $b$ to obtain
$m'=\pm 1$ is equal to
\begin{eqnarray}
&P_{QM}^{V_2}(m,m'|\phi_{a}^{i},\phi_{b}^{j})={1\over 4}
\left(1-V_2mm'\cos(\phi_{a}^{i}+\phi_{b}^{j})\right),&
\label{quantum}
\end{eqnarray}
where $V_2$ takes into account the possible less than perfect
visibility (see the discussion of the notion of visibility in Introduction
- equation \cyt{noise0} and the discussion below).

The question is if this set of probabilities is reproducible
by local hidden variables, i.e., by
\begin{eqnarray}
&P_{HV}(m,m'|\phi_{a}^{i},\phi_{b}^{j})=
\int_{\Lambda}d\lambda\rho(\lambda)P_{a}(m|\lambda,\phi_{a}^{i})
P_{b}(m'|\lambda,\phi_{b}^{j})&
\label{hidden}
\end{eqnarray}
(see also \cyt{stochLHV}).

The set of quantum probabilities \cyt{quantum} as well as
the set of local hidden variable probabilities \cyt{hidden}
can be treated as the components of $4\times N_a\times N_b$ dimensional
real vectors $\hat{P}_{QM}^{V_2}$ and $\hat{P}_{HV}$ with components
\be
&&P_{QM}^{V_2}(mm';ij)={1\over 4}\left(1-V_2mm'\cos(\phi_{a}^{i}
+\phi_{b}^{j})\right)\nonumber\\
&&P_{HV}(mm';ij)=\int_{\Lambda}d\lambda\rho(\lambda)
P_{a}(m|\lambda,\phi_{a}^{i})P_{b}(m'|\lambda,\phi_{b}^{j})
\ee
where $i,j,m,m'$ enumerate the components of the vectors.
Both $\hat{P}_{QM}^{V_2}$ and $\hat{P}_{HV}$ can be considered as
vectors belonging to a real 
Hilbert space with
the following scalar product
\begin{eqnarray}
&(\hat{F}|\hat{G})=\sum_{m=-1}^{1}\sum_{m'=-1}^{1}\sum_{i=1}^{N_a}\sum_{j=1}^{N_b}
F(mm';ij)G(mm';ij),&
\end{eqnarray}
where $\hat{F}$ and $\hat{G}$ are arbitrary vectors from
this space.

In every real Hilbert space any two vectors $\hat{F},\hat{G}$
are equal, i.e. $\hat{F}=\hat{G}$, if and only if
\be
&&(\hat{F}|\hat{G})=||\hat{F}||^2=
||\hat{G}||^2.
\label{F=G}
\ee
Thus, if one has two vectors $\hat{F},\hat{G}$ and one knows
the norm of, say, the vector $\hat{F}$, 
and $(\hat{F}|\hat{G})<||\hat{F}||^2$ then $\hat{F}\neq\hat{G}$.
This simple
observation is especially useful for us because we can always calculate
the norm of the vector $\hat{P}_{QM}^{V_2}$, which is not possible for the
vector $\hat{P}_{HV}$. However, as we will see next, one can
estimate the scalar product of $\hat{P}_{QM}^{V_2}$ and $\hat{P}_{HV}$.
If $(\hat{P}_{QM}^{V_2}|\hat{P}_{HV})<||\hat{P}_{QM}^{V_2}||^2$ one has the Bell 
inequality.

As the simplest example let us show how one can obtain within this approach
the standard threshold value of the visibility (obtained using
the CHSH inequality) for the situation
in which both observers have two local settings to choose from. 
To this end let us consider the case in which $N_a=N_b=2$
and let us choose $\phi_{a}^{1}=0$, $\phi_{a}^{2}={\pi\over 2}$,
and $\phi_{b}^{1}=-{\pi\over 4}$, $\phi_{b}^{2}=+{\pi\over 4}$. 
For such local settings one has
\be
&&P^{V_2}_{QM}(m,m';1,1)={1\over 4}(1-mm'{V_2\over\sqrt 2})\nonumber\\
&&P^{V_2}_{QM}(m,m';1,2)={1\over 4}(1-mm'{V_2\over\sqrt 2})\nonumber\\
&&P^{V_2}_{QM}(m,m';2,1)={1\over 4}(1-mm'{V_2\over\sqrt 2})\nonumber\\
&&P^{V_2}_{QM}(m,m';2,2)={1\over 4}(1+mm'{V_2\over\sqrt 2})
\label{chochlik}
\ee
and
the norm of $||\hat{P}_{QM}^{V_2}||^2$ equals 
$1+{1\over 2}V_2^2$. The next step is to write down
explicitly the scalar product $(\hat{P}_{HV}|\hat{P}_{QM}^{V_2})$
\begin{eqnarray}
&&(\hat{P}_{HV}|\hat{P}_{QM}^{V_2})=\int d\lambda\rho(\lambda)[
1-{1\over 4\sqrt2}V_2\{I_{a}(\phi_{a}^{1},\lambda)
I_{b}(\phi_{b}^{1},\lambda)+
I_{a}(\phi_{a}^{1},\lambda)I_{b}(\phi_{b}^{2},\lambda)\nonumber\\
&&+
I_{a}(\phi_{a}^{2},\lambda)I_{b}(\phi_{b}^{1},\lambda)-
I_{a}(\phi_{a}^{2},\lambda)I_{b}(\phi_{b}^{2},\lambda)\}],
\label{A6}
\end{eqnarray}
where, for instance, 
$I_{a}(\phi_{a}^{2})=\sum_{m=-1}^{1}mP_{a}(m|\lambda,\phi_{a}^{2})$.
We see that $|I_{a}(\phi_{a}^{2})|\leq 1$ and this immediately 
implies that the modulus of expression in the curly brackets 
never exceeds $2$.
Thus,
\begin{eqnarray}
&&(\hat{P}_{HV}|\hat{P}_{QM}^{V_2})\leq 1+V_2
{\sqrt2\over 4}.
\label{12}
\end{eqnarray}
Comparing \cyt{12} with the norm of $||\hat{P}_{QM}^{V_2}||^2$ one obtains
that the necessary condition for $\hat{P}_{HV}$ to be equal to
$\hat{P}^{V_2}_{QM}$ is $V_2\leq{1\over\sqrt 2}$.

\section{Quantum and local realistic description of the ge\-dan\-ken experiment}
We consider a source emitting $M$ qubits each of
which propagates towards one of $M$ spatially separated measuring devices.
The generic form of a GHZ $M$ qubit state is 
\begin{equation}
|\Psi(M)\rangle=\frac{1}{{\sqrt{2}}}(|0\rangle_{1}\dots|0\rangle_{M}+
|1\rangle_{
1}\dots
|1\rangle_{M}).
\label{kukurydza1}
\end{equation}
Let as assume that the operation of each of the measuring apparatus is
controlled by a knob which sets a parameter $\phi_{l}$, and the $l$-th
apparatus measures a dichotomic observable $O_{l}(\phi_{l})$ with two
eigenvalues $\pm 1$ and the eigenstates defined by
$|\pm,\phi_{l}\rangle_{l}=\frac{1}{\sqrt{2}}\left(|0\rangle_{l}\pm
e^{(i\phi_{l})}|1\rangle_{l}\right).$ The quantum prediction for obtaining
specific results at the $M$ measurement stations (for the idealised, perfect,
experiment) reads
\begin{eqnarray}
&P_{QM}^{(M)}(r_{1},r_{2},\dots,r_{M}|\phi_{1},\dots,\phi_{M})
&\nonumber\\
&={1\over 2^M}\left[1+\prod_{l=1}^{M}r_{l}
\cos\left(\sum_{k=1}^{M}\phi_{k}\right)\right]&,
\label{kukurydza4}
\end{eqnarray}
($r_{l}$ equal to $-1$ or $+1$).
The GHZ correlation function is defined as
\begin{eqnarray}
&E^{(M)}(\phi_{1},\dots,\phi_{M})&\nonumber\\
&=\sum_{r_{1},r_{2},\dots,r_{M}=-1}^{1}\prod_{l=1}^{M}r_{l}P^{(M)}(r_{1},\dots,
r_{M}|\phi_{1},\dots,\phi_{M}),&
\label{kukurydza5}
\end{eqnarray}
and in the case of quantum mechanics, i.e. for $P^{(M)}=P_{QM}^{(M)}$, it
reads $ E_{QM}^{(M)}=\cos(\sum_{l=1}^{M}\phi_{l}).  $
\par

Local realism implies the following structure of probabilities
of specific results (compare with \cyt{stochLHV})
\begin{eqnarray}
&P_{HV}^{(M)}(r_{1},\dots,r_{M}|\phi_{1},\dots,\phi_{M})&\nonumber\\
&=\int_{\Lambda}d\lambda\rho(\lambda)\prod_{l=1}^{M}P_{l}(r_{l}|\lambda,
\phi_{l})&
\label{kukurydza7}
\end{eqnarray}
where $P_{l}(r_{l}|\lambda,\phi_{l})$ is the probability to obtain result
$r_{l}$ in the $l$-th apparatus under the condition that the hidden state is
$\lambda$ and the macroscopic variable defining the locally
measured observable is set to the value $\phi_{l}$\footnote{
We present our results only for the case of deterministic
local hidden variable theories, i.e., $P_l(r_l|\lambda,\phi_l)=1,0$, 
but generalisation to
stochastic ones is obvious.}. The 
locality
of this description is guaranteed by independence of $P_{l}$ on $\phi_{i}$
for all $i\neq l$.

\section{Derivation of Bell inequalities via the geometrical method}
We shall now derive a series of inequalities for the $M$ qubit GHZ
processes based on the already mentioned geometric method
\cite{ZUK93}. 

We assume that each of the $M$ spatially separated observers has {\it three}
measurements to choose from.  The local phases that they are allowed to set
are $\phi_{1}^{1}=\pi/6,
\phi_{2}^{1}=\pi/2,\phi_{3}^{1}=5\pi/6$ 
(for the first observer) and for all the other $M-1$ observers they are
$\phi_{1}^{i}=0, \phi_{2}^{i}=\pi/3,\phi^{i}_{3}=2\pi/3$,
($i=2,\dots,M$).

Out of the quantum predictions for the $M$ qubit correlation function at
these settings one can construct a matrix endowed with $M$ indices
\begin{eqnarray}
&E_{QM}^{(M)}(\phi_{i_{1}}^{1},\dots\phi_{i_{M}}^{M})=\cos\left(\sum_{k=1}^{M}
\phi_{i_{k}}^{k}\right)
=Q_{i_{1},\dots,i_{M}}^{(M)}
\label{kukurydza8}
\end{eqnarray}
($i_{k}=1,2,3$).

All that we know about local hidden variable theories is that their
predictions (for the same set of settings as above) must have the following
form:
\begin{eqnarray}
&E_{HV}^{(M)}(\phi_{i_{1}}^{1},\dots\phi_{i_{M}}^{M})& \nonumber\\
&=\int
d\lambda\rho(\lambda)\prod_{k=1}^{M}I_{k}(\lambda
,\phi_{i_{k}}^{k})=H_{i_{1},\dots,i_{M}}^{(M)},&
\label{kukurydza10}
\end{eqnarray}
where
\begin{eqnarray}
&I_{k}(\lambda,\phi_{i_{k}}^{k})=
\sum_{r_{k}}r_{k}P_{k}(r_{k}|\lambda,\phi_{i_{k}}^{k}).&
\label{kukurydza11}
\end{eqnarray}
Of course, in the case of a {\it deterministic} \LHV\ theory
$I_{k}(\lambda,\phi)=\pm1$.  $H^{(M)}$ is our test matrix.
Please note, that all one knows about $H^{(M)}$ is its structure.

The scalar product of two real matrices is defined by
\begin{equation}
(H^{(M)},Q^{(M)})=\sum_{i_{1},\dots,i_{M}}H_{i_{1},\dots,i_{M}}^{(M)}
Q_{i_{1},\dots,i_{M}}^{(M)}.
\label{kukurydza12}
\end{equation} 
Our aim is to show the incompatibility of the local hidden variable
description with the quantum prediction.  To this end, we shall show that,
for two or more qubits,
\begin{eqnarray} 
&({Q}^{(M)},{H}^{(M)})&\nonumber\\ &\leq
2^{M-1}\sqrt{3}<||{Q}^{(M)}||^2=\frac{3^M}{2}&.
\label{rw1} 
\end{eqnarray} 

First, we show that $||{Q}^{(M)}||^2=3^M/2$.  This can be reached in
the following way:
\begin{eqnarray}
&||{Q}^{(M)}||^2=\sum_{i_{1},\dots,i_{M}}
\cos^2\left(\sum_{k=1}^{M}\phi_{i_{k}}^{k}\right)&\nonumber\\
&=\frac{1}{2}\sum_{i_{1},\dots,i_{M}}
\left[1+\cos\left(2i\sum_{k=1}^{M}\phi_{i_{k}}^{k}\right)\right]
&\nonumber\\ &=Re\left\{\sum_{i_{1},\dots,i_{M}}
\left[1+\exp\left(2\sum_{k=1}^{M}\phi_{i_{k}}^{k}\right)\right]\right\}&
\nonumber\\
&=3^M/2+Re\left(\prod^{M}_{k=1}
\sum_{i_{k}=1}^{3}\exp(2i\phi_{i_{k}}^{k})\right),&
\end{eqnarray}
where $Re$ denotes the real part.  Since
$\sum_{l=1}^{3}e^{i(l-1)(2/3)\pi}=0$, the last term vanishes.

The scalar product $(H^{(M)},Q^{(M)})$ is bounded from above by the maximal
possible value of
\begin{equation}
S^{(M)}_{\lambda}=\sum_{i_{1},\dots,i_{M}}\left[
\cos\left(\sum^{M}_{k=1}\phi_{i_{k}}^{k}\right)
\prod_{l=1}^{M}I_{l}(\lambda, \phi_{i_{l}}^{l})\right],
\label{kukurydza13}
\end{equation}
and for $M\geq2$
\begin{equation}
S^{(M)}_{\lambda}\leq {2^{M-1}\sqrt{3}}.
\label{S}
\end{equation}

To show (\ref{S}), let us first notice that 
\begin{eqnarray} 
S_{\lambda}^{(M)}=Re
\left[\prod_{k=1}^{M}\sum_{i_{k}=1}^{3}I_{k}(\lambda|\phi_{i_{
k}}^{k})\exp(i\phi_{i_{k}}^{k}))\right].
\label{rw8} 
\end{eqnarray}
For $k=2,\dots,M$, one has $e^{i\phi_{l}^{k}}=e^{i[(l-1)/3]\pi}$ whereas
for $k=1$, $e^{i\phi_{l}^{1}}=e^{i(\pi/6)}e^{i[(l-1)/3]\pi}$.
Thus, since $I(\lambda|\cdot)=\pm 1$, the possible values for
\begin{equation}
z_{1}^{\lambda}=\sum_{i_{1}=1}^{3}I_{1}(\lambda|\phi_{i_{1}}^{1})
\exp(i\phi_{i_{1}}^{1})
\label{rw11}
\end{equation}
are $0$, $\pm 2e^{i\pi/2}$, $\pm 2e^{-i(\pi/6)} $, or
finally $\pm 2e^{i(\pi/6)} $, whereas for $k=2,\dots,M$ the possible
values of
\begin{equation} 
z_{k}^{\lambda}=\sum_{i=1}^{3}I_{k}(\lambda|\phi_{i_{k}}^{k})
\exp(i\phi_{i_{k}}^{k})
\label{rw12} 
\end{equation} 
have their complex phases shifted by $\pi/6$ with respect to the
previous set; i.e., they are $0$, $\pm 2e^{i(2\pi/3)}$, 
$\pm 2$, or finally $\pm 2e^{i(\pi/3)}$.  Since
$|z_{1}^{\lambda}\prod_{k=2}^{M}z_{k}^{\lambda} |\leq 2^{M}$ and the minimal
possible overall complex phase (modulo $2\pi$) of 
$z_{1}^{\lambda}\prod_{k=2}^{M}z_{k}^{\lambda}$ is 
$\pi/6$, one has $Re(z_{1}^{\lambda}
\prod_{k=2}^{M}z_{k}^{\lambda})\leq 2^{M}\cos(\pi/6).$
Thus inequalities (\ref{S}) and (\ref{rw1}) hold.

The left inequality of (\ref{rw1}) is a Bell inequality for the $M$
qubit
experiment. If one replaces ${H}^{(M)}$ by the quantum prediction 
${Q}^{(M)}$
(compare (\ref{kukurydza8}))
the inequality is violated since
\begin{equation}
({Q^{(M)}},{Q^{(M)}})=\frac{3^{M}}{2}>2^{M-1}\sqrt{3},
\label{row1}
\end{equation}
i.e., (\ref{rw1}) is violated by the factor
$(3/2)^M/\sqrt3$ (compare \cite{MERMIN90}).

\subsection{Critical visibility and quantum efficiency of detectors}
The magnitude of violation of a Bell inequality is not a parameter which
can objectively define to what extent local realism is violated. 
It is rather the visibility of the
$M$ qubit interference fringes which can be directly observed.  Further,
the significance of all Bell-type experiments depends on the efficiency of
the collection of the qubits. Below a certain threshold value for this
parameter experiments cannot be considered as tests of local realism. They
may confirm the quantum predictions but are not falsifications of the
hypothesis of local hidden variables. Therefore we will search for the
critical minimal visibility of $M$ qubit fringes and collection efficiency, which
do not allow anymore a local realistic model.

In a real experiment (under the assumption that quantum mechanics gives
idealised, but correct predictions), the visibility of the $M$ qubit
fringes, $V_2(M)$, would certainly be less than 1. Also the probability of
registering all potential events $\eta_2(M)$ 
would be reduced by the overall collection
efficiency. If one assumes that all $M$ local apparata have the same
collection efficiency $\eta$, 
and takes into account that these operate
independently of each other\footnote{The parameter $\eta$ describes
here the efficiency of a single detector. The assumption of 
the independency of detectors gives $\eta_2(M)=\eta^M$.}, 
one can model the expected experimental results
by
\begin{eqnarray}
&P_{expt}^{(M)}(r_{1},\dots,r_{M}|\phi^{1},\dots,\phi^{M})&\nonumber\\
&=\eta^{M}{\left(\frac{1}{2}\right)^{M}}\left(1+V_2(M)\prod_{l=1}^{M}r_{l}
\cos\sum_{k=1}^{M}\phi
^{k}\right)&.
\label{24}
\end{eqnarray}

The full set of events at a given measuring station consists now of the
results $+1$ and $-1$, when we succeed to measure the dichotomic observable,
and a non-detection event (which is, in principle observable, if one uses
event-ready state preparation \cite{YS}) for which one can introduce the
value $0$. The local realistic description requires that the 
probabilities of
the possible events should be given by
\begin{eqnarray}
&P_{expt}^{HV}(m_{1},\dots,m_{M}|\phi^{1},\dots,\phi^{M})&\nonumber\\ &=\int
d\lambda\rho(\lambda)\prod_{k=1}^{M}P_{k}(m_{k}|\lambda,
\phi^{k})&,
\label{row4}
\end{eqnarray}
with $m_{i}=+1, -1$ or $0$. The local hidden
variable correlation function for the experimental results (at the chosen
settings) is now given by
\begin{equation}
{E_{expt}^{HV}}_{i_{1},\dots,i_{M}}=\int
d\lambda\rho(\lambda)\prod_{k=1}^{M}I_{k}'(
\lambda, \phi_{i_{k}}^{k}),
\label{row8}
\end{equation}
with
\begin{equation}
I_{k}'(\lambda,\phi_{i_{k}}^{k})=\sum_{m_{k}=-1,0,+1}m_{k}
P_{k}^{HV}(m_{k}|\lambda
,\phi_{i_{k}}^{k}).
\label{row9}
\end{equation}
For deterministic models one has now $I_{k}'(\lambda,\phi_{i_{k}}^{k})
=1,0,-1$.

One can impose several symmetries on ${P_{expt}^{HV}}$. These symmetries are
satisfied by the quantum prediction (\ref{24}), and we can expect them to
be satisfied in real experiments, within experimental error. The one that we
impose here is that:
\begin{quote}
For all sets of results, $\{m_{1},\dots,m_{M}\}$, that have equal number of
zeros (one zero or more) the probability $P_{expt}^{HV}(m_{1},\dots,m_{M})$ 
has
the same value, and this value is independent of the settings of the local
parameters $\{\phi_{i_{1}}^{1},\dots,\phi_{i_{M}}^{M}\}$.
\end{quote}
One can define a function $f_{M}(m)$ which for $m=+1,-1,0$ has the following
values: $f(\pm 1)=\pm 1$, $f(0)=-1$ (compare \cite{GM}) and introduce
auxiliary correlation function
\begin{eqnarray}
&\tilde{E}_{i_{1},\dots,i_{N}} =\int
d\lambda\rho(\lambda)\sum_{m_{1},\dots,m_{N}=-1,0,+1}&\nonumber\\
&\times\prod_{k=1}^{M}[f(m_{k})P_{k}(m_{k}|\lambda,
\phi_{i_{k}}^{k})]={\tilde{H}}^{(M)}_{i_{1},\dots,i_{M}}&.
\label{row3}
\end{eqnarray}
Since, due to the symmetry conditions, one has, e.g., $\sum_{m_{2}=1,-1}
f(m_{2})P_{expt}^{HV}(0,m_{2},\dots,m_{M})=0$, the
following relation results:
\begin{equation}
\tilde{E}_{i_{1},\dots,i_{M}}={E^{HV}_{expt}}_{i_{1},\dots,i_{M}}+[f(0)]^
{M} P(0,\dots,0),
\label{row6}
\end{equation}
where
$P(0,\dots,0)$
is the probability that all detectors would fail to register qubits,
and under our assumptions
it is independent of the settings, and equals $(1-\eta)^M$.

The auxiliary correlation function must satisfy the original inequality
(\ref{rw1}); i.e., one has
\begin{equation}
(Q^{(M)},{\tilde{H}}^{(M)})\leq{2^{M-1}}\sqrt{3}.
\label{row10}
\end{equation}
However, this implies that
\begin{eqnarray}
&-{2^{M-1}}\sqrt{3}-f(0)^MP(0,\dots,0)q_{(M)}&\nonumber\\ &\leq
(Q^{(M)},E_{expt}^{HV})&\nonumber\\
&\leq{2^{M-1}}\sqrt{3}-f(0)^MP(0,\dots,0)q_{(M)},
\label{row11}
\end{eqnarray}
where
\begin{equation}
q_{(M)}=\sum_{i_{1},\dots,i_{M}}Q_{i_{1},\dots,i_{M}}^{(M)}.
\label{row12}
\end{equation}
Therefore, since if $x$ is a possible value for $(Q^{(M)},E_{expt}^{HV})$ 
then so
is $-x$, one has
\begin{eqnarray}
&|({Q^{(M)}},E_{expt}^{HV})|\nonumber\\
&\leq{2^{M-1}}\sqrt{3}-P(0,\dots,0)|q_{(M)}|.
\label{a}
\end{eqnarray}
Thus, we have obtained Bell inequalities of a form which is more suitable
for the analysis of the experimental data.

The prediction (\ref{24}) leads to the following correlation function
\begin{equation}
E_{expt}^{QM}=\eta^{M}V_2(M)E^{QM},
\label{row13}
\end{equation}
which, when put into (\ref{a}) in the place of $E_{expt}^{HV}$, gives the 
following relation between the threshold visibility, $V^{tr}_2(M)$,
and the threshold collection efficiency, $\eta^{tr}$, for the $M$-qubit
experiment:
\begin{equation}
{\eta^{tr}}^M\frac{3^M}{2}
V^{tr}_2(M)={2^{M-1}}\sqrt{3}-|q_{(M)}|(1-\eta^{tr})^M.
\label{c}
\end{equation}

The value of the expression $q_{(M)}$ can be found in the following way:
\begin{eqnarray}
&q_{(M)}=\sum_{i_{1},\dots,i_{M}}
\cos\left(\sum_{k=1}^{M}\phi_{i_{k}}\right)&\nonumber\\
&=Re\left(\sum_{i_{1},\dots,i_{M}}\prod_{k=1}^{M}\exp(i\phi_{i_{k}}^{k})
\right)&\nonumber\\
&=Re\left(\prod_{k=1}^{M}\sum_{i_{k}}\exp(i\phi_{i_{k}}^{k}\right)&
\nonumber\\
&=Re\left[2^{M}i\exp\left(i(M-1)
\frac{\pi}{3}\right)\right]=-2^{M}\sin\left((M-1)\frac{\pi}{3}\right).&
\end{eqnarray}
\section{Results}
The threshold value of the visibility of the multi-qubit fringes decreases
now faster than in the earlier approaches \cite{MERMIN90}. For perfect
collection efficiency, ($\eta=1$), it has the lowest value, which is
\begin{equation}
V^{tr}_2(M)={\sqrt{3}}\left(\frac{2}{3}\right)^{M},
\end{equation}
and, if $M\geq4$, it is lower than $(\frac{1}{\sqrt{2}})^{M-1}$.  The
specific values for several qubits are $V^{tr}_2(2)=76.9\%$,
$V^{tr}_2(3)=51.3\%$, $V^{tr}_2(4)=34.2\%$, $V^{tr}_2(5)=22.8\%$ and 
$V^{tr}_2(10)=3\%$ (see also figure \cyt{newV}),
whereas the standard methods lead to $V^{old}_2(2)=70.7\%$,
$V^{old}_2(3)=50.0\%$, $V^{old}_2(4)=35.4\%$, $V^{old}_2(5)= 25.0\%$ and
$V^{old}_2(10)=4.4\%$.  

The threshold efficiency of the qubit collection also decreases with growing
$M$ (see figure \cyt{newE}), and for perfect visibilities it reads $\eta^{tr}(2)=87.0\%$,
$\eta^{tr}(3)=79.8\%$, $\eta^{tr}(4)=76.5\%$, $\eta^{tr}(5)=74.4\%$ (here the number in the
brackets indicates the number of entangled particles $M$). The gain over
the inequalities \cite{MERMIN90} is in this respect very small, and begins
again at $M=4$. However, for very big $M$ the critical efficiency is close
to $\frac{2}{3}$ (compared with $\frac{1}{\sqrt{2}}$ for \cite{MERMIN90}). 
\begin{center}
\begin{figure}[htbp]
\includegraphics[width=16cm, height=16cm, angle=270]{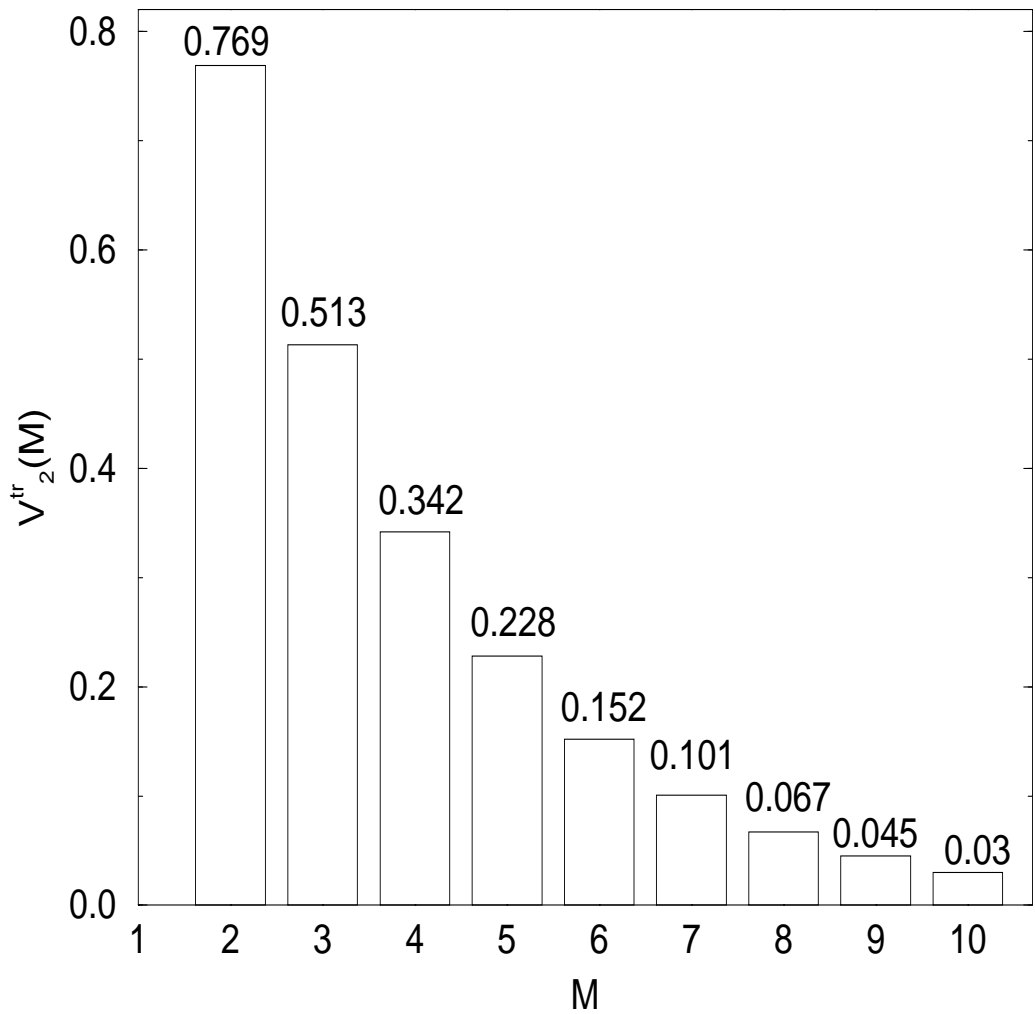}
\caption{The values of critical visibility $V_2^{tr}(M)$ 
versus the number of qubits $M$.\label{newV}}
\end{figure}
\end{center}
\begin{figure}[htbp]
\begin{center}
\includegraphics[width=16cm, height=16cm, angle=270]{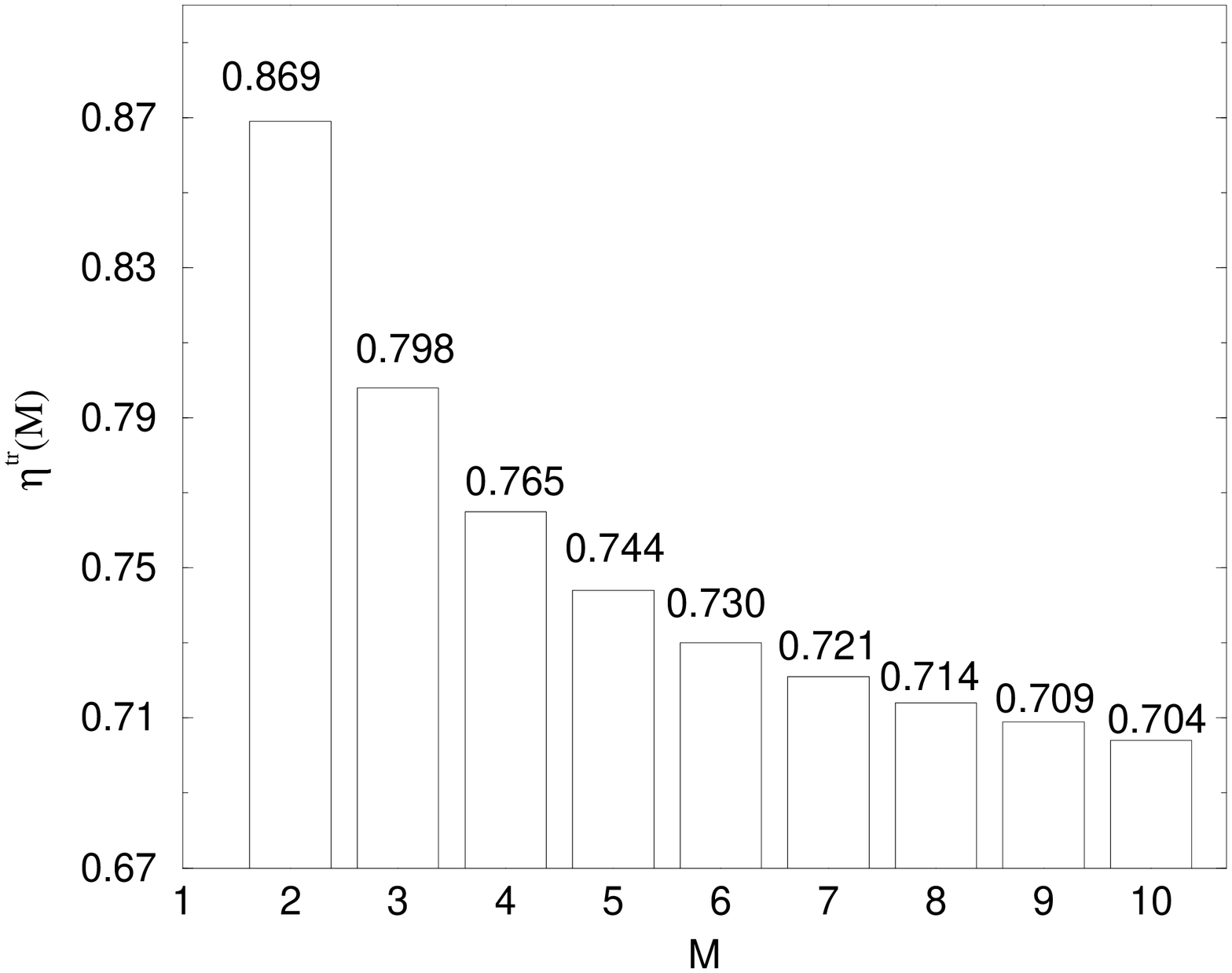}
\caption{The values of critical efficiency $\eta^{tr}(M)$ 
versus the number of qubits $M$.\label{newE}}
\end{center}

\end{figure}

\section{Conclusions}
We conclude that for the original GHZ problem (four
qubits) one should rather aim at making experiments which allow for three
settings at each local observation station.  Surprisingly, the measurements
should not be performed for the values for which we have perfect GHZ-EPR
correlations (i.e the values for which the correlation function equals to
$\pm1$).


\chapter{Bell inequality for all possible local settings [4]}
\section{Introduction}
The Bell theorem is usually formulated with the help of the Clauser-Horne
\cite{CH} or the CHSH inequality \cite{CHSH69}. These 
inequalities are satisfied
by any
local realistic theory and are violated by quantum mechanical predictions.
They involve two apparatus settings at each of the two sides of the experiment.
However, generalisation to more than two settings at each side are possible
\cite{GARUCCIO80,BR,ZUK93,GISIN99}.

There are several motivations for such generalisations. First of all, 
new Bell inequalities 
may be more appropriate in some experimental situations, e.g., the chained 
Bell inequalities
can reveal violation of local realism for the Franson type experiment 
\cite{AERTS}.
Also, the academic question, why only two settings at each side, is
that always necessary, is interesting in itself. Further, many of
the currently performed quantum interferometric Bell tests did not
involve stabilisation of the interferometers at specified settings
optimal for the standard Bell inequalities, but rather involved
sample scans of the entire interferometric patterns. Thus it is useful
to have inequalities that are {\em directly} applicable to such data.

Here we present a Bell-type inequality that involves all possible settings
of the local measuring apparatus for a pair of two qubits, which is
always
equivalent to two spin ${1\over 2}$ particles.
The method applied is a development of the one given in \cite{ZUK93}.
However, here we do not restrict ourselves to pairs of coplanar settings
(in the meaning appropriate for two Stern-Gerlach apparatuses).

Our method has two characteristic traits. The first one is that it
indeed involves the entire range of the measurement parameters. 
By this, e.g., it distinguishes itself from the limits of infinitely many
settings at each side of the so-called chained inequalities \cite{BR}, in
which not every {\em pair} of possible settings is utilised. The second one
is that the method involves the quantum prediction from the very beginning. 
As we shall see the quantum prediction determines the structure of our
Bell inequality.
\section{Quantum mechanical and local realistic description of the
gedanken experiment}
As usual one has a source emitting two qubits
each of which propagates towards one of two spatially separated
observers $a$ and $b$. The qubits are described by the maximally entangled
state \cyt{singlet}.

Let as assume that every observer has a Stern-Gerlach
apparatus, which measures the observable $\vec{n}\cdot\vec{\sigma}$, 
where $n=a,b$,
$\vec{n}$ is a unit vector representing direction at which observer 
$n$ makes a measurement
and $\vec{\sigma}$ is a vector the components of which are standard Pauli
matrices. 
The family of observables $\vec{n}\cdot\vec{\sigma}$ 
covers all possible dichotomic observables for a two qubit system, endowed
with a spectrum consisting of $\pm 1$. 

In each run of the experiment every observer obtains one of the two possible
results of measurement, $\pm 1$. The probability of
obtaining by the observer $a$ the result $m=\pm 1$, when 
measuring the observable
$\vec{\sigma}\cdot\vec{a}$, and the result 
$m'=\pm 1$
by the observer $b$, when measuring the observable 
$\vec{\sigma}\cdot\vec{b}$ is equal to
\begin{eqnarray}
&P_{QM}^{V_2}(m,m'|\vec{a},\vec{b})={1\over4}(1-V_2mm'\vec{a}\cdot\vec{b}).&
\label{predictions}
\end{eqnarray} 
where $0\leq V_2 \leq 1$ stands for the visibility.

The structure of \LHV\ gives
\begin{eqnarray}
&P_{HV}(m,m'|\vec{a},\vec{b})=\int_{\Lambda}d\lambda\rho(\lambda)P_{a}(m
|\lambda,\vec{a})P_{b}(m'|\lambda,\vec{b}),
\label{hiddenprob}
\end{eqnarray}
with the standard meaning of the used symbols (see \cyt{stochLHV}).
\section{Derivation of the in\-equal\-ity via the ge\-o\-met\-ri\-cal method}
To apply the geometrical method we must define appropriate
Hilbert space. Because we deal with functions 
$P_{QM}^{V_2}(m,m'|\theta_{a},\phi_{a},\theta_{b},\phi_{b})$ and 
$P_{HV}(m,m'|\theta_{a},\phi_{a},\theta_{b},\phi_{b})$
that 
depend on discrete numbers $m,m'$
and continuous variables $\theta_{n},\phi_{n}$, where
$\vec{n}=(\sin\theta_{n}\cos\phi_{n},\sin\theta_{n}
\sin\phi_{n},\cos\theta_{n})$ it is convenient to define 
the scalar product of certain two real functions $f$ and $g$ as
\begin{eqnarray}
&&\langle f| g\rangle
=\sum_{m=-1}^{1}
\sum_{m'=-1}^{1}\nonumber\\
&&\int d\Omega_{a}
\int d\Omega_{b}
f(m,m';\theta_{a},\phi_{a},\theta_{b},\phi_{b})
g(m,m';\theta_{a},\phi_{a},\theta_{b},\phi_{b}),
\label{prodscal}
\end{eqnarray}
where $d\Omega_{n}=\sin\theta_{n}d\theta_{n}d\phi_{n}$ 
is the rotationally invariant measure
on the sphere of radius one. Our {\it known} vector is $P_{QM}^{V_2}$, whereas
the {\it test} one is $P_{HV}$ (compare with the section describing
geometrical method).

One has
\begin{eqnarray}
&&||P_{QM}^{V_2}||^{2}=\langle P_{QM}^{V_2}|P_{QM}^{V_2}\rangle\nonumber\\
&&=(2\pi)^{2}+V_2^2{{4\pi}^{2}\over3}.
\label{qnorm}
\end{eqnarray} 
To estimate the scalar product $\langle P_{QM}^{V_2}|P_{HV}\rangle$ one has to
use the specific structure of probabilities that are
described by local hidden variables (LHV) 
(\ref{hiddenprob}). Since $P_{HV}$ is a weighted average 
over the hidden parameters one can make the following estimate
\begin{eqnarray}
&&\langle P_{QM}^{V_2}|P_{HV}\rangle\leq 
\max_{\lambda\in\Lambda}\left [\sum_{m,m'=-1}^{1}
\int d\Omega_{a}\int d\Omega_{b}
P_{a}(m|\lambda,\vec{a})\right.\nonumber\\
&&\left.\times P_{b}(m'|\lambda,\vec{b})
{1\over4}(1-mm'V_2\vec{a}\cdot\vec{b})\right].
\label{estmean}
\end{eqnarray}
Since 
\begin{eqnarray}
&\sum_{m=-1}^{1}
P_{a}(m|\lambda,\vec{a})=
\sum_{m'=-1}^{1}P_{b}(
m'|\lambda,\vec{b})=1,&
\label{8}
\end{eqnarray}
the first term of (\ref{estmean}) satisfies
\begin{eqnarray}
&&{1\over4}\sum_{m,m'=-1}\int
d\Omega_{a}
\int d\Omega_{b}P_{a}(m|\lambda,\vec{a})
P_{b}(m'|\lambda,\vec{b})\nonumber\\
&&=(2\pi)^{2}.
\label{firstexp}  
\end{eqnarray}  
We transform the other term of (\ref{estmean}) to a more convenient form
\begin{eqnarray}
&&{1\over4}\sum_{m,m'=-1}^{1}\int
d\Omega_{a}\int d\Omega_{b}mm'
P_a(m|\lambda,\vec{a})
P_b(m'|\lambda,\vec{b})V\vec{a}\cdot\vec{b}\nonumber\\
&&={1\over4}\int
d\Omega_{a}\int
d\Omega_{b}
I_{a}(\vec{a},\lambda)
I_{b}(\vec{b}
,\lambda)V_2\vec{a}\cdot\vec{b},
\label{secondexp}
\end{eqnarray}
where 
\begin{eqnarray}
&I_{n}(\vec{n},\lambda)
=\sum_{m=-1}^{1}mP_{n}(m|\lambda,\vec{n}),&
\label{11}
\end{eqnarray}
and 
one has $|I_{n}(\vec{n},\lambda)|\leq1$ ($n=a,b$). 

The scalar product of two three dimensional vectors $\vec{a}$ and
$\vec{b}$ that appears in (\ref{secondexp}) can be written 
as $\vec{a}\cdot\vec{b}=
\sum_{k=1}^{3}a_{k}(\theta_{a},\phi_{a})
b_{k}(\theta_{b},\phi_{b})$,
where
\begin{eqnarray}
&\vec{n}=(n_{1},n_{2},n_{3})&\nonumber\\
&=(\sin\theta_{n}\cos\phi_{n},\sin\theta_{n}
\sin\phi_{n},\cos\theta_{n}).&
\end{eqnarray}
Therefore
(\ref{secondexp}) reads
\begin{eqnarray}
&&{V_2\over4}\sum_{k=1}^{3}\int
d\Omega_{a}I_{a}(\theta_{a},\phi_{a},\lambda) 
a_{k}(\theta_{a},\phi_{a})\nonumber\\
&&\times\int d\Omega_{b}
I_{b}(\theta_{b},\phi_{b},\lambda)
b_{k}(\theta_{b},\phi_{b}).
\label{tutaj}
\end{eqnarray}
We notice here that our expression is a sum of three terms,
each of which is a product of two integrals.
 
The functions in (\ref{tutaj}) are square integrable, i.e.
integrals 
\be
&&\int d\Omega_{n}|I_{n}(\theta_{n},\phi_{n},\lambda)|^2
\ee
and
\be
&&\int d\Omega_{n}|n_{k}(\theta_{n},\phi_{n})|^2
\ee
exist (we remind that 
$|I_{n}(\theta_{n},\phi_{n},\lambda)|\leq 1$ which guarantees the
existence of the first integral). This allows us to use formalism of
Hilbert space of square integrable functions on the unit sphere, which
we denote as $L^{2}(S^3)$.

The functions $n_{k}(\theta_{n},\phi_{n})$ fulfil the orthogonality relation
$\int d\Omega_{n}n_{k}(\theta_{n},\phi_{n})n_{l}(\theta_{n},\phi_{n})=
{4\pi\over 3}\delta_{kl}$. 
Thus, if we normalise $n_{k}$ (i.e. we divide them
by their norm, which is $\sqrt{4\pi\over3}$) we can interpret the integral
$\alpha_{k}^{n}(\lambda)=\sqrt{3\over4\pi}\int d\Omega_{n}
I_{n}(\theta_{n},\phi_{n},\lambda)n_{k}(\theta_{n}
,\phi_{n})$ as
a k-th coefficient of the {\it projection} of 
$I_{n}(\theta_{n},\phi_{n},\lambda)$ into a three dimensional 
subspace of $L^2(S^{3})$ spanned by the (normalised) basis functions 
$\sqrt{3\over4\pi}n_{k}(\theta_{n},\phi_{n})$ ($k=1,2,3$).
For later reference we will call this space $\Sigma(3)$. 
Therefore (\ref{secondexp}) transforms into
\begin{eqnarray}
&&V_2{\pi\over3}\sum_{k=1}^{3}\alpha^{a}_{k}(\lambda)
\alpha^{b}_{k}(\lambda).
\label{transformed}
\end{eqnarray}
Denoting the projection of $I_{n}(\theta_{n},\phi_{n},\lambda)$ into 
$\Sigma(3)$ by 
$I_{n}^{||}(\theta_{n},\phi_{n},\lambda)$ and 
using the Schwartz inequality we arrive at
\begin{eqnarray}
{\pi\over3}\sum_{k=1}^{3}\alpha_{k}^{a}(\lambda)
\alpha_{k}^{a}(\lambda)\leq {\pi\over3}||I_{a}^{||}(\cdot,\lambda)||
||I_{b}^{||}(\cdot,\lambda)||.
\label{almost}
\end{eqnarray} 

Therefore, our last step is to calculate the maximal possible value of
the norm $||I_{n}^{||}(\cdot,\lambda)||$. Since the length (norm) of a projection of a 
vector into a certain subspace
is equal to the maximal value of its scalar product with any normalised
vector belonging to this subspace,
 the norm $||I_{n}^{||}(\cdot,\lambda)||$ is given by
\begin{eqnarray}
&||I_{n}^{||}(\cdot,\lambda)||=\max_{|\vec{c}|=1}[\sqrt{3\over4\pi}\int 
d\Omega_{n}
I_{n}(\theta_{n},\phi_{n},\lambda)\sum_{k=1}^{3}c_{k}
n_{k}(\theta_{n},
\phi_{n})],
\label{mistake}
\end{eqnarray}
where $\vec{c}=(c_{1},c_{2},c_{3})$ and $|\vec{c}|^2=\sum_{k=1}^{3}c_{k}^2=1$. 
 Because 
$|I_{n}(\vec{a},\lambda)|\leq 1$ one has
\begin{eqnarray}
&||I_{n}^{||}(\cdot,\lambda)||\leq \max_{|\vec{c}|=1}[\sqrt{3\over4\pi}\int 
d\Omega_{n}
|\sum_{k=1}^{3}c_{k}n_{k}(\theta_{n},
\phi_{n})|].
\label{par}
\end{eqnarray}

Every vector
$\vec{c}$ can be obtained by a certain rotation of the versor $\vec{z}$.
Such a rotation is represented by an
orthogonal matrix $\hat{O}$ belonging to the rotation group $SO(3)$. 
Therefore, (\ref{par}) can be
rewritten as
\begin{eqnarray}
&||I_{n}^{||}(\cdot,\lambda)||\leq \max_{\hat{O}}[\sqrt{3\over4\pi}\int 
d\Omega_{n}
|\hat{O}\vec{z}\cdot\vec{n}(\theta_{n},\phi_{n})|],
\label{integral1}
\end{eqnarray}
where the maximum is taken over all possible rotation matrices $\hat{O}$. Since
$|O\hat{z}\cdot\vec{n}(\theta_{n},\phi_{n})|$ is the modulus of the scalar product of 
two ordinary three dimensional vectors, it is equal to 
$|\vec{z}\cdot\hat{O}^{-1}\vec{n}(\theta_{n},\phi_{n})|$.
An active rotation of the vector $\vec{n}$ is equivalent to a (passive) 
change of the spherical coordinates. Utilising the fact that the measure 
$d\Omega_{n}$ is rotationally invariant we see that 
\begin{eqnarray}
&&||I_{n}^{||}||\leq\int d\Omega_{n}
|\sqrt{3\over4\pi}\cos\theta_{n}|=2\pi\sqrt{3\over4\pi}.
\end{eqnarray}

Therefore (\ref{almost}) is not greater then ${1\over4}(2\pi)^2$, which
with (\ref{qnorm}) and (\ref{firstexp}) gives us the following
inequalities
\begin{eqnarray}
||P_{QM}^{V_2}||^{2}=(2\pi)^{2}+{V_2^2\over 3}(2\pi)^{2}>(2\pi)^{2}+ 
{V_2\over 4}(2\pi)^{2}\geq \langle P_{QM}^{V_2}|P_{HV}\rangle.
\label{damian}
\end{eqnarray}
\section{Results}
The inequality \cyt{damian} is violated by quantum predictions provided that
the visibility $V_2$ is higher then $75\%$. Please notice that
the right hand inequality is a form of a "functional" Bell inequality.
It simply gives the upper bound for the value of a certain functional
defined on the local realistic probability functions $P_{HV}$. The left
hand inequality shows that the insertion of $P_{QM}^{V_2}$ into the
functional Bell inequality leads to its violation provided $V_2>0.75$.
The characteristic trait of our functional Bell inequality
is that its form is defined by the quantum prediction $P_{QM}^{V_2}$.
\section{Conclusions}
The threshold visibility for two qubit
interference to violate the inequality (\ref{damian}) is lower than
in the case of coplanar settings \cite{ZUK93}, for which the critical 
visibility is
${8\over \pi^2}$. Also, it is lower than the one given recently
by Gisin \cite{GISIN99}. For
his inequalities involving arbitrary many settings the threshold 
visibility equals $V_2={\pi\over 4}$. The chained inequalities \cite{GARUCCIO80}, 
\cite{BR},
for evenly spaced settings, with the number of settings going to infinity,
have the property that the critical visibility approaches 1 in the limit 
of infinitely many settings. 

Taking into account the fact that the necessary condition for the existence
of a local hidden variable model for two local settings at each side of
the experiment implied by the CHSH inequality is $V_2\leq {1\over\sqrt 2}$,
the critical visibility of any hidden variable model
that aims at reproducing quantum correlations for the gedanken experiment
described above for any number of local settings (infinite or not) cannot 
be greater than 
${1\over\sqrt 2}$. Therefore, the numerical value of the critical visibility
obtained by the above inequality is overestimated. This is due to the fact
that this inequality is only a necessary condition for the existence
of \LHV . However, the inequality \cyt{damian} may be the first step towards finding
the threshold visibility of the local hidden variable model
reproducing quantum mechanical predictions for all positions of
measuring apparatus.
The presented inequality also solves the academic problem of
finding a Bell inequality that involves all possible settings
of the local apparata.

In \cite{ZUK93} \.Zukowski has shown that for GHZ states 
involving four or more
particles (and employing all coplanar settings) the functional
Bell inequality approach leads to much more stringent conditions
on the critical visibilities than the approaches of Mermin \cite{MERMIN90} and
Ardehali \cite{ARDEHALI92}. However, the extension of the presented approach
to GHZ states does not give any improvement with respect to the
standard one. This is due to the fact that original GHZ paradox is
obtainable for co-planar settings only.


\chapter{Greenberger-Horne-Zeilinger paradoxes for qu$N$its [5,6]}
\section{Introduction}
The Greenberger-Horne-Zeilinger correlations, discovered in 1989,
started a new chapter in the research related with entanglement. To
a great extent this discovery was responsible for the sudden renewal 
in the interest in this field, both in theory and experiment.
All these developments finally led to the first actual observation
of three qubit GHZ correlations in 1999 \cite{INNSBRUCK}, 
and as a by product, since
the experimental techniques involved were of the same kind, 
to the famous teleportation experiment \cite{BOUW}.

In this chapter we would like to examine whether GHZ-type paradoxes exist
also in the case of correlations expected in gedanken experiments
involving multiport beam splitters \cite{KLYSHKO88,ZBGHZ93,ZZHBG94}, 
i.e. for a specific
case of non dichotomic observables (which have properties distinctive
to the dichotomic ones \cite{GLEASON57,Bell66,KOCHEN67}). 
To this end, we shall study a
GHZ-Bell type experiment in which one has a source emitting
$M$ qu$N$its in a specific  entangled state of the property, that the
qu$N$its propagate towards one of $M$ spatially separated
non conventional measuring devices operated by independent observers.
Each of the devices consists of an unbiased symmetric multiport beam splitter
\cite{ZZH} (with $N$ input and $N$ exit ports), $N$ phase shifters
operated by the observers (one in front of each input), and $N$
detectors (one behind each exit port).

\section{Unbiased multiport beamsplitters}
An unbiased symmetric $2N$-port beam splitter is defined as an $N$-input and
$N$-output interferometric device which has the property that a beam
of light entering via single port is evenly split between all output
ports (see \cyt{port-rys1}). I.e., the unitary matrix defining such a device has the
property that
the modulus of all its elements equals ${1\over \sqrt N}$. 
\begin{figure}[htbp]
\begin{center}
\includegraphics[width=5cm, height=10cm, angle=270]{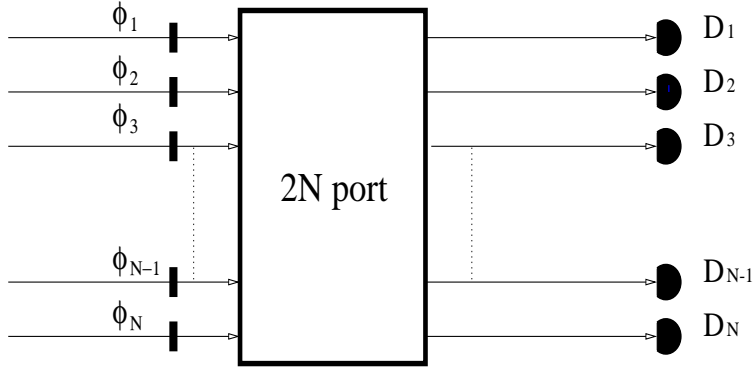}
\caption{The schematic picture of an $2N$ port. There are $N$ phases
in front of each input port and $N$ detectors behind each output ports.\label{port-rys1}}
\end{center}

\end{figure}
An extended
introduction to the physics and theory of such devices is given in
\cite{ZZH}, and therefore the reader not acknowledged with those
concepts is kindly asked to consult this reference. Multiport
beam splitters were introduced into the literature on the EPR paradox
in \cite{KLYSHKO88,ZBGHZ93,ZZHBG94} in order to extend two 
qubit Bell-phenomena to
observables described as operators in Hilbert spaces of dimension
higher than two. In contradistinction to the higher than 1/2 spin
generalisations of the Bell-phenomena \cite{MERMIN80,GARG82,MERMIN82,
ARDEHALI91,AGARWAL93,WODKIEWICZ94,WODKIEWICZ95}, this type of
experimental devices generalise the ideas of beam-entanglement
\cite{HORNE85,ZUKOWSKI88,HORNE89,RARITY90}. 
Unbiased symmetric multiport beam splitters are performing unitary
transformations between "mutually unbiased" bases in the Hilbert space
\cite{SCHWINGER60,IVANOVIC81,WOOTERS86}. 
They were tested in several recent experiments
\cite{MMWZZ95,RECKPHD96}, and also various aspects of such devices were
analysed theoretically \cite{RECK94,JEX95}.

We shall use here only multiport beam splitters which have the property
that the elements of the unitary transformation which describes their
action are given by
\begin{equation}
U_{m,m'}^{N}={1\over \sqrt N}\gamma_{N}^{(m-1)(m'-1)},
\end{equation}
where $\gamma_{N}=\exp(i{2\pi\over N})$ and the indices $m$, $m'$
denote the input and exit ports. Such devices were called in
\cite{ZZH} the Bell multiports.

\section{Quantum mechanical predictions}
One assumes that the initial $M$ qu$N$it state that feeds $M$ spatially
separated multiports, each of which has $N$ inputs and $N$ outputs,
has the following form:

\begin{eqnarray}
\label {eq1}
&&|\psi(M)\rangle={1\over \sqrt
N}\sum_{m=1}^{N}\prod_{l=1}^{M}|m\rangle_{l},
\label{2}
\end{eqnarray}
where $|m\rangle_{l}$ describes the $l$-th qu$N$it being in the
$m$-th beam, which leads to the $m$-th input of the $l$-th multiport.
Please note, that only one qu$N$it enters each multiport. 
However, each of the qu$N$its itself is in a mixed state (with equal
weights), which gives it equal probability to enter the local
multiport via any of the input ports.
\begin{figure}[htbp]
\begin{center}
\includegraphics[width=14cm, height=14cm, angle=0]{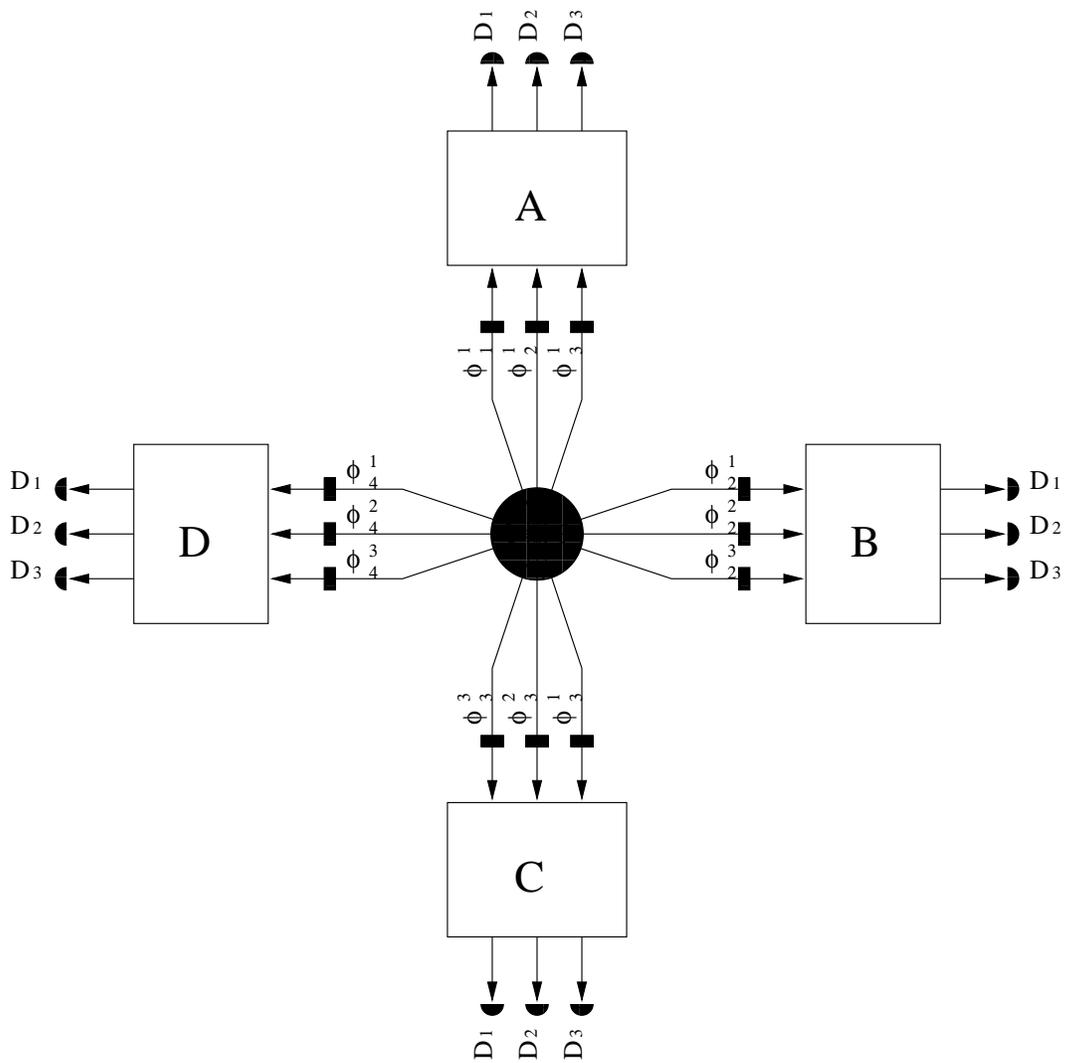}
\caption{The scheme of the experiment with four tritters.\label{port-rys2}}
\end{center}

\end{figure}

The state (\ref{2}) seems to be the most straightforward
generalisation of the GHZ states to the new type of observables. In
the original GHZ states the number of their components (i.e., two) is
equal to the dimension of the Hilbert space describing the relevant
(dichotomic) degrees of freedom of each of the qu$N$its. This
property is shared with the EPR-type states proposed in \cite{ZZH} for
a two-multiport Bell-type experiment - in this case the number of
components equals the number of input ports of each of the multiport
beam splitters. We shall not discuss here the possible methods to
generate such states. However, we briefly mention that the recently
tested entanglement swapping \cite{EVENT,ZZW,Pan98} 
technique could be used for
this purpose.

As it was mentioned earlier, in front of every input of each multiport
beam splitter one has a tunable phase shifter. The initial state is
transformed by the phase shifters into

\begin{eqnarray}
\label {eq2}
&&|\psi(M)'\rangle={1\over \sqrt N}\sum_{m=1}^{N}\prod_{l=1}^{M}
\exp(i\phi_{l}^{m})|m\rangle_{l},
\end{eqnarray}
where $\phi_{l}^{m}$ stands for the setting of the phase shifter in
front of the $m$-th port of the $l$-th multiport.

The quantum prediction for  probability to register the first photon
in the output $k_{1}$ of an $2N$ - port device, the second one in the
output $k_{2}$ of the second such device ,..., and the $M$-th one in
the output $k_{M}$ of the $M$-th device is given by:
\begin{eqnarray}
 &P_{QM}(k_{1},\dots,k_{M}|\vec{\phi_{1}},\dots,\vec{\phi_{M}})=&\nonumber\\
 &({1\over N})^{M+1}|\sum_{m=1}^{N}\exp(i\sum_{l=1}^{M}\phi_{l}^{m})
 \prod_{n=1}^{M}\gamma_{N}^{(m-1)(k_{n}-1)}|^{2}=&\nonumber\\
 &=({1\over N})^{M+1}\left[N+2\sum_{m>m'}^{N}
 \cos\left(\sum_{l=1}^{M}\Delta\Phi_{l,k_{l}}^{m,m'}
 \right)\right],&
\label{5}
\end{eqnarray}
where
$\Delta\Phi_{l,k_{l}}^{m,m'}=\phi_{l}^{m}-\phi_{l}^{m'}+{2\pi\over
N}(k_{l}-1)(m-m')$. The shorthand symbol
$\vec{\phi}_{k}$ stands for the full set of phase settings in front of
the $k$-th multiport, i.e.
$\phi_{k}^{1},\phi_{k}^{2},\dots,\phi_{k}^{N}$.

\subsection{Bell number assignment}
To efficiently describe the local detection events 
let us employ a specific value assignment method (called Bell
number assignment; for a detailed
explanation see again \cite{ZZH}), which ascribes to the detection event
behind the $m$
- th output of a multiport the value  $\gamma_{N}^{m-1}$, where
$\gamma_{N}=\exp(i{2\pi\over N})$.
With such a value assignment to the detection events, the Bell-type
correlation function, which is the average of the product of the
expected results, is defined as

\begin{eqnarray}
&E(\vec{\phi_{1}},\cdots,\vec{\phi_{M}})=&\nonumber\\
&=\sum_{k_{1},\cdots,k_{M}=1}^{N}
\prod_{l=1}^{M}\gamma_{N}^{k_{l}-1}P(k_{1},\cdots,k_{M}|\vec{\phi_{1}},\cdots,\vec{\phi_{M}})&
\label{eq4}
\end{eqnarray}
and as we shall see for the quantum case it acquires particularly simple
and universal form\footnote{This is the main purpose for using this
non-conventional value assignment.}.

The easiest way to compute the correlation function for the quantum
prediction employs the mid formula of (\ref{5}):
\begin{eqnarray}
&E_{QM}(\vec{\phi_{1}},\cdots,\vec{\phi_{M}})=&\nonumber\\ &=({1\over
N})^{M+1}\sum_{k_{1},\cdots,k_{M}=1}^{N}\sum_{m,m'=1}^{N}
\exp\left(i\sum_{n=1}^{M}(\phi_{n}^{m}-\phi_{n}^{m'})\right)&\nonumber\\
&\times \prod_{l=1}^{M}\gamma_{N}
^{(k_{l}-1)(m-m'+1)}=&\nonumber\\
&=({1\over N})^{M+1}\sum_{m,m'=1}^{N}
\exp\left(i\sum_{n=1}^{N}(\phi_{n}^{m}-\phi_{n}^{m'})\right)&\nonumber\\
&\times\prod_{l=1}^{M}
\sum_{k_{l}=1}^{N}\gamma_{N}
^{(k_{l}-1)(m-m'+1)}.&
\label{corr1}
\end{eqnarray}
Now, one notices that
$\sum_{k_{l}=1}^{N}\gamma_{N}^{(k_{l}-1)(m-m'+1)}$
differs from zero
(and equals to N)
only if $m-m'+1=0$, modulo N. Therefore we can finally write:

\begin{eqnarray}
&E_{QM}(\vec{\phi_{1}},\cdots,\vec{\phi_{M}})\nonumber&\\ 
&={1\over
N}\sum_{m=1}^{N}\exp(i\sum_{l=1}^{M}\phi^{m,m+1}_{l}),&
\label{corr2}
\end{eqnarray}
where $\phi^{m,m+1}_{l}=\phi^{m}_{l}-\phi^{m+1}_{l}$ and the above
sum is
understood modulo N, which means that
$\phi^{N+1}_{l}=\phi^{1}_{l}$.

One can notice here a striking simplicity and symmetry of this quantum
correlation function (\ref{corr2}). It is valid for all possible
values of M (number of qu$N$its) and for all possible values of
$N\geq 2$ (number of ports). For $N=2$, it reduces itself to the usual
two qubit, and for $N=2$, $M\geq2$ the standard GHZ type
multi-qubit correlation function for beam-entanglement experiments,
namely $\cos(\sum_{l=1}^{M}\phi_{l}^{1,2})$ \cite{MERMIN90}. The Bell -
EPR phenomena discussed in
\cite{ZZH} are described by (\ref{corr2}) for $M=2, N\geq 3$.

Even for $N=2$, $M=1$ the function (\ref{corr2}) describes
the following process. Assume that
a traditional four-port 50-50 beam splitter, is fed a single photon
input in  a state in which is an equal superposition of being in each
of the two input ports. The value of (\ref{corr2}) is the average of
expected photo counts behind the exit ports (provided the click at one
of the detectors is described as $+1$ and at the other one as $-1$),
and of course it depends on the relative phase shifts in front of the
beam splitter. In other words, this situation describes a Mach-Zehnder
interferometer with a single photon input at a chosen input port. For
$N=3$, $M=1$ the same interpretation applies to the case of a
generalised three input, three output Mach-Zehnder interferometer
described in \cite{WEIHS96}, provided one ascribes to firings of
the three detectors respectively
$\gamma_{3}=\alpha\equiv\exp(i{2\pi\over 3})$, $\alpha^{2}$ and
$\alpha^{3}$.

\subsection{Perfect correlations}
The described set of gedanken experiments is rich in EPR-GHZ
correlations (for $M\geq2$). To reveal the above, let us first analyse
the conditions (i.e. settings) for such correlations. As the
correlation function (\ref{corr2}) is an average of complex numbers of
unit modulus, one has
$|E_{QM}(\vec{\phi_{1}},\cdots,\vec{\phi_{M})}|\leq 1$. The equality
signals a perfect EPR-GHZ correlation. It is easy to notice that this
may happen only if
$$\exp(i\sum_{l=1}^{M}\phi_{l}^{1,2})=
\exp(i\sum_{l=1}^{M}\phi_{l}^{2,3})=\cdots=
\exp(i\sum_{l=1}^{M}\phi_{l}^{M,1})=\gamma_{N}^{k},$$ where k is an arbitrary
natural number. Under this condition
$E(\vec{\phi_{1}},\cdots,\vec{\phi_{M}})=\gamma_{N}^{k}$. This means that only
those sets of $M$ spatially separated detectors may fire, which are
ascribed such Bell numbers which have the property that their product
is $\gamma_{N}^{k}$. Knowing, which detectors fired in the set of
$M-1$ observation stations, one can predict with certainty which
detector would fire at the sole observation station not in the set.
\section{Paradoxes for $N+1$ maximally entangled qu$N$its}
\subsection{Four tritters}
We shall now present the simplest GHZ-type paradox for such systems.
We take $N=3$ and $M=4$. That is we consider now, the experimental
situation in which one has the source producing the ensemble of four
three-state particles (qutrits) described by the state $|\psi(4)\rangle$
(compare, (\ref{2})) that feeds four three-port beam splitters (i.e.,
tritters \cite{ZZH}). In this case the quantum correlation function
has the form:

\begin{eqnarray}
&E_{QM}(\vec{\phi_{1}},\vec{\phi_{2}},\vec{\phi_{3}},\vec{\phi_{4}})
&\nonumber\\
&={1\over 3}\sum_{k=1}^{3}\exp\left(i\sum_{l=1}^{4}(\phi_{l}^{k}-\phi_{l}^{k+1})\right).&
\end{eqnarray}

The (deterministic) \LHV\ correlation
function for this type of experiment must have the following structure
\cite{Bell64}:

\begin{eqnarray}
&&E_{HV}(\vec{\phi_{1}},\vec{\phi_{2}},\vec{\phi_{3}},\vec{\phi_{4}})
=\int_{\Lambda}\prod_{k=1}^{4}I_{k}(\vec{\phi_{k}},\lambda)\rho(\lambda)d\lambda.
\end{eqnarray}
The hidden variable function $I_{k}(\vec{\phi_{k}},\lambda)$, which
determines the firing of the detectors behind the k-th multiport,
depends only upon the local set of phases, and takes one of the three
possible values $\alpha$, $\alpha^{2}$, $\alpha^{3}=1$ (these values
indicate which of the detectors is to fire), and $\rho(\lambda)$ is
the distribution of hidden variables.

Consider four gedanken experiments. In the first one our observers,
each of whom operates one of the spatial separated devices,  choose
the following phases in front of their three-port beam splitters:

\begin{eqnarray}
&\vec{\phi_{1}}\equiv(\phi_{1}^{1},\phi_{1}^{2},\phi_{1}^{3})=(0,{2\pi\over 9},{4\pi\over 9})=\vec{\phi_{2}}
=\vec{\phi_{3}}=\vec{\phi}&
\nonumber\\
&\vec{\phi_{4}}\equiv(\phi_{4}^{1},\phi_{4}^{2},\phi_{4}^{3})=(0,0,0)=\vec{\phi'}.&
\end{eqnarray}
In the second experiment, the third observer  sets
$\vec{\phi_{3}}=\vec{\phi'}$ whereas the other ones set $\vec{\phi}$.
We repeat this swapping of the settings procedure in  the next two
experiments until the first observer sets $\vec{\phi'}$ and the other
set $\vec{\phi}$. Quantum mechanics predicts that in all four such
experiments 
the correlation function
is equal to $\alpha^{2}$ (i.e. we have perfect GHZ correlations).
Namely we have
\begin{eqnarray}
&E_{QM}(\vec{\phi},\vec{\phi},\vec{\phi},\vec{\phi}')
=E_{QM}(\vec{\phi},\vec{\phi},\vec{\phi}',\vec{\phi})&\nonumber\\
&=E_{QM}(\vec{\phi},\vec{\phi}',\vec{\phi},\vec{\phi})=
E_{QM}(\vec{\phi}',\vec{\phi},\vec{\phi},\vec{\phi})=\alpha^2.&
\label{q}
\end{eqnarray}

However, this immediately implies that for any \LHV\ theory that aims
at describing these phenomena one must have for every $\lambda$

\begin{eqnarray}
&I_{k}(\vec{\phi'},\lambda)\prod_{{l=1},{l\neq
k}}^4I_{l}(\vec{\phi},\lambda)
=\alpha^{2},&
\label{hidden1}
\end{eqnarray}
and this must hold for all $k=1,2,3,4$. But, since
$I_{l}(\vec{\phi},\lambda)^3=\alpha^{3k}=1$ (where, $k$ represents a
certain integer), then after multiplying these four equations side by
side, one has for every $\lambda$

\begin{eqnarray}
&\prod_{l=1}^{4}I_{l}(\vec{\phi'},\lambda)=\alpha^{2}.&
\end{eqnarray}
Therefore, if the local hidden variable theory is to agree with the
earlier mentioned quantum predictions (\ref{q}), then one must have

\begin{eqnarray}
&E(\vec{\phi'},\vec{\phi'},\vec{\phi'},\vec{\phi'})=\alpha^{2}=\alpha^{*}.&
\label{contradiction}
\end{eqnarray}
However, the quantum prediction is
$E_{QM}(\vec{\phi'},\vec{\phi'},\vec{\phi'},\vec{\phi'})=1$. Thus we
have a GHZ-type contradiction that $1=\alpha^{*}$. I.e., hidden
variables predict a different type perfect EPR-GHZ correlation.
I.e. we have a realisation of the GHZ paradox for non-dichotomic 
observables.

\subsection{General case: $N+1$ maximally entangled qu$N$its}

We will extend the reasoning to the case when one has $M=N+1$ qu$N$its
(described by the state of the form (\ref{2})) beamed into multiport
beam splitters with $N$ input and output ports. The quantum prediction for the Bell
correlation function is given by (\ref{corr2}), with the
appropriate value of $M$.

The \LHV\ correlation function must have
the following structure

\begin{eqnarray}
&\int\prod_{k=1}^{M}I_{k}(\vec{\psi_{k}},\lambda)\rho(\lambda)d\lambda,
\label{mulcorr}
\end{eqnarray}
where $k$ now extends from $1$ to $M$ and $\vec{\psi_{k}}$ stands for the
full set of settings in front of the $k$-th multiport, i.e.
$\psi_{k}^{1},
\psi_{k}^{2},\cdots,\psi_{k}^{M-1}$, and $I_{k}(\vec{\psi_{k}},\lambda)$
is a  hidden variable function depending on the local phase settings,
which has the property that its value, which can be any integer power
of $\gamma_{M-1}$, indicated which local detector is to fire.

Now, as it was in the previous case, we must choose appropriate phases
for each of observers that will be taken in the first experiment. The
appropriate choice is the following one:

\begin{eqnarray}
&\vec{\psi}_{1}=\cdots=\vec{\psi}_{M-1}=
(0,\delta,2\delta,\cdots,(M-2)\delta)=\vec{\psi}&\nonumber\\
&\vec{\psi}_{M}=(0,\cdots,0)=\vec{\psi'},&
\end{eqnarray}
where $\delta={2\pi\over (M-1)^2}$. In the next $M-1$ experiments one
applies previously described swapping of the settings procedure until
the first observer sets $\vec{\psi'}$ and the other ones set
$\vec{\psi}$. For such choice of phases the quantum correlation
function for every of the $M$ experiment, is equal to
$\gamma_{M-1}^{M-2}=\gamma_{M-1}^*$.
(i.e. we have perfect GHZ correlations of the same type for each of
the experiments). But this implies that, for any \LHV\ theory that
aims at describing these phenomena one must have, for every $\lambda$,

\begin{eqnarray}
&I_{k}(\vec{\psi'},\lambda)\prod_{l\neq k}I_{l}(\vec{\psi},\lambda)
=\gamma_{M-1}^{*},&
\label{hidden2}
\end{eqnarray}
and that this must hold for all $k=1,\cdots,M$. However, after
multiplying these $M$ equations one has:

\begin{eqnarray}
&\prod_{l=1}^{M}I_{l}(\vec{\psi'},\lambda)=\gamma_{M-1}^{*},&
\end{eqnarray}
where we have used the property of the Bell numbers generated by
$\gamma_{M-1}$, that each of them to the $M-1$-th power gives 1, and
therefore that $I_{k}(\vec{\psi},\lambda)^{M}=1$. Thus, the local
hidden variable implies that
$$E_{QM}(\vec{\psi'},\cdots,
\vec{\psi'})=\gamma_{M-1}^{*}.$$
However, the quantum prediction is 1. Thus we have the GHZ
contradiction that $1=\gamma_{M-1}^{*}$. I.e. hidden variables predict
a different type perfect EPR - GHZ correlations. In other words the
EPR idea of elements of reality makes no sense for the discussed
experiments, and this hold for an arbitrary number of qu$N$its $M$,
and for suitably related ($M-1$), but in principle arbitrarily high
number of input and exit ports of symmetric multiport beam splitters.

\section{Paradoxes for $N$ maximally entangled qu$N$its.}
In this section we show that the above reasoning can be
as well applied to the case where the number of observers
$M$ equals the number of input ports $N$ of the $2N$ port Bell
multiports.
\subsection{Three tritters} 
As the simplest example let us consider the 
gedanken experiment with three observers
each of which having tritters (3 input and 3 output ports) as
a measuring device. The correlation function for such
an experiment reads
\begin{eqnarray}
&&E_{QM}(\vec{\phi}_{1},\vec{\phi}_{2},\vec{\phi}_{3})\nonumber\\
&&={1\over 
3}\sum_{k=1}^{3}
\exp\left(i\sum_{l=1}^{3}(\phi_{l}^{k}-
\phi^{k+1}_{l})\right).
\label{3trit}
\end{eqnarray}

In the first run of the experiment we allow the observers 
to choose the following settings of the measuring apparatus
\begin{eqnarray}
&&\vec{\phi}_{1}=(0,{\pi\over 3},{2\pi\over 3})=\vec{\phi}_{2}=\vec{\phi}\nonumber\\
&&\vec{\phi}_{3}=(0,0,0)=\vec{\phi'}
\end{eqnarray}
in the second run they choose
\begin{eqnarray}
&&\vec{\phi}_{1}=\vec{\phi'}\nonumber\\
&&\vec{\phi}_{2}=\vec{\phi}=\vec{\phi}_{3}\nonumber\\
\end{eqnarray}
whereas in the third run they fix the local settings
of their tritters on 
\begin{eqnarray}
&&\vec{\phi}_{1}=\vec{\phi}=\vec{\phi}_{3}\nonumber\\
&&\vec{\phi}_{2}=\vec{\phi'}.
\end{eqnarray}

Now, let us calculate the numerical values of
the correlation function for each experimental
situation. We easily find that for all three
experiments this value, due to the special form
of the correlation function, is the same and reads
\begin{eqnarray}
&&E_{QM}(\vec{\phi},\vec{\phi},\vec{\phi'})=
E_{QM}(\vec{\phi'},\vec{\phi},\vec{\phi})=
E_{QM}(\vec{\phi},\vec{\phi'},\vec{\phi})\nonumber\\
&&=\exp(-i{2\pi\over 3})=\alpha^2,
\end{eqnarray}
i.e., we observe perfect correlations.

Proceeding in exactly the same way as in the previous
section we can write the equations that must be fulfilled
by \LHV\ for every $\lambda$, i.e.,
\begin{eqnarray}
&&I_{1}(\vec{\phi},\lambda)I_{2}(\vec{\phi},\lambda)I_{3}(\vec{\phi'}
,\lambda)=
\alpha^{2}\nonumber\\
&&I_{1}(\vec{\phi'},\lambda)I_{2}(\vec{\phi},\lambda)I_{3}(\vec{\phi}
,\lambda)=
\alpha^{2}\nonumber\\
&&I_{1}(\vec{\phi},\lambda)I_{2}(\vec{\phi'},\lambda)I_{3}(\vec{\phi}
,\lambda)=
\alpha^{2}
\label{telephone1}
\end{eqnarray}
After multiplication of the above equations one arrives at:
\begin{eqnarray}
&&\prod_{k=1}^{3}I_{k}(\vec{\phi'},\lambda)\prod_{k=1}^{3}I_{k}(\vec{\phi},
\lambda)^{2}=(\alpha^{2})^{3}=1,
\label{telephone2}
\end{eqnarray}
which can be also written in the following form
\begin{eqnarray}
&&\prod_{k=1}^{3}I_{k}(\vec{\phi'},\lambda)=\prod_{k=1}^{3}I_{k}(\vec{\phi},
\lambda),
\label{telephone3}
\end{eqnarray}
where we have used the property of the hidden variable functions,
namely that
$I_{k}(\vec{\phi},\lambda)^{2}=I_{k}(\vec{\phi},\lambda)^{*}$ for
every $\lambda$. However, because $E_{QM}(\vec{\phi'},\vec{\phi'},
\vec{\phi'})=1$ we must also have
\begin{eqnarray}
&&\prod_{k=1}^{3}I_{k}(\vec{\phi'},
\lambda)=1,
\label{maja}
\end{eqnarray}
which, because of \cyt{telephone3}, gives
\begin{eqnarray}
&&\prod_{k=1}^{3}I_{k}(\vec{\phi},
\lambda)=1
\label{maja2}
\end{eqnarray}
for every $\lambda$. This in turn implies that
\be
&&E_{QM}(\vec{\phi},\vec{\phi},
\vec{\phi})=1,
\ee
which means that local hidden variables predict perfect correlations for
the experiment when all observers set their local settings at 
$\vec{\phi}$. However, the true quantum prediction is that 
\be
&&E_{QM}(\vec{\phi},\vec{\phi},\vec{\phi})=-{1\over 3}.
\label{Maja3}
\ee
Therefore we have the contradiction: $1=-{1\over 3}$.

This contradiction is of the different
type than the one derived in the previous section although it has been
obtained in the similar way, i.e., the perfect correlations have been
used to derive it- equations \cyt{telephone1} and \cyt{telephone2}. 
Here local hidden variables
imply a certain perfect correlation, which is not predicted by quantum
mechanics.

\subsection{General case: N maximally entangled qu$N$its.}
Now, we employ the above procedure for the case when we have an arbitrary odd
number of multiports and qu$N$its $N=2m+1$. As we have seen in
the previous section the crucial point is to find the
proper phases for the multiports. Let us choose for the
first gedanken experiment the following ones

\begin{eqnarray}
&&\vec{\phi}_{1}=(0,{\pi\over 2m+1},{2\pi\over
2m+1},...,{2m\pi\over 2m+1})=\vec{\phi}\nonumber\\
&&\vec{\phi}_{2}=\vec{\phi}_{3}=...=\vec{\phi}_{2m}=\vec{\phi}\nonumber\\
&&\vec{\phi}_{2m+1}=(0,0,...,0)=\vec{\phi'}.
\end{eqnarray}

As before, in the next run of the experiment we choose the same
phases but we change the role of observers such that in the second run 
the first
one chooses $\vec{\phi'}$ while the rest of them choose $\vec{\phi}$, 
in the third run the third one chooses $\vec{\phi'}$ while the rest 
of them choose
$\vec{\phi}$, etc. Again, the value of the correlation
function for each experiment is the same 
\begin{eqnarray}
&&E_{QM}(\vec{\phi},...,\vec{\phi'})=
E_{QM}(\vec{\phi'},\dots,\vec{\phi})=\dots=
E_{QM}(\vec{\phi},\dots,\vec{\phi'},\vec{\phi})\nonumber\\
&&={1\over 2m+1}\left[\exp\left(-i{2m\pi\over 2m+1}\right)+\exp\left(-i{2m\pi\over 
2m+1}\right)\right.\nonumber\\
&&\left. +...+
\exp\left(-i{2m\pi\over 2m+1}\right)+\exp\left(i{4m^{2}\pi\over 2m+1}\right)\right]\nonumber\\
&&={1\over 2m+1}\left[2m\exp\left(-i{2m\pi\over 2m+1}\right)+\exp\left(i{4m^{2}\pi\over 
2m+1}\right)\right]=\gamma_N^{-m}
\label{vdots}
\end{eqnarray} 
Using \cyt{vdots}, the structure of the hidden
variables correlation function \cyt{mulcorr} 
and multiplying the above equations
by each other we arrive at
\begin{eqnarray}
&&\prod_{k=1}^{2m+1}I_{k}(\vec{\phi'},\lambda)=
\prod_{k=1}^{2m+1}I_{k}(\vec{\phi},\lambda)^{-2m}=
\prod_{k=1}^{2m+1}I_{k}(\vec{\phi},\lambda)
\end{eqnarray}
for every $\lambda$. Because $E_{QM}(\vec{\phi'},\dots,\vec{\phi'})=1$
one must have
\be
&&\prod_{k=1}^{2m+1}I_{k}(\vec{\phi'},\lambda)=1,
\ee
which gives
\be
&&\prod_{k=1}^{2m+1}I_{k}(\vec{\phi},\lambda)=1
\ee
for every $\lambda$. Thus, local hidden variables imply
the following perfect correlation
\be
&&E_{QM}(\vec{\phi},...,\vec{\phi})=1,
\ee
which is untrue because quantum mechanics gives
\be
&&E_{QM}(\vec{\phi},...,\vec{\phi})\nonumber\\
&&={1\over 2m+1}\left[2m\exp\left(-i\frac{2m+1}{2m+1}\pi\right)\right. 
\nonumber\\
&&\left.+\exp\left(i\frac{(2m+1)2m}{2m+1}\pi\right)\right]={-2m+1\over 2m+1}.
\ee
Therefore, for each $m$ one obtains 
the untrue identity $1=\frac{-2m+1}{2m+1}$,
which in the limit of $m\longrightarrow \infty$ becomes
$1=-1$.

Now, let us see what happens when one has the even number of qu$N$its
and the multiports $N=2m$ ($m\geq 2$).
As before we must find appropriate phases. Let us make the following 
choice:
\begin{eqnarray}
&&\vec{\phi}_{1}=(0,{\pi\over 2m-1},{2\pi\over
2m-1},...,{(2m-1)\pi\over 2m-1})=\vec{\phi}\nonumber\\
&&\vec{\phi}_{2}=\vec{\phi}_{3}=...=\vec{\phi}_{N-1}=\vec{\phi}\nonumber\\
&&\vec{\phi}_{N}=(0,0,...,0)=\vec{\phi'}.
\end{eqnarray}
One easily finds that
\begin{eqnarray}
&&E_{QM}(\vec{\phi},\dots,\vec{\phi'})=E_{QM}(\vec{\phi'},\dots,\vec{\phi})
=\dots=E_{QM}(\vec{\phi},\dots,\vec{\phi'},\vec{\phi})\nonumber\\
&&={1\over 2m}\left[(2m-1)\exp\left(-i\frac{2m-1}{2m-1}\pi\right)\right.\nonumber\\
&&\left.+\exp\left(i\frac{(2m-1)^{2}}
{2m-1}\pi\right)\right]=-1
\end{eqnarray}
and
\begin{eqnarray}
&&E_{QM}(\vec{\phi},..,\vec{\phi})\nonumber\\
&&={1\over 2m}
\left[(2m-1)\exp\left(-i\frac{2m}{2m-1}\pi\right)+
\exp\left(i\frac{2m(2m+1)}{2m+1}\pi\right)\right]\nonumber\\
&&={1\over 2m}\left[(2m-1)\exp\left(-i\frac{2m}{2m-1}\pi\right)+1\right].
\end{eqnarray}
Applying the same reasoning as above
one has 
\begin{equation}
1={1\over 2m}\left[(2m-1)\exp\left(-i\frac{2m}{2m-1}\pi\right)+1\right].
\end{equation}
Again, for each $m>2$ one has a contradiction, which 
in the limit of $m\longrightarrow\infty$ becomes
$1=-1$.

According to the old quantum rules of thumb one approaches the classical
limit with the growing quantum numbers. The above results (for both the odd
and the even case) once more show
the contrary behaviour. 

\section{Conclusions}
The derived paradoxes can be divided into two groups.
To the first group belong the series of paradoxes where one has more
observers than input ports in the unbiased multiport beamsplitters. Within this
group the contradiction manifests itself in different perfect correlations
predicted by quantum mechanics and local realism. As a special case
one obtains the original GHZ paradox.

In the second group one has the series of paradoxes in which the number of 
observers and input ports in the unbiased multiport beamsplitters is the same.
Within this group the derivation of the paradoxes relies on the perfect
correlations (as in the first group) but the final result is different.
The resulting contradiction between quantum mechanics and local realism
manifests itself in the fact that local realism predicts a certain perfect
correlation whereas quantum mechanics does not.

An interesting feature of paradoxes within the second group is that they naturally
split into two parts: the even and the odd number of observers (input ports). In each
part the final contradiction has different numerical values.
Furthermore, for the case of two observers one cannot derive the GHZ paradox
with the method presented here.

All paradoxes do not vanish with the growing dimension $N$ of the Hilbert space
describing each qu$N$it.

In conclusion we state that the multiport beam splitters, and the idea
of value assignment based Bell numbers, lead to a strikingly
straightforward generalisation of the GHZ paradox for entangled
qu$N$its. 
These properties may possibly find an application in
future quantum information and communication schemes (especially as
GHZ states are now observable in the lab \cite{INNSBRUCK}).


\part{Extension of Bell Theorem via numerical approach}

\newpage
~~~
\thispagestyle{empty}
\newpage

\chapter*{}
\begin{center}
\Large{***}
\end{center}
\phantom{~~~}

In the present part of the dissertation a novel approach
to the Bell theorem, via numerical linear optimisation,
will be presented.
This approach enables one to first of all
solve the old question\footnote{See for instance
\cite{MERMIN80,ARDEHALI91,WODKIEWICZ91,GISIN92,
WODKIEWICZ94,GISIN99,LARSSON99}.} 
of generalisation of the Bell theorem
to
\begin{itemize}
\item more possible settings at each side of the experiment (for two
and three qubits)
\item entangled pairs of qu$N$its ($N > 2$) (in earlier
literature called spins higher than $\frac{1}{2}$)
\end{itemize}

The early Bell theorem involved only two entangled qubits and two
observers, who could choose among two local mutually incompatible 
observables. Generalisations involving more than two observers
were considered by many authors and here in the previous sections.
However, one can ask the following question what the necessary
and {\it sufficient} conditions for local realistic description
of situations in which each observer can choose from more than two
observables, and for the case when two observables are not of a 
dichotomic nature are.

There were several attempts to unite Bell inequalities
for such problems, however only the stage of necessary
conditions for local realism was reached.
No one has been able to construct the full set of Bell inequalities
for such problems which would provide the sufficient condition
for local realistic description.

This was a serious issue, since necessary conditions are
{\it per se} weaker than sufficient ones.
Several papers were written discussing the exploding algebraic
difficulty in finding sufficient sets of Bell inequalities
\cite{PERESBELL}. For example Peres gives as a number of the so called
Farkas vectors (effectively coefficients in generalised Clauser-Horne
inequalities) for a two qu$N$it experiment with only two observables 
(non-degenerate) allowed to be measured at each side \cite{PERESBELL}. 
It reads $N_F(N)=(2^N-2)^4$, i.e., one has $N_F(2)=16,N_F(3)=1296$ etc.
This causes that the computational time grows extremely fast not
allowing
one to calculate anything in reasonable time.

However if one asks {\it directly} the question whether the
probabilities given by quantum mechanics are describable via a local
realistic theory, it turns out that this question can be formulated
as a typical linear optimisation problem. Since excellent
numerical methods exist it turns out
that for the number of settings of the order of 10 for two qubits and of
the order of 4 for three qubits, and for
qu$N$its with $N$ up to 10, such problems are solvable
on a standard work-station with the computing time of the 
order of one day.
The algorithm to tackle these problems together with 
the results will be
presented here. 

The threshold noise, 
or in other words visibility, below which\footnote{In terms of the
visibility above which.} there is no
local realistic model for two and three maximally entangled qubits 
is of the same value as in the case of standard
Bell inequalities (involving only two settings at each side).
Nevertheless, the presented approach gives a new method of a {\it direct}
analysis of experimental data. Such data can be directly tested for
a possibility of a local realistic model of them (without any of the usual
hypothesis about the curves best fitting to the data).

In the case of the second problem, entangled qu$N$its, new surprising 
results have been obtained. The computer data reveal growing discrepancy
of the quantum predictions with local realism with the increasing
dimensionality of the Hilbert space of the sub-systems. Earlier
approaches suggested contrary behaviour (see , for instance,
\cite{MERMIN80,MERMIN82}).
\chapter{Necessary and sufficient conditions to
violate local realism for two maximally entangled qubits- 
extension to more than two local settings [7]}
\section{Introduction}
It is a common wisdom in the quantum optical community that the
threshold visibility of the sinusoidal two-qubit interference
pattern beyond which the Bell inequalities are violated is (for the
case of perfect detectors) $\frac{1}{\sqrt{2}}$ (see, e.\ g.\ 
\cite{CLAUSER78}). Most of the experiments exceed that limit (with the usual
``fair sampling assumption'') \cite{FREEDMAN72,MANDEL88,RARITY90,
KWIAT93,PITTMAN95,TITTEL98}. 
Some difficulties to
reach this threshold were observed in the very early experiments
\cite{CLAUSER78}, as well as in some recent ones involving novel techniques.
Thus far, in atomic interferometry EPR experiments \cite{Hagley97} and
for the phenomenon of entanglement swapping \cite{ZZW,Pan98}, the
resulting visibility is less than the magic $71\%$.

It is also well known that the Clauser-Horne inequality and the CHSH
inequality are not only necessary conditions for the existence of
local realistic models but also sufficient \cite{Fine} (in the case of
the CHSH inequality, this requires some simplifying assumptions
\cite{Garg87}). However, the sufficiency proofs used involve only
{\em two pairs of settings} of the local macroscopic parameters (e.g.
orientations of the polarisers) that define the measured local
observables. Thus, the constructions are valid for precisely those
settings and nothing more, and there is no guarantee that the models
can be extended to more settings.  Consequently one may ask what is
the maximal visibility for a model applicable to {\em all possible
settings} of the measuring apparata, that returns {\em sinusoidal}
two qubit interference fringes. It is already known that for
perfect detectors, this value cannot be higher than
$\frac{1}{\sqrt{2}}$ or lower than $2/\pi$ (this is the visibility of
the recent {\it ad hoc} model by Larsson \cite{LARSSON99}; for earlier
models returning visibilities of $50\%$ see e.g. \cite{WODKIEWICZ91}).

The knowledge of the maximal visibility of sinusoidal two-qubit
fringes in a Bell-type experiment that still can be fully modelled in a
local realistic way, may help us to distinguish better between `local'
and `nonlocal' density matrices.  For two two-state systems one can
find precise conditions which have to be satisfied by density matrices
describing the general state, pure or mixed, of the full system, that
enable violation of the CHSH inequalities \cite{HOROD95}. States
fulfilling such conditions are often called ``nonlocal''. However,
since the CHSH inequality is necessary and sufficient only for two
pairs of settings, it is not excluded {\it a priori} that some states
that satisfy such inequalities for all possible sets of two pairs of
local dichotomic observables, nevertheless give predictions that in
their entirety cannot be modelled by local hidden variables. Such
models must first of all reproduce the full continuous sinusoidal
variation of two-qubit interference fringes, as well as the other
predictions.

It is clear that the full solution of the question would require a
construction, or a proof of existence, of local hidden variable models
which return sinusoidal fringes of the maximal possible visibility,
that are applicable for all possible settings of the measuring
apparata.  Since this seems to be very difficult, we chose a numerical
method of point wise approximation at a finite number of settings at
each side of the experiment. Due to exponential growth of the
computation time when the number of settings increases, we managed to
reach up to 10 settings on each side, i.e. up to 100 measurements points
(which due to a certain symmetry, about which we will say more later,
effectively can be transformed into $20\times 20=400$ points).  The
exponential growth hinders any substantial increase in this
number. Such numerical models cannot give a definite
answer concerning the critical visibility of sinusoidal fringes,
however our calculations enable us to put forward a strong conjecture
that this value must be indeed ${1\over\sqrt2}$ (see below). The
numerical method presented in this chapter is applied to the correlation
function for two qubits. The alternative approach, much better suited for
general problems is to apply the method to probabilities
of all events that are observed in the experiment. However, for maximally
entangled two qubit state, with the interference visibility reduced by some
noise (the generic problem studied here) application of numerical
approach to correlation function is equivalent to a similar procedure
with probabilities, and what is important due to the fact that
it leads to a less complicated linear optimisation problem
it allows to have numerical results for much more settings on each 
side than the method involving probabilities. Nevertheless the algorithm 
involving probabilities was also written, and it returns identical
results. It will be presented in the chapter on two entangled qu$N$it
problem.

Experimentally our problem can be formulated in the following way: the
two qubit state produced by the source does not allow for single
qubit interference, and in the experiment less-than-perfect
two-qubit fringes are obtained, due to some fundamental limitations
(like those present in the case of entanglement swapping, e.g.
\cite{ZZW,Pan98}) or due to imperfections of the devices. What is the
critical two qubit interference visibility beyond which the
observed process falsifies local realism? We shall ask these questions
assuming, for simplicity, perfect detection efficiencies, which is
possible theoretically, and experimentally thus far amounts to the
usual ``fair sampling assumption''.  

Furthermore, the problem may be investigated without the use of the
assumption that the observed fringes are of a sinusoidal nature even
though the observed two-qubit fringes in experiments with high
photon counts follow almost exactly the sinusoidal curves
\cite{FREEDMAN72,MANDEL88,RARITY90,
KWIAT93,PITTMAN95,TITTEL98}. In experiments with lower count rates, still with
relatively good level of confidence the recorded data have
approximately the same character, and it is customary to fit them with
sinusoidal curves. It is now a standard procedure to perform the
two-qubit interference experiments by recording many points of the
interference pattern, rather than stabilising the devices at
measurement settings appropriate for the best violation of some Bell
inequality.  Further, in some of the experiments, e.g. those involving
optical fibre interferometers it was not possible to
stabilise the phase differences and what is observed is just the
interference pattern changing in time, and the visibility of the
sinusoidal two-qubit fringes is used as the critical parameter
\cite{TAPSTER94}.

Even though the numerical calculations presented here only reach
$20\times 20$ points, this is more than enough in comparison to the
experimental data. The usual experimental scans rarely
involve more than 20 points. Further, our 
algorithm can be applied {\it directly to the
measurement data}, and in that way one can even avoid the standby
hypothesis that the fringes follow a sinusoidal pattern.  The
algorithm can directly answer the question: are the data compatible
with local realism or not?  Since physics is an experimental science,
the questions about Nature get their final answers solely in this way.

\section{Quantum mechanical and local realistic 
description of the gedanken experiment}
We consider the following gedanken experiment. Two observers
$a$ and $b$ perform measurements of observables 
$\vec{a}\cdot\vec{\sigma}$ (observer $a$) and 
$\vec{b}\cdot\vec{\sigma}$ (observer $b$) on the state
defined by the equation \cyt{singlet}. Observer $a$ can choose
between $N_a$ settings of the measuring apparatus, which
are defined by vectors $\vec{a}_i$ ($i=1,\dots,N_a$)
whereas observer $b$ can choose between $N_b$ settings
of the measuring apparatus defined by vectors $\vec{b}_j$
($j=1,\dots,N_b$). Therefore, they perform $N_a\times N_b$
mutually exclusive experiments. For each experiment
one has the following quantum probabilities (see
the equation \cyt{predictions})
\be
&&P_{QM}^{V_2}(l,m|\vec{a}_i,\vec{b}_j)={1\over 4}
(1-lmV_2\vec{a}_i\cdot\vec{b}_j)
\label{predictions2}
\ee
and the quantum correlation function
\be
&&E_{QM}^{V_2}(\vec{a}_i,\vec{b}_j)=-V_2\vec{a}_i\cdot\vec{b}_j.
\label{EQM}
\ee
There is no one qubit interference, i.e.,
$P_{QM}^{V_2}(l|\vec{a}_i)=P_{QM}^{V_2}(m|\vec{b}_j)={1\over 2}$
for all $i,j$.

If one assumes that the unit vectors which
define the measured observables are always coplanar the correlation
function can be simplified to $E^{V_2}_{QM}(\alpha_i,
\beta_j)=-V_2\cos{(\alpha_i+\beta_j)}$ 
(with the obvious definition of $\alpha_i$
and $\beta_j$).

Please note that in the considered here 
gedanken experiment there is one-to-one equivalence\footnote{
This is possible due to "well" defined values which we ascribe to
the results of measurements (here $\pm 1$) and symmetries
exhibited by the quantum probabilities. In the case of qu$N$its
with $N=3$ we will see that the Bell number assignment (see the chapter
on GHZ paradoxes for qu$N$its) also enables us to use the 
quantum correlation
function \cyt{3trit} instead of the probabilities.}
between the quantum correlation function and quantum 
probabilities (see
\cyt{predictions2} and \cyt{EQM}).
Therefore,
we can use either the quantum correlation function
or the quantum probabilities to describe the experiment.
However, the description of the experiment in terms
of the quantum correlation function is more convenient
from the numerical point of view, 
which we will be seen
further.  

From the numerical values of the quantum correlation
function we form a certain matrix $\hat{Q}_{V_2}$ 
of quantum predictions
with the entries: 
$Q_{ij}(V_2)=E^{V_2}_{QM}(\vec{a}_i,\vec{b}_j)$.

Within the local hidden variables formalism the correlation 
function must have
the following structure
\begin{equation} 
E_{LHV}(\vec{a}_i, \vec{b}_j)= 
\int d\lambda\rho(\lambda)A(\vec{a}_i,\lambda)B(\vec{b}_j,\lambda), 
\end{equation}
where for dichotomic measurements 
\be
&&A(\vec{a}_i,\lambda)=\pm1 
\ee
and
\be
&&B(\vec{b}_j,\lambda)=\pm1,
\ee
and they represent the values of local
measurements predetermined by the \LHV , denoted by $\lambda$, for the
specified local settings. This expression is an average over a certain
\LHV\ distribution $\rho(\lambda)$ of certain {\it factorisable} matrices, 
namely those with elements given by
$M_{ij}(\lambda)=A(\vec{a}_i,\lambda)B(\vec{b}_j,\lambda)$.  The
symbol $\lambda$ may hide very many parameters.  However, since the
only possible values of $A(\vec{a}_i,\lambda)$ and
$B(\vec{b}_j,\lambda)$ are $\pm1$ there are only $2^{N_a}$ {\it
different} sequences of the values of $(A(\vec{a}_1,\lambda),
A(\vec{a}_2,\lambda),..., A(\vec{a}_{N_a},\lambda))$, and only $2^{N_b}$
different sequences of $
(B(\vec{b}_1,\lambda),B(\vec{b}_2,\lambda),...,B(\vec{b}_{N_b},\lambda))$
and consequently they form only $2^{N_a+N_b}$ matrices $\hat{M}(\lambda)$.

Therefore the structure of \LHV\ models of $E_{LHV}(\vec{a}_i,
\vec{b}_j)$ reduces to discrete probabilistic models involving the
average of all the $2^{N_a+N_b}$ matrices $\hat{M}(\lambda)$.  In other
words, the \LHV\ can be replaced, without any loss of generality\footnote{See, for instance,
\cite{Wigner70,Belinfante73}.}, by a
certain pair of variables $k$ and $l$ that have integer values
respectively from $1$ to $2^{N_a}$ and from $1$ to $2^{N_b}$.  To each $k$
we ascribe one possible sequence of the possible values of
$A(\vec{a}_i,\lambda)$, denoted from now on by $A(\vec{a}_i,k)$,
similarly we replace $B(\vec{b}_j,\lambda)$ by $B(\vec{b}_j,l)$.
With this notation the possible \LHV\ models of the correlation function
$E_{LHV}(\vec{a}_i, \vec{b}_j)$ acquire the following simple form
\begin{equation} 
  E_{LHV}(\vec{a}_i, \vec{b}_j)=
  \sum_{k=1}^{2^{N_a}}\sum_{l=1}^{2^{N_b}}p_{kl}A(\vec{a}_i,k)B(\vec{b}_j,l),
  \label{MODEL}
\end{equation}
with, of course, the probabilities satisfying $p_{kl}\geq0$ and
$\sum_{k=1}^{2^{N_a}}\sum_{l=1}^{2^{N_b}}p_{kl}=1$.

The special case that we study here enables us to simplify the
description further. To satisfy the additional requirement that the
\LHV\ model returns the quantum prediction of equal probability of the
results at the local observation stations, that $P_{QM}^{V_2}(l|\vec{a})=
P_{QM}^{V_2}(m|\vec{b})={1\over 2}$, one can use the following observation.
For each $k$, there must exist a $k'\neq k$ with the property that
$A(\vec{a}_i,k')=-A(\vec{a}_i,k)$, and similarly for each $l$, there
must exist an $l'\neq l$ for which
$B(\vec{b}_j,l')=-B(\vec{b}_j,l)$. Then
$A(\vec{a}_i,k)B(\vec{b}_j,l)=A(\vec{a}_i,k')B(\vec{b}_j,l')$, and
thus they give exactly the same matrix of \LHV\ predictions. By assuming
$p_{kl}=p_{k'l'}$ the property of total randomness of local results
will always be reproduced by the \LHV\ models, and the generality will
not be reduced since the contributions of $p_{kl}$ and $p_{k'l'}$ to
(\ref{MODEL}) cannot be distinguished. 
Hence, we will always take only one
representative of the two pairs, reducing in this way the number of
probabilities and matrices of LHV predictions in (\ref{MODEL}) by a
factor of two\footnote{The method involving directly probabilities
of all pairs of events avoids this artificiality. However, the
price paid is a much longer computational time. Nevertheless,
we have performed also such a calculation (see the last chapter).
Of course the results of the presented calculation were confirmed.}. 
This is equivalent to taking only half of different
$B(\vec{b}_j,l)$ (or $A(\vec{a}_i,k)$ for the situation 
is symmetrical), say, first $2^{N_b-1}$ ones. Thus, (\ref{MODEL})
acquires the form
\be
&&E_{LHV}(\vec{a}_i, \vec{b}_j)=
\sum_{k=1}^{2^{N_a}}\sum_{l=1}^{2^{N_b-1}}\tilde{p}_{kl}A(\vec{a}_i,k)B(\vec{b}_j,l),
\label{MODEL2}
\ee
where we still assume that $\sum_{k=1}^{2^{N_a}}\sum_{l=1}^{2^{N_b-1}}
\tilde{p}_{kl}=1$. 

Another reduction by a factor of four is given by the fact that in the
coplanar case, the choice of the settings may be limited on each side
to ranges not greater than $\pi$ (i.e. $\phi \leq \alpha_{i} \leq \phi
+\pi, \psi \leq\beta_{j} \leq \psi +\pi$). This is due to the simple
observation that a model of the type (\ref{MODEL}) once established
for such settings can be easily extended to settings
$\alpha_{i}^{'}=\alpha_{j}+\pi, \beta_{j}^{'}=\beta_{j}+\pi$ by
putting $A(\alpha_{i}^{'},l)=-A(\alpha_{i},l)$ and
$B(\beta_{j}^{'},l)=-B(\beta_{j},l)$. 

The conditions for \LHV\ to reproduce the quantum prediction with a
final visibility $V_2$ can be simplified to the problem of maximising a
parameter $V_2$ for which exists a set of $2^{N_a+N_b-1}$ probabilities 
$\tilde{p}_{kl}$, such that 
\begin{equation}
  \sum_{k=1}^{2^{N_a}}\sum_{l=1}^{2^{N_b-1}}
  \tilde{p}_{kl}A(\vec{a}_i,k)B(\vec{b}_j,l)
  =Q_{ij}(V_2).
  \label{MODEL1}
\end{equation}
Because, for the given local settings
(\ref{MODEL1}) 
imposes linear constraints on the probabilities and the
visibility\footnote{The visibility should also
fulfil $V_2\leq 1$.}, 
and we are looking for the maximal $V_2$, the problem
can be solved by means of linear programming-
a certain method of optimisation.
\section{Linear programming and Downhill Simplex Me\-thod}
Let us briefly describe the idea of linear programming sending more
interested readers to the excellent book by Gass \cite{GASS}.

The set of linear equations (\ref{MODEL1}) constitute a certain region
in a $D=2^{N_a+N_b-1}+1$ dimensional real space- $2^{N_a+N_b-1}$ probabilities
plus the visibility. The border of the region
consists of hyper planes each defined by one of the equations belonging
to (\ref{MODEL1}); thus, if the equations do not contradict each other 
the region is a convex set with a certain number of vertices. 
On this convex set we define a linear function (cost function)
$f(p_1,\cdots,p_{2^{N_a+N_b-1}},V_2)=V_2$, which maximum we seek.

The fundamental
theorem of linear programming states that {\it the cost function 
reaches its maximum at one of the vertices}. Hence, it suffices to
find numerical values of the cost function calculated at the vertices and
then pick up the largest one. Of course, the algorithmic implementation
of this simple idea is not so easy for we must have a method of finding
the vertices for which the value of the function continually 
increases\footnote{We 
must also know how to find a starting vertex.} so 
that the program
reaches the optimal solution in the least possible number of steps. 
Calculating the value of the cost function
at every vertex would take too much time as there may be too many
of them.

There are lots of excellent algorithms which solve the above optimisation
problem. Here we have used the algorithm invented by Gondzio \cite{GONDZIO}
and implemented in the commercial code HOPDM 2.30 (Higher Order Primal-Dual
Method) written in C programming language.

However, finding the maximal visibility for the given local settings
of the measuring apparata is not enough. We should remember that our main goal
is to find such local setting for which the threshold visibility is the lowest
one. The maximal visibility $V^{max}_2$ returned by the HOPDM 2.30 procedure 
depends on the local settings entering right hand side of \cyt{MODEL1}. 
Thus, returned $V^{max}_2$ 
can be treated as the many variable function, which depends on $N_a+N_b$
angles in the coplanar case and two times more in the non coplanar one, i.e.,
$V_2^{max}=V_2^{max}(\vec{a}_1,\dots,\vec{a}_{N_a},\vec{b}_1,\dots,\vec{b}_{N_b})$.

Hence, we should also have a numerical procedure which finds the minimum
of $V^{max}_2$. 
Because we do not know much about the structure of $V^{max}_2$ as a function
of the local settings
the only reasonable
method of finding the $V^{max}_2$ minimum is the Downhill Simplex Method (DSM) 
\cite{DSM}. The way it looks for the extremum is the following. If the 
dimension of the domain of a function is $Dim$ the DSM randomly generates
$Dim+1$ points. This way it creates a starting simplex, which vertices are 
these points. Then it calculates the value of a function at the vertices and
starts exploring the space by stretching and contracting the simplex. In every
step when it 
finds a vertex where the value of the function is lower then in others it 
"goes" in this direction. For more elaborate discussion of this issue see
\cite{DSM}.
\subsection{Numerical difficulties}
Obviously there are some numerical limitations
of our computer program. First of all it is obvious that the time
needed for computation grows with the number of local settings $N_a, N_b$.
For the given $N_a, N_b$ we have $2^{N_a+N_b-1}+1$ unknowns and $N_a\times N_b+2$
linear constraints imposed on them. Furthermore, the dimension
of the domain of $V^{max}_2$ is in general $2N_a+2N_b$, which makes it
harder for the DSM procedure to find a global minimum of $V^{max}_2$. 

Secondly, the amount of the memory needed to store "hidden"
matrices grows exponentially with $N_a$ and $N_b$. All these problems 
taken together has limited our research
to the cases where $N_a+N_b\leq 20$.

It is worth
mentioning at this point about the advantages of using the
quantum correlation function instead of quantum probabilities as the 
description of the gedanken experiment. If we are to use
probabilities\footnote{For the details concerning this approach see
the chapter about pairs of entangled qu$N$its.} we will have
$2^{N_a+N_b}+1$ unknowns ($2^{N_a+N_b}$ probabilities plus the
visibility $V_2$) and $4\times N_a\times N_b+1$ constraints. 
Because we struggle with
the memory capacity and the computation time every reduction in the number
of unknowns and constraints is a great advantage.

Recently A. Peres \cite{PERESBELL} has discussed the
algorithms which search for so-called Farkas vectors,
which in turn define coefficients in generalised Bell-inequalities,
the set of which is a sufficient and necessary condition for
classical probabilistic model (here, essentially, local realistic)
to reproduce a certain set of probabilities for pairs of
experiments.  However, his method explodes numerically much much
faster then ours.  Simply, our method is applied directly to a
certain finite set of specified quantum probabilities\footnote{
Instead of the probabilities we use the correlation function but 
this does not change anything because
of the mentioned equivalence of these two ways of description of
the considered gedanken experiment.}.  Whereas 
inequalities based on the Farkas lemma apply to {\it
all possible sets of probabilities}.

\section{Results}
First of all, we have performed calculations in the coplanar case. 
We have checked situations
in which the number of settings for observer $a$ and $b$ is the same, i.e.,
$N_a=N_b=2,3,4,\dots,10$  
as well as the cases
in which the number of settings for observer $a$ and $b$ is different, i.e.,
$N_a\neq N_b$. 
Because of mentioned limitations 
of the computer program we could
only reach the limit of $N_a+N_b=20$ in both cases, i.e., for the equal 
and unequal number of the settings at each side.

Because of the richness of $N_a+N_b$ dimensional space (coplanar case)
to
find a global minimum of $V^{max}_2$ for the cases where
$N_a+N_b\leq 12$ we have run the DSM
procedure $30$ times with varied starting points. For the higher 
dimensional ones, i.e., where
$13\leq N_a+N_b\leq 15$ we have run the DSM only a few times whereas 
for the extremal cases, i.e., where
$16\leq N_a+N_b\leq 20$ we have not used the DSM but we simply have 
calculated $V^{max}_2$ on some
sets of angles containing (or not) the Bell angles (the 
definition of the Bell angles is given below). 

For $N_a+N_b\leq 15$ the DSM has always converged to
$V^{max}_2={1\over\sqrt2}$ within the given numerical precision
of computation.
Furthermore, within the angles 
$\alpha_i, \beta_j$ ($i=1\dots N_a,
j=1\dots N_b$) for which the optimum has been achieved 
there has been always the
subset of four ones (we call the angles belonging to this subset 
the Bell angles)
for which the maximal violation of the CHSH inequality
occurs. In other words it means that in the quantum matrix $\hat{Q}$
(matrix of results for the visibility equal one) 
with entries calculated for the optimal local settings found by the DSM, 
i.e., settings for
which $V_2^{max}$ reaches its minimum,
a $2\times 2$ sub matrix $\hat{B}$ appears with moduli of all its
elements equal to ${1\over\sqrt 2}$, and three of them of the same sign,
e.g.
\be
&\hat{B}=\left(
\begin{array}{cc}
-{1\over\sqrt 2} & -{1\over\sqrt 2} \\[2mm]
-{1\over\sqrt 2} & +{1\over\sqrt 2} 
\end{array}
\right).&
\ee
Of course, the elements of $\hat{B}$ can be scattered throughout
the matrix $\hat{Q}$ but by relabelling the indices it is always possible,
for instance,
to have $\hat{B}$ in the left hand corner of $\hat{Q}$. 

For the combinations of the number of local settings at each side
of the gedanken experiment such that $16\leq N_a+N_b\leq 20$ 
we have calculated $V^{max}_2$ on the sets of
local settings including the Bell angles with exactly the same result, i.e.,
$V_2^{max}={1\over\sqrt2}$ whereas for the sets of the settings not including
the Bell angles $V_2^{max}>{1\over\sqrt2}$.

We have also tested non coplanar settings with exactly the same result.
The visibility was higher than ${1\over\sqrt 2}$ if among the
local settings were no the Bell angles (no sub matrix $\hat{B}$)
and it was exactly ${1\over\sqrt 2}$  otherwise (sub matrix
$\hat{B}$ present).
\subsection{Exemplary numerical model}
Let us show an example of the computer solution for
the $3\times 3$ coplanar case with  
the optimisation
over local settings of the measuring apparata (the DSM has been used). The threshold
visibility for the optimal settings found by the program is $V^{max}_2=0.707$ 
whereas the quantum matrix
is given below
\be
&&\hat{Q}_{V^{max}_2}=V^{max}_2\hat{Q}=V^{max}_2\left(
\begin{array}{ccc}
 0.107    & -0.994 & -0.399 \\
-0.707    & -0.707 & -0.964 \\
-0.707    &  0.707 & -0.266
\end{array}
\right).
\ee
The computer hidden variable model reproducing the 
above matrix reads
\be
&&\tilde{p}(-1-1-1;+1+1+1)=0.250\nonumber\\
&&\tilde{p}(+1+1-1;+1-1+1)=0.060\nonumber\\
&&\tilde{p}(+1+1-1;+1-1-1)=0.135\nonumber\\
&&\tilde{p}(-1-1+1;+1+1+1)=0.157\nonumber\\
&&\tilde{p}(+1-1+1;+1+1+1)=0.026\nonumber\\
&&\tilde{p}(+1-1+1;+1+1-1)=0.067\nonumber\\
&&\tilde{p}(-1+1-1;+1-1-1)=0.055\nonumber\\
&&\tilde{p}(+1-1-1;+1-1+1)=0.217\nonumber\\
&&\tilde{p}(+1-1-1;+1-1-1)=0.033,
\ee
where, for instance, $\tilde{p}(+1+1-1;+1-1+1)$ denotes
the probability of appearing the factorisable matrix $\hat{M}$
\be
&&\hat{M}=\left(
\begin{array}{ccc}
 +1    & -1 & +1 \\
 +1    & -1 & +1 \\
 -1    & +1 & -1
\end{array}
\right)=\left(
\begin{array}{c}
+1  \\
+1  \\
-1  
\end{array}
\right)\left(
\begin{array}{c}
 +1  \\
 -1  \\
 +1  
\end{array}
\right)^{T}
\ee  
($T$ denotes ordinary transposition).

This example clearly exhibits the characteristic 
trait which has 
been mentioned already, namely that in the matrix $\hat{Q}$ 
there is
a sub matrix $\hat{B}$ corresponding to the matrix obtained for the Bell 
angles- 
here the elements $Q_{21},Q_{31},Q_{32},Q_{22}$. 

The numerical precision of the entries
of the matrix $\hat{Q}$ and the visibility $V^{max}_2$ is in this 
example $10^{-3}$.
Of course the precision can be increased when necessary but this
makes the time of computation longer. However, in some cases we
have run the program with the numerical precision of order $10^{-6}$
with the same result, i.e., the $V_2^{max}=0.707107\pm 10^{-6}$.

\section{Application to experimental data}
To apply our computer program for analysis of experimental data we replace
$Q_{ij}(V_2)$ in (\ref{MODEL1}) by the measured values
$E^{QM}_{ij}(exp)$, and perform the same task. If the critical $V_2$
returned by the program\footnote{In this case $V_2$
does not have the direct interpretation of visibility. Its
value tells us by what factor the observed values of the correlation
function have to be reduced, so that a local hidden variables model
exists.} is less than $1$, the data cannot
be reproduced by any local hidden variable model. Note that one even
does not have to know what the settings are.

In a recent Bell-type experiment Weinfurter and Michler\footnote{The experimental
results were reported in a different context in \cite{MICHLER2000}.} 
have
obtained the following matrix of results 
\begin{equation} 
  \hat{Q}^{exp}=\left(
    \begin{array}{ccc}
      -0.894 &  -0.061 &  0.761 \\
      -0.851 &  0.343 &  0.765 \\
      -0.625 &  0.688 &  0.516 \\
      -0.251 &  0.860 &  0.103 \\
      0.226 &  0.921 &  -0.389 \\
      0.530 &  0.651 &  -0.648 \\
      0.855 &  0.323 &  -0.832 \\
      0.852 &  -0.092 &  -0.843 \\
      0.785 &  -0.539 &  -0.638 \\
      0.397 &  -0.795 &  -0.253 \\
    \end{array}
  \right).
 \label{Michl} 
\end{equation} 
The program gives the verdict that the values of all entries to the
matrix of results have to be reduced by the factor of $0.796$ to be
describable by local hidden variables. 

Some explanation is needed here. The entries of (\ref{Michl}) 
give the values of the correlation function - there were
three different setting on side A of the experiment and 27 settings
on side B, however only 10 of them are shown here.  In the actual
experiment only data from a pair of detectors were collected.  To
obtain the matrix we have used the usual assumption that
$E(\alpha,\beta)=4P(+,+;\alpha, \beta)-1.$ We have also renormalised
the numbers of photon pairs counted, so that the average of the
counts over approximately two periods of settings at side B
represents the probability of $\frac{1}{4}$ (in concurrence with the
quantum prediction). To this end we have used data for all 27 settings on
side B.

In the recent long-distance EPR-Bell experiment the following set of
values of the correlation function was obtained\footnote{The numerical
values of the entries
to the matrix were provided by G. Weihs (private communication).}:
\begin{equation} 
  \hat{Q}^{exp}=
  \left(
    \begin{array}{cc}
      0.960 &  -0.102 \\
      0.903 &  -0.375 \\
      0.733 &  -0.660 \\
      0.479 &  -0.809 \\
      0.191 &  -0.903 \\
      -0.120 &  -0.923 \\
      -0.429 &  -0.807 \\
      -0.666 &  -0.656 \\
      -0.842 &  -0.395 \\
      -0.951 &  -0.152 \\
      -0.953 &   0.171 \\
    \end{array}
  \right).
\end{equation}
This matrix has to be reduced by the factor of $0.737$ to have a
local realistic description.

\section{Conclusions}
In his paper Peres \cite{PERESBELL} has conjectured that if the CH
inequalities are satisfied for all possible subsets of two
settings on one side and two settings on the other (out of all
$N_a\times N_b$ settings) then this is sufficient for a local realistic
model for all $N_a\times N_b$ settings. Our numerical calculations
strongly support this conjecture.

Furthermore, for the situations characterised by the condition
$N_a+N_b\leq12$ we have found the ultimate value of the visibility still
admitting the existence of \LHV . Surely, we must be aware of the fact 
that the DSM may have not found
a global minimum of $V^{max}_2$, in which the case the threshold $V_2$ would be
lower than ${1\over\sqrt2}$. However, we think that this is highly 
improbable, which is strongly supported 
by calculations for $N_a+N_b>12$. 

As it has been shown on two examples our computer program
can be also used to analyse directly the raw experimental data.
Here the analysis in terms of the correlation function
has been performed. However, this is not necessary. In the
chapter concerning two maximally entangled qu$N$its (including
also qubits) we
will present a computer program entirely based on the quantum
probabilities rather then on the correlation function. This is
important because the one-to-one equivalence between the correlation 
function and probabilities is valid only if one assumes that certain
symmetries concerning probabilities are present 
(see \cyt{predictions2}). In the real experiment one
may expect that these symmetries are not always fulfilled. The program
based on probabilities utterly solves this problem allowing us to
analyse real experiments basing solely on the observed clicks 
of detectors.

To conclude, the performed calculations enable us to put forward the
following conjecture:
sinusoidal two-qubit fringes of visibility up to
$\frac{1}{\sqrt{2}}$ are describable by local realistic theories.  At
this stage we are not able to give an analytic proof of the above.
However, for finite sets of measurement points the results of data
analysis with the use of our program fully concur with this
hypothesis. One can even say that this is more than enough, since
one cannot experimentally test exact sinusoidal nature of the two
qubit fringes. The real output of an experiment are count rate
sequences at finite number of the measurement points which
follow sinusoidal-like pattern. To such data our program
can be directly applied giving verdict whether the data
admit a local realistic model or not.
\chapter{Necessary and sufficient conditions to
violate local realism for three maximally entangled qubits- 
extension to more than two local settings}
Exactly the same question
on the threshold visibility allowing local and realistic
description for all positions of measuring apparata
can be posed in the case of correlations involving three qubits instead
of two ones. Up to now, the threshold visibility for three qubit
correlations (GHZ correlations) is known to be $V_2(3)={1\over2}$ (in
the brackets we show that we deal with the visibility for three qubits to
distinguish it from the visibility for two qubits $V_2$.) 
\cite{MERMIN90}.
However, this limit has been established for the case when there are two local
settings of the measuring apparatus at each side
of the experiment and it has been obtained in a standard 
way, i.e., as
a condition for the violation of an appropriate Bell inequality.

From the previous section we know that there is a more direct method of finding
the numerical value of the threshold visibility, which in addition 
gives necessary 
and sufficient
conditions for the existence of \LHV\ for the given set of 
local settings for each
observer.
In this section we show the application of the presented numerical method to
the GHZ correlations, i.e., to the maximally entangled state of three 
qubits. Surprisingly, the obtained results bear great resemblance to 
those for two qubits.

\section{Description of the method}
To this end let us consider the following maximally entangled state of
3 qubits 
\be
&&|\psi\rangle={1\over\sqrt 2}(|0\rangle_{1}|0\rangle_{2}|0\rangle_{3}+
|1\rangle_{1}|1\rangle_{2}|1\rangle_{3})
\ee
where $|i\rangle_j$ is the $i$-th state of the $j$-th qubit. 

Each observer measure the observable
$\vec{n}\cdot\vec{\sigma}$, where $n=a,b,c$ ($a$ for the first
observer, $b$ for the second one and $c$ for the third one), $\vec{n}$ 
is a unit vector
characterising the observable which is measured by observer $n$ and
$\vec{\sigma}$ is a vector the components of which are standard Pauli
matrices. As in the case of two qubits the family of observables 
$\vec{n}\cdot\vec{\sigma}$
covers all possible dichotomic observables for a qubit
system, endowed with a spectrum consisting of $\pm 1$. 

The probability of obtaining
the result $m=\pm 1$ for the observer $a$, when measuring the
observable characterised by the vector
$\vec{a}$, the result $l=\pm 1$ for the observer $b$, when measuring the
observable characterised by the vector
$\vec{b}$ and the result $k=\pm 1$ for the observer $c$, when measuring the
observable characterised by the vector
$\vec{c}$ is equal to
\begin{eqnarray}
&P_{QM}(m,l,k;\vec{a},\vec{b},\vec{c})=
{1\over 8}(1+mla_3b_3+mka_3c_3+lkb_3c_3&\nonumber\\
&+mlk\sum_{r,p,s=1}^{3}M_{rps}a_r b_p c_s),&
\label{GHZpredictions}
\end{eqnarray}
where $a_r, b_p, c_s$ are components of vectors $\vec{a},\vec{b},\vec{c}$ 
and where nonzero elements of the tensor $M_{rps}$ are $M_{111}=1, M_{122}=-1, 
M_{212}=-1, M_{221}=-1$. In spherical coordinates vectors $\vec{a}, 
\vec{b}, \vec{c}$ read
\be
&&\vec{n}=(\cos\phi_n\sin\theta_n,\sin\phi_n\sin\theta_n,\cos\theta_n),
\label{GHZspherical}
\ee
where $\theta_n\in [0,\pi],\phi_n\in [0,2\pi]$. From now on 
we will be
considering only the measurement of the observables characterised
by vectors with the zero third component, which is equivalent to
putting $\theta_{n}=\pi/2$. Thus, the formula (\ref{GHZpredictions})
acquires simpler form (we have replaced $\phi_a,\phi_b,\phi_c$ 
by $\alpha,\beta,\gamma$ respectively)
\begin{eqnarray}
&P_{QM}(m,l,k;\alpha,\beta,\gamma)=
{1\over 8}(1+mlk\sum_{r,p,s=1}^{3}M_{rps}a_r b_p c_s)&
\label{GHZpredictionsbis}
\end{eqnarray}
in which only a term responsible for three qubit correlations
is present. 

The probabilities of obtaining one of the results in the local
stations reveal no dependence on the local parameters,
$P_{QM}(l|\alpha)=P_{QM}(m|\beta)=P_{QM}(n|\gamma)
={1\over2}$. Similarly, the 
probabilities describing two qubit correlations do not
reveal dependence on the local parameters either, 
$P_{QM}(l,m|\alpha,\beta)=P_{QM}(m,n|\beta,\gamma)=
P_{QM}(l,n|\alpha,\gamma)=
{1\over4}$. 

As usual, if $V_2(3)<1$ we replace (\ref{GHZpredictionsbis}) by
\be
&P_{QM}^{V_2(3)}(m,l,k|\alpha,\beta,\gamma)=
{1\over 8}(1+mlkV_2(3)\sum_{r,p,s=1}^{3}M_{rps}a_r b_p c_s).&
\label{GHZpredictions2}
\ee

The quantum prediction
for the correlation function with reduced visibility reads:
\begin{eqnarray}
&E_{QM}^{V_2(3)}(\alpha, \beta,\gamma)
=\sum_{m,l,k=-1}^{1} mlkP(m,l,k;\alpha,\beta,\gamma)
\nonumber\\
&=V_2(3)\sum_{r,p,s=1}^{3}M_{rps}a_r b_p c_s=V_2(3)\cos(\alpha+\beta+\gamma).&
\label{EQM1}
\end{eqnarray}
and there is no single and two qubit interference\footnote{We again
use the quantum correlation function instead of quantum probabilities. In
the case considered here these two approaches are equivalent- see the
discussion in the previous chapter.}. 

Now, we proceed analogously to the case of two qubits. Observer
$a$ chooses between $N_a$ settings of the measuring 
apparata $\alpha_1,\dots, \alpha_{N_a}$, observer
$b$ between $N_b$ settings $\beta_1,\dots, \beta_{N_{b}}$ 
and ,finally, observer
$c$ between $N_c$ settings $\gamma_1,\dots, \gamma_{N_{c}}$. 
For each triple of local 
settings we calculate
the quantum correlation function (\ref{EQM1}) $E_{QM}^{V_2(3)}(\alpha_i,
\beta_j,\gamma_k)$, where $i=1,\dots,N_a,
j=1,\dots, N_b, k=1,\dots,N_c$. Thus we have a tensor\footnote{We call it
a tensor for it has three indices. This has nothing in common
with the transformation properties of this object.} 
$Q_{ijk}(V_2(3))=E_{QM}^{V_2(3)}(\alpha_i,
\beta_j,\gamma_k)$ 
of quantum predictions. 

Within the local hidden variables formalism the correlation 
function must have
the following structure
\begin{equation} 
E_{LHV}(\alpha_i, \beta_j, \gamma_k)= 
\int d\rho(\lambda)A(\alpha_i,\lambda)B(\beta_j,\lambda)
C(\gamma_k,\lambda), 
\end{equation}
where for dichotomic measurements 
\be
&&A(\alpha_i,\lambda)=\pm1\nonumber\\
&&B(\beta_j,\lambda)=\pm1\nonumber\\  
&&C(\gamma_k,\lambda)=\pm1,
\ee 
and they represent the values of local
measurements predetermined by the local hidden variables, 
denoted by $\lambda$, for the
specified local settings. This expression is an average over a certain
\LHV\ distribution $\rho(\lambda)$ of certain {\it factorisable} tensors, 
namely those with elements given by
$T_{ijk}(\lambda)=A(\alpha_i,\lambda)B(\beta_j,\lambda)
C(\gamma_k,\lambda)$.
Since the
only possible values of $A(\alpha_i,\lambda)$,
$B(\beta_j,\lambda)$ and $C(\gamma_k,\lambda)$ are $\pm1$ there are 
only $2^{N_{a}}$ {\it
dif\-fer\-ent} sequences of the values of $(A(\alpha_1,\lambda),
..., A(\alpha_{N_{a}},\lambda))$, $2^{N_{b}}$ 
different sequences of 
$(B(\beta_1,\lambda),...,B(\beta_{N_{b}},\lambda))$,
$2^{N_{c}}$ 
different sequences of 
$(C(\gamma_1,\lambda),
..., C(\gamma_{N_{c}},\lambda))$
and consequently they form only $2^{N_{a}+N_{b}+N_{c}}$ 
tensors $T_{ijk}(\lambda)$.

Therefore the structure of \LHV\ models of $E_{LHV}(\alpha_i,
\beta_j,\gamma_k)$ reduces to discrete probabilistic models involving the
average of all the $2^{N_{a}+N_{b}+N_{c}}$ tensors $T_{ijk}(\lambda)$.  
In other
words, the \LHV\ can be replaced, without any loss of generality, by a
certain triple of variables $l,m,n$ that have integer values
respectively from $1,\dots,2^{N_{a}},1,\dots,2^{N_{b}},1,\dots,
2^{N_{c}}$.  To each $l$
we ascribe one possible sequence of the possible values of
$A(\alpha_i,\lambda)$, denoted from now on by $A(\alpha_i,l)$,
similarly we replace $B(\beta_j,\lambda)$ by $B(\beta_j,m)$ and
$C(\gamma_k,\lambda)$ by $C(\gamma_k,n)$ .
With this notation the possible \LHV\ models of the correlation function
$E_{LHV}(\alpha_i, \beta_j, \gamma_k)$ acquire the following simple form
\begin{equation} 
  E_{LHV}(\alpha_i, \beta_j,\gamma_k)=
  \sum_{l=1}^{2^{N_{a}}}\sum_{m=1}^{2^{N_{b}}}\sum_{n=1}^{2^{N_{c}}}p_{lmn}
  A(\alpha_i,l)B(\beta_j,m)C(\gamma_k,n),
  \label{MODELGHZ}
\end{equation}
with, of course, the probabilities satisfying $p_{lmn}\geq0$ and
$\sum_{l=1}^{2^{N_{a}}}\sum_{m=1}^{2^{N_{b}}}
\sum_{n=1}^{2^{N_{c}}}p_{lmn}=1$. 

Not all of tensors $T_{ijk}(lmn)$
are different. It is easy to check that
only one fourth of them are. The
situation is similar to that with two
qubits and enables one to simplify the actual computer program. 

For the given local settings of the
measuring apparatus at each side of the experiment we want to find 
the maximal $V_2(3)$ still
admitting the local realistic description in
the form \cyt{MODELGHZ}. Then we want to find such local
settings for which this maximal $V_2(3)$ reaches its
minimum. From the previous section we know how to cope with 
such a problem. We
use exactly the same numerical approach, i.e., the HOPDM 2.30 and 
the DSM procedures.
\section{Results}
We have checked 
three experimental situations: $N_{a}=N_{b}=N_{c}=2,3,4,5$ with
the result that the threshold visibility admitting \LHV\ is $V_2(3)={1\over 2}$.
This result is in concurrence with 
the threshold visibility obtained earlier in \cite{MERMIN90}
with the usage of appropriate Bell inequalities.

Again, because of the complexity of the space being the domain
of the $V^{max}_2(3)$ function (defined in analogy to that for
two qubits)- to find a global minimum- for the case $N_{a}=N_{b}
=N_{c}=2,3$ we 
have run the amoeba procedure
30 times with varied starting points. For $N_{a}=N_{b}=N_{c}=4$ 
we calculated $V^{max}_2(3)$
on $9000$ randomly chosen sets of the local settings whereas in the
case $N_{a}=N_{b}=N_{c}=5$ we have calculated $V^{max}_2(3)$ on the following
set of the local settings: $\alpha_1=0,\alpha_2=\pi/8,
\alpha_3=\pi/4,\alpha_4=3\pi/8,\alpha_5=\pi/2,\beta_1=\gamma_1=-\pi/4,
\beta_2=\gamma_2=-\pi/8,\beta_3=\gamma_3=0,\beta_4=\gamma_4=\pi/8,
\beta_5=\gamma_5=\pi/4$.

An interesting feature of the results is that, as for
two qubits, the 
threshold visibility
$V^{max}_2(3)={1\over2}$ is always achieved for such 
settings of the measuring
apparata which include as a subset the settings 
giving maximal violation of the inequalities presented in 
\cite{MERMIN90}. This is in analogy with the two qubit
case in which the threshold visibility of ${1\over\sqrt 2}$
is always obtained if among the settings one has
a subset which lead to maximal violation of
the CHSH inequality (for the maximally entangled state). 

\section{Conclusions}
The presented numerical approach to the three qubit GHZ correlations
gives the sufficient and necessary conditions for the existence
of \LHV\ for the given experimental situation, i.e., for the fixed number
of positions of the measuring apparata at each side of the experiment. 

For the cases where $N_a=N_b=N_c=2,3$ we have found such numerical
values of the local settings for which the visibility admitting \LHV\
has the lowest possible value. Up to the possibility that the DSM 
procedure has not succeeded in finding the global minimum of
$V^{max}_2(3)$ the visibility $V_2(3)={1\over2}$ is the ultimate limit drawing
the borderline between \LHV\ and \QM\ for these cases, i.e., for
2 and 3 settings of the measuring apparatus at each side of the experiment. 
As
far as I know these are the first results giving the necessary and
sufficient conditions for violation of local realism in the GHZ
case\footnote{The inequalities found in \cite{MERMIN90}
give only necessary condition for the existence of local realism.}.

For $N_a=N_b=N_c=4$ the visibility returned by the program for every
random choice of local settings has been always higher\footnote{Sometimes it was
very close to $\frac{1}{2}$. Obviously, it is extremely
difficult to find at random such local settings for which visibility
equals exactly $\frac{1}{2}$.} 
than $\frac{1}{2}$.

In the extremal case, i.e., for $N_a=N_b=N_c=5$ we have found
the threshold value for local settings including as a subset
settings giving maximal violation of Mermin's inequality 
\cite{MERMIN90} with the result that $V^{max}_2(3)={1\over 2}$ (the DSM
has not been used).

Unfortunately, due to the computer time and memory limitations 
(of the same origin as in the two qubit case) we
could not check more settings of the measuring apparatus. Nevertheless, 
we suppose
that increasing the number of settings will not lead to visibility
lower than $V_2(3)={1\over 2}$. 

The important aspect of the presented analysis of the GHZ
correlations is that, just like in the case of two qubits, the
numerical approach can be directly applied to measurement data.

We should also mention that the program based on the quantum probabilities
(the idea is presented in the chapter 10)
has also been written for considered here three maximally entangled qubits. 
Some calculations
have been performed but due to the time limitations we could not test
so much cases as we did with the program based on the correlation function.
As one expects, the results for the tested cases ($N_a=N_b=N_c=2,3$ with 
the DSM
optimisation) 
are the same, i.e., the threshold visibility equals $\frac{1}{2}$.

\chapter{Entangled pairs of qu$N$its: the violation of local 
realism increases with $N$ [8]}
\section{Introduction}
John Bell  has shown
that no local realistic models 
can agree with all quantum mechanical predictions
for the maximally
entangled states of two qubits.
After some years researchers started to ask questions about the
Bell theorem for more complicated systems. The most spectacular
answer came for multiple qubits in the form the
GHZ theorem \cite{GHZ89}: the conflict
between local realism and quantum mechanics 
is much sharper than for two qubits, 
and can be shown even at the level
of perfect EPR-type correlations. 

The other possible extension are entangled states
of pairs of qu$N$its ($3\leq N$). First results, in 1980-82, suggested that
the conflict between local realism and quantum mechanics
diminishes with growing N \cite{MERMIN80,MERMIN82,GARG82}. 
This was felt to be in concurrence with the old
quantum wisdom of higher quantum numbers leading to a
quasi-classical behaviour.
However, the early research was confined to Stern-Gerlach type
measurements performed on pairs of maximally entangled ${N-1\over2}$
spins \cite{MERMIN80,MERMIN82,GARG82}. 
Since operation of a Stern-Gerlach device
depends solely on the orientation of the quantisation axis,
i.e. on only two parameters, devices of this kind
cannot make projections into arbitrary orthogonal
bases of the subsystems. That is, they cannot make full use of 
the richness of the $N$-dimensional Hilbert space.

W\'odkiewicz \cite{WODKIEWICZ94} proposed to employ the measurement of 
the observable of a different kind (projection onto a coherent
state) and the Clauser-Horne inequality. Even in this case
in the limit of $N\longrightarrow\infty$ the violation vanishes.

In early 1990's Peres and Gisin \cite{PERES92,GISIN92} 
have shown, that if one
considers certain {\em dichotomic} observables applied to maximally
entangled pairs of qu$N$its, the violation of local realism, or
more precisely of the CHSH inequalities, survives
the limit of $N\rightarrow\infty$ and is maximal there. 
However, for any dichotomic
quantum observables the CHSH inequalities give violations
bounded by the 
Tsirelson limit \cite{TSIRELSON80}, i.e. limited by the 
factor of $\sqrt{2}$.  
Therefore, the question 
whether the violation of local realism increases with growing $N$ 
was still left open.

There are some reasons to suspect that violations of local realism 
should get stronger with increasing $N$.
For systems described by  observables which are at least three valued the
Bell-Kochen-Specker theorem \cite{GLEASON57,KOCHEN67,Bell66}
on non-contextual hidden variable
theories can be applied. This means that any realistic theory of local
observations must be inevitably contextual.
In contradistinction, the original Bell
theorem is formulated for subsystems for which such problems do
not arise.

In this chapter we show that violation of local realism indeed
increases with growing $N$ if one uses non-degenerate observables- already
introduced
unbiased symmetric $2N$ ports.

As a "measure" of the strength of violation of
local realism in this chapter the threshold noise 
admixture $F_N$ (see \cyt{noise0}) still allowing
a local and realistic description of the gedanken experiment
will be used. Its link with the visibility $V_N$ is simple, namely
$F_N=1-V_N$. However, as the results for objects belonging
to Hilbert spaces of different dimensions will be compared, using
the parameter $F_N$ is more objective.

\section{Description of the gedanken experiment}
The general idea is the same as in the previous chapters. 
We analyse a Bell-type experiment with two \dits\ flying towards
two spatially separated observers A and B. We assume that the \dits\
are prepared in the mixed state $\hat{\rho}(F_N)$
\be
&&\hat{\rho}(F_N)=F_N\hat{\rho}_{noise}+(1-F_N)\hat{\rho}_{max},
\ee
where $\hat{\rho}_{noise}={1\over N^2}I\otimes I$ 
($I$ is an $N\times N$ identity matrix)
and 
$\hat{\rho}_{max}$ is
the projector $\ket{\psi}\bra{\psi}$ on the maximally entangled state
\be
&&\ket{\psi}={1\over\sqrt{N}}\sum_{k=1}^{N}\ket{k}_A\ket{k}_B.
\label{pure}
\ee
\begin{figure}[htbp]
\begin{center}
\includegraphics[angle=0, height=4.cm, width=12.5cm]{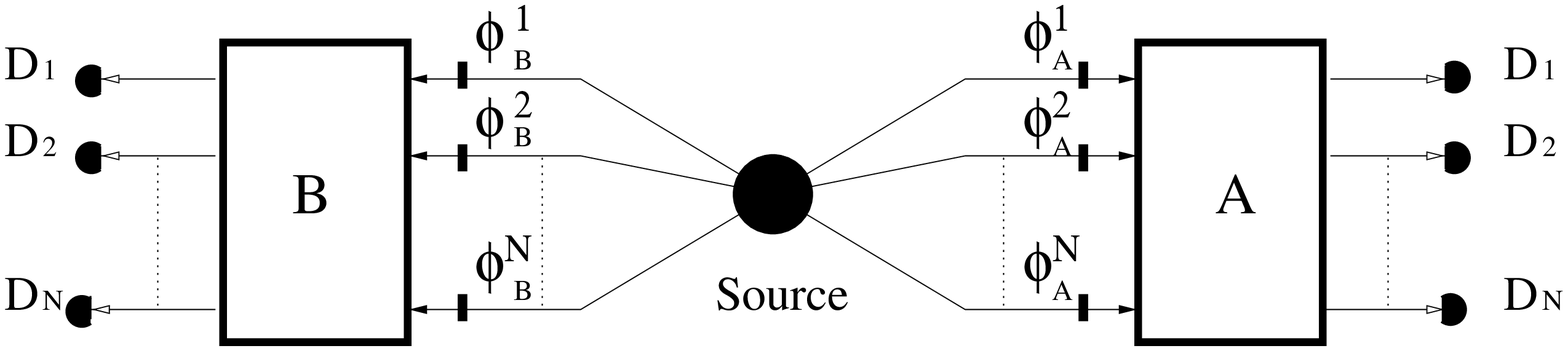}
\caption{The experiment of observer A and B with two maximally 
entangled qu$N$its. 
Each of their measuring apparata consist of a set of $N$
phase shifters just in front of an $2N$ port Bell multiport, and
$N$ photon detectors $D_k, D_l$ (perfect, in the gedanken situation described
here) which register photons in the output ports of the device. 
The phase shifters serve the role of the devices which
set the free macroscopic, classical parameters that can be
controlled by the experimenters. The source 
produces a beam-entangled two qu$N$it
state.\label{plot1}}  
\end{center} 
\end{figure}

Observer A can choose between the non commuting observables
$A_1,A_2$ and 
observer B also can choose between the non commuting observables
$B_1,B_2$ (each observable for observer A and B is 
characterised by some set of
local parameters (knobs)) . We assume that the spectrum 
of each observable consists of
$N$ points, which we enumerate by subsequent natural numbers $k,l=1,2,\dots,
N$, where index $k$ refers to observer A and index $l$ to observer B. 
Thus, the observers can perform $2\times 2$ 
mutually exclusive global experiments. 

The quantum probability distribution for the specific pairs of results $k$ 
and $l$, 
provided a specific pairs of local observables is
chosen ($A_i$ and $B_j$), will be denoted by 
$P_{QM}^{F_N}(k,l|{A}_i,{B}_j)$. In the case considered
here the quantum probabilities read 
\begin{eqnarray}
&P_{QM}^{F_N}(k,l|A_i,B_j)& \nonumber \\
&=\frac{1}{N^2}F_N+(1-F_N)P_{QM}^{max}(k,l|A_i,B_j),&
\end{eqnarray} 
where $P_{QM}^{max}(k,l|A_i,B_j)$ is the probability for the given pair 
of events for the pure maximally entangled state \cyt{pure}. Of course, the
exact form of $P_{QM}^{max}(k,l|A_i,B_j)$ depends on the specific choice of
the observables $A_i,B_j$.
According to \QM\ the set of $4N^2$ 
such probabilities is the only
information available to the observers.
\section{Local realism and joint probability distribution}
It is well known (see, e. g. \cite{Fine}, \cite{PERESBELL}) that
the hypothesis of  local hidden variables
is equivalent to the existence of a (non-negative) joint
probability distribution involving all 
four observables from which it should be possible to 
obtain all the quantum predictions as marginals\footnote{The simple
proof of this statement is given in the Appendix A.}.
Let us denote this hypothetical distribution by
$P_{HV}(k,m;l,n|A_1,A_2,B_1,B_2)$, where $k$ and $m$,
represent the outcome values for observer A observables ($l$ and $n$
for observer B). 
In \QM\ one cannot even define such objects, since 
they involve mutually incompatible measurements.
The local hidden variable probabilities 
$P_{HV}(...)$ are defined as
the marginals
\be
&P_{HV}(k,l|A_1, B_1)=
\sum_{m}
\sum_{n}
P_{HV}(k,m;l,n),&\nonumber \\
&P_{HV}(k,n|A_1,B_2)=
\sum_{m}
\sum_{l}
P_{HV}(k,m;l,n),&\nonumber \\
&P_{HV}(m,l|A_2,B_1)=
\sum_{k}
\sum_{n}
P_{HV}(k,m;l,n),&\nonumber \\
&P_{HV}(m,n|A_2,B_2)=
\sum_{k}
\sum_{l}
P_{HV}(k,m;l,n),&\nonumber \\ 
\label{sumation2}
\ee
where $P_{HV}(k,m;l,n)$ is a short hand notation for
$P_{HV}(k,m;l,n|A_1,A_2,B_1,B_2)$. 
The $4\times N^2$ 
equations (\ref{sumation2})
form the full set of necessary and sufficient conditions
for the existence of  
local and realistic
description of the experiment, i.e., for the joint
probability distribution $P_{HV}(k,m;l,n)$.
The Bell theorem says that there are quantum predictions, which for $F_N$ 
below a certain
threshold cannot be 
modelled by (\ref{sumation2}), i.e. there exists a critical $F_N^{tr}$
below  which one cannot have any local realistic model with 
$P_{HV}(k,l|A_i,B_j)=P_{QM}^{F_N}(k,l|A_i,B_j)$.
Our goal is to find observables for the two qu$N$its 
returning the highest possible critical $F_N^{tr}$.
\section{Linear programming}
The set of conditions (\ref{sumation2}) with $P_{QM}^{F_N}(k,l|A_i,B_j)$
replacing $P_{HV}(k,l|A_i,B_j)$ imposes linear constraints on the $N^{4}$
``hidden probabilities" $P_{HV}(k,m;l,n)$ 
and on the parameter $F_N$, which are the nonnegative unknowns. We have more  
unknowns ($N^{4}+1$) than equations 
($4 N^2$), and we want to find the
minimal $F_N$ for which the set of constraints can still be satisfied. Then
we want to find such local settings characterising the observables that the found
minimal $F_N$ reaches its highest possible value\footnote{Please, notice
that here we are looking for the threshold value of the noise
admixture $F_N$ and not the threshold $V_N$ as in the chapters
concerning entangled qubits. Because $V_N=1-F_N$ to maximal
visibility refers the minimal noise admixture.}.
From the previous chapters we know how to handle such a problem. We
use our sledge hammer: HOPDM 2.30 and the DSM procedures.
\section{Observables}
In our numerical calculations we have used the observables
defined by unbiased multiport beamsplitters (for the definition
of such a device see chapter devoted
to the GHZ paradoxes for qu$N$its). 

The
quantum prediction for the joint probability $P_{QM}^{max}(k,l|A_i,B_j)$ to
detect a photon at the $k$-th output of the multiport A characterised 
by phase shifters $\vec{\phi}^i_A=
(\phi^1_A(i),...\phi^N_A(i))$ ($i=1,2$) and
another one at the $l$-th output of the multiport B characterised
by phase shifters $\vec{\phi}^j_B=
(\phi^1_B(j),...\phi^N_B(j))$ ($j=1,2$) 
for the maximally entangled state \cyt{pure} can be derived from
\cyt{5} with $M=2$ and reads:
\begin{eqnarray}
&P_{QM}^{max}(k,l|\vec{\phi}^i_A,\vec{\phi}^j_B)\nonumber\\ 
&=(\frac{1}{N^3})\left(N+
2\sum^N_{m>n}\cos{({\bf \Phi}^m_{kl}(ij)-{\bf \Phi}^n_{kl}(ij))}\right),& 
\label{25a}
\end{eqnarray}
where
${\bf \Phi}^m_{kl}(ij) \equiv \phi^m_A(i)+\phi^m_B(j)+
[m(k+l-2)] \frac{2\pi}{N}$.
The counts at a single detector, of course, do not depend upon
the local phase settings:
$
P_{QM}(k|A_i)=P_{QM}(l|B_j)={1}/{N}$ for all $i,j=1,2$.  

\section{Results}
The results are depicted in \cyt{Aston}.
\begin{figure}[htbp]
\begin{center}
\includegraphics[ height=15cm, width=13cm, angle=270]{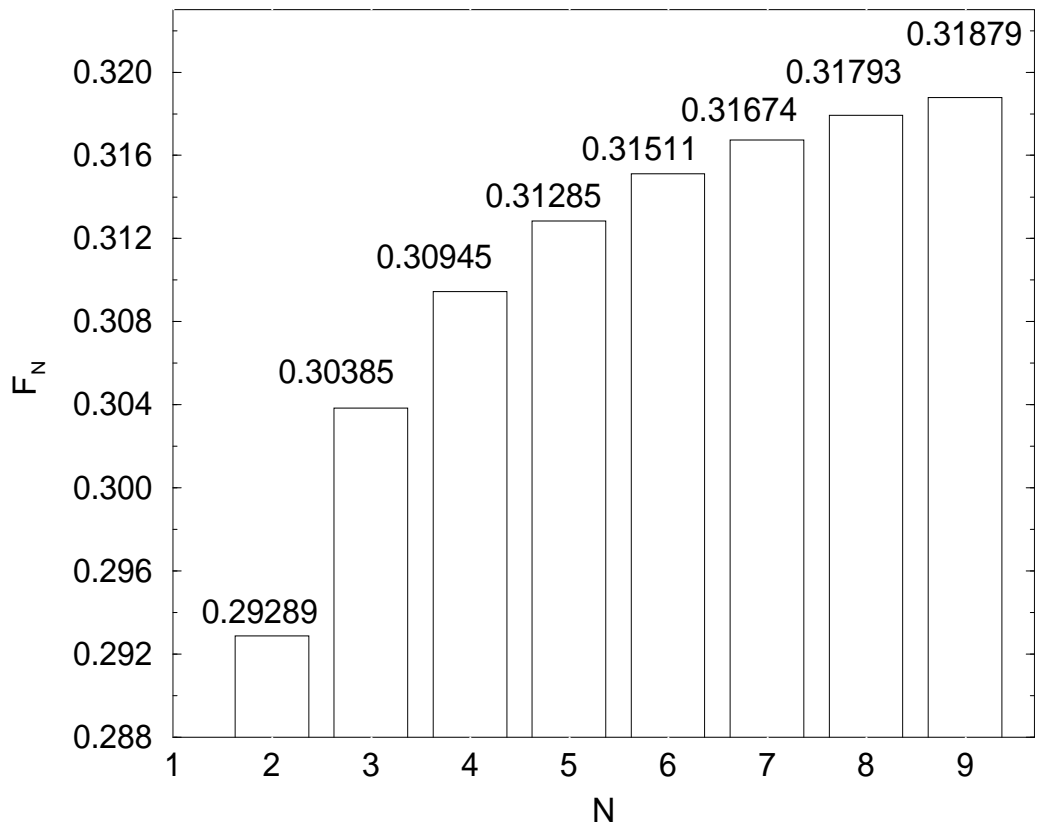}
\caption{The threshold noise admixture $F_N$ versus the 
dimension of the Hilbert space
of qu$N$its. For $N=3$ basing on the computer calculations we
were able to find the analytical expression for the threshold noise
admixture $F_3=\frac{11-6\sqrt 3}{2}$.\label{Aston}}
\end{center}

\end{figure}
We see that $F_N$ continuously increases with growing $N$ exhibiting
opposite behaviour to the results obtained in earlier works
\cite{MERMIN80,MERMIN82,GARG82,PERES92,GISIN92,WODKIEWICZ94}.

For instance, let us compare \cyt{Aston} with the results
obtained in \cite{GISIN92}, which are depicted in \cyt{GisPer}.
\begin{figure}[htbp]
\begin{center}
\includegraphics[ height=15cm, width=13cm, angle=270]{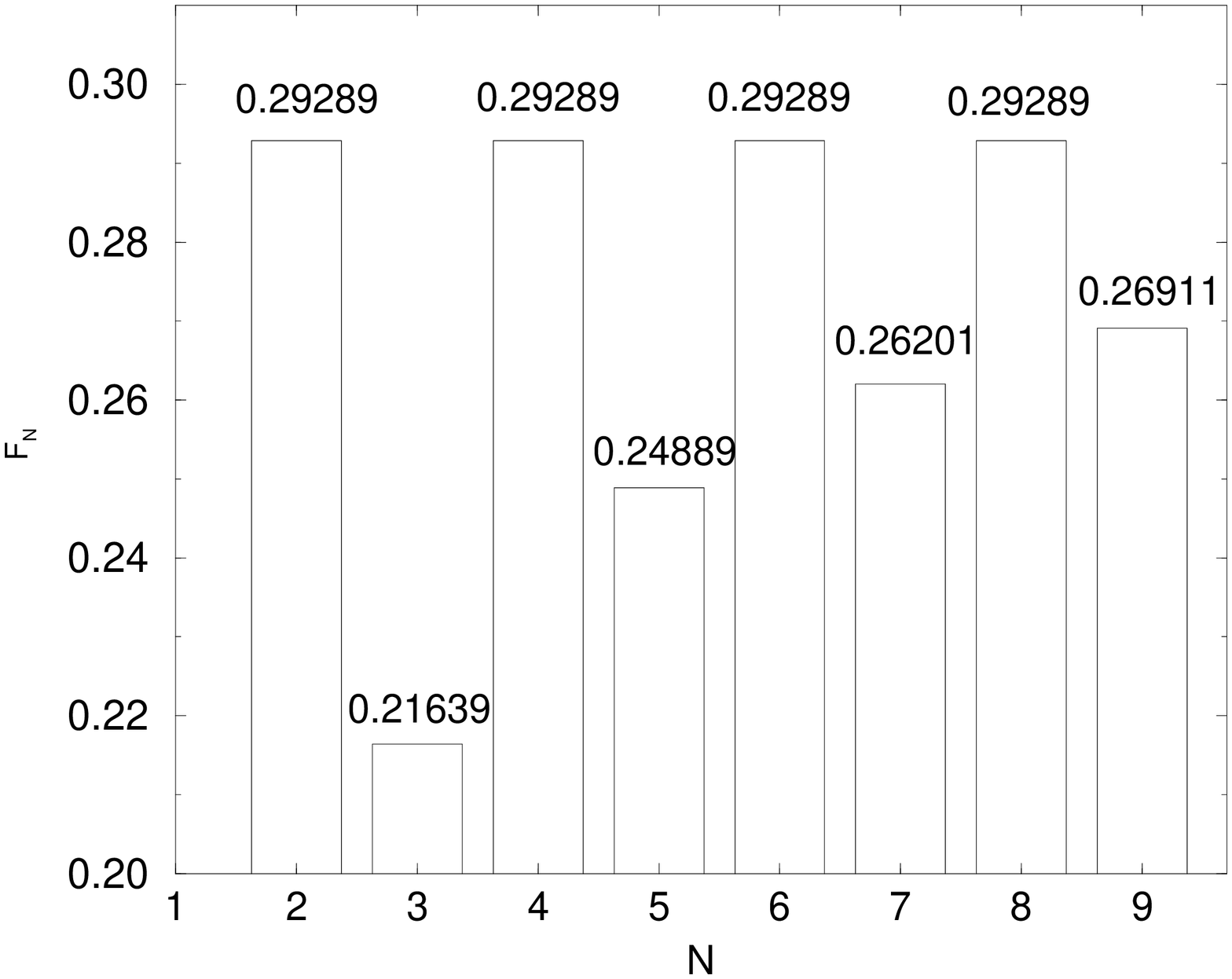}
\caption{The threshold noise admixture $F_N$ versus the dimension
of the Hilbert space of qu$N$its obtained by Gisin and Peres 
\cite{GISIN92}.\label{GisPer}}
\end{center}

\end{figure}
We see that for $N=2$ both results are identical. However, starting from
$N\geq 3$ the results in \cyt{GisPer} never exceed the 
$1-{1\over\sqrt 2}$. This is due to the fact that in \cite{GISIN92}
dichotomic observables were used and the threshold $F_N$ was obtained as
a condition for violation of CHSH inequality. This behaviour
is in agreement with the theorem proved by Tsirelson 
\cite{TSIRELSON80}, which states that the CHSH inequality
for dichotomic observables can be violated only if $F_N$ is
lower than $1-{1\over\sqrt 2}$.

A few words of comment are needed.  
One may argue that because of a quite large number
of local macroscopic parameters (the phases)
defining the function to be maximised with the DSM procedure 
we 
could have missed the global minimum. 
While this argument cannot be ruled out in principle, we 
stress that in that case the ultimate violation would even be larger.
This would only strengthen our conclusion that two entangled qu$N$it
systems are in stronger conflict with local realism than two entangled
qubits.

An important question is whether unbiased multiports provide us with 
a family of observables in maximal conflict with 
local realism. For a check of this question
we have also calculated the threshold value
of $F_3$ for the case where both observers apply to
the incoming qutrit ($N=3$) the most general
unitary transformation belonging to a full SU(3) group (i.e. 
we have {\it any} trichotomic observables on each side). Again
we have assumed that each observer chooses between two sets of
local settings. However, in this case each set consists of 8
local settings rather than the three (effectively two) in the tritter case.
The result appears to be the same as for two tritters, which suggests 
that tritters (an perhaps generally unbiased 
multiports) are optimal devices to test \QM\ against
local realism  for $N=3$ (for all $N$).

\section{Exemplary analytical model for $N=3$}
The presented above approach based on the idea of marginals
and joint probability distribution compatible with them
is the most general one, i.e., it does not
assume any symmetries within the tested quantum 
probabilities (marginals)\footnote{Therefore it can be applied for any 
marginals not necessarily
quantum ones.}.
However, this approach has one
drawback. Namely, it involves lots of linear constraints imposed
on the joint distribution probability. In the most general
case, i.e., when observer A chooses $N_A$ observables $A_i$
and observer B chooses $N_B$ observables $B_j$ (each observable has
$N$ point spectrum) there are 
$N_{A}\times N_{B}\times N^2$ 
linear constraints
that must be fulfilled by the joint probability distribution
(we does not count the constraint that $F_N\leq 1$). This
considerably lengthens computation time.

On the
other hand the quantum probabilities (marginals) that
we consider here exhibit some symmetries. Therefore, one
can ask the question if, as it has been with two
and three maximally entangled qubits (see previous chapters),
it is possible to use the properly defined quantum 
correlation function
instead of quantum probabilities.

In this section we show that the answer is {\it yes} but only 
for\footnote{We have also performed calculations for $N\geq 4$ with the use of
the quantum correlation function \cyt{corr2} with $M=2$. The correlation
function can be reproduced for growing $N$ for lower and lower values
of the noise fraction $F_N$. Of course, this  
effect has nothing to
do with the possibility of a local realistic description for such
cases. Simply the correlation functions defined with Bell numbers
for $N\geq 4$ start to wash out the details of the full set
of probabilities describing the experiment.}
$N=3$.
Considering the quantum
correlation function will benefit in calculating the threshold $F_3$
for more local settings of the measuring apparata than with
the program presented in this chapter.

To this end let us consider two maximally entangled qutrits
(qu$N$its with $N=3$) and the quantum correlation function 
defined in 
\cyt{corr2} with $M=2$ and $N=3$. We remember that
it has been obtained by ascribing to the local results 
of measurements subsequent powers of 
$\alpha=\exp(\frac{2\pi i}{3})$. We can prove that
there is one-to-one equivalence between quantum probabilities
\cyt{5} and the quantum correlation function \cyt{corr2}.

According to \cyt{5} for $N=3$ and for
any local settings $\vec{\phi}_{A},\vec{\phi}_{B}$
quantum probabilities $P^{QM}(kl|\vec{\phi}_{A},\vec{\phi}_{B})$
can be divided into three groups
\be
&&P_{QM}^1=P_{QM}(12)=P_{QM}(21)=P_{QM}(33)\nonumber\\
&&P_{QM}^2=P_{QM}(11)=P_{QM}(23)=P_{QM}(32)\nonumber\\
&&P_{QM}^3=P_{QM}(13)=P_{QM}(31)=P_{QM}(22),
\label{Probmodulo}
\ee
where we do not show the dependence on local settings to shorten
the notation.
This allows us to write the correlation function
$E_{QM}(\vec{\phi}^i_A,\vec{\phi}^j_B)$ in the form (for
convenience we put $F_3=0$)
\be
&&E_{QM}=
3(P_{QM}^1+\alpha^2 P_{QM}^2+\alpha P_{QM}^3),
\label{Corrmodulo}
\ee
where again we do not show the dependence of the correlation
function on local settings.
Using the identity ${1\over 3}=P_{QM}^1+P_{QM}^2+P_{QM}^3$ and
$\alpha+\alpha^2
+\alpha^3=0$ we arrive at
\be
&&{1\over 3}(E_{QM}-1)=(\alpha-1)P_{QM}^3-(2+\alpha)P_{QM}^2,
\label{Corrmoduloreduced}
\ee
which allows us to express all quantum probabilities through
the correlation function (now $F_3$ is arbitrary)
\be
&&P_{QM}^1={1\over 9}+{2\over 9}ReE_{QM}^{F_3}\nonumber\\
&&P_{QM}^2={1\over 9}-{1\over 9}(\sqrt 3 ImE_{QM}^{F_3}+
ReE_{QM}^{F_3})\nonumber\\
&&P_{QM}^3={1\over 9}+{1\over 9}(\sqrt 3 ImE_{QM}^{F_3}-ReE_{QM}^{F_3}).
\label{corrprob}
\ee
Therefore, we can apply the approach based on the quantum 
correlation function presented in the chapter concerning
two and three qubits. 

For every measurement of the pair of observables $A_i$ and $B_j$ 
($i=1,\dots,N_a; j=1,\dots,N_b$)
defined
by the tritter and the phase shifters we calculate the quantum 
correlation function obtaining this way the 
quantum matrix of predictions
$\hat{Q}$ with entries
$Q_{ij}(F_3)=E_{QM}^{F_3}(\vec{\phi}^i_A,\vec{\phi}^j_B)$. 
Note that now the observer A has $N_A$ observables to choose
from (B has $N_B$).

The hidden variable model 
in this case reads
\be
&&E_{LHV}(\vec{\phi}^i_A,\vec{\phi}^j_B)=\sum_{k=1}^{N^{N_a}}
\sum_{l=1}^{N^{N_b-1}}p_{kl}A(\vec{\phi}^i_A,k)B(\vec{\phi}^j_B,l),
\label{Nqmodel}
\ee
where the functions $A(\vec{\phi}^i_A,k)=\alpha^m$ and 
$B(\vec{\phi}^j_B,l)=\alpha^n$ ($m,n=1,2,3$). If it is to
reproduce the quantum correlation function it must fulfil
the following set of $N_a\times N_b$ linear constraints\footnote{The number
of relevant $p_{kl}$ can be reduced by a factor of $3$ using a development
of the trick that has led to the representation of the correlation function
for two qubits in the form of \cyt{MODEL2}.}
\be
&&Q_{ij}(F_3)=\sum_{k=1}^{N^{N_a}}
\sum_{l=1}^{N^{N_b-1}}p_{kl}A(\vec{\phi}^i_A,k)B(\vec{\phi}^j_B,l).
\label{Nmodel}
\ee

Using HOPDM 2.30 and the DSM we can find the threshold $F_3$ still
admitting a local and realistic description of the gedanken experiment.
The results are identical with those obtained by the previous
method. However, since the algorithm is now much faster
we were able to perform calculations involving up to 5
different phase settings on each side (i.e. allowing
each observer to have a choice of up to 5
different observables to measure). Just as in the
two qubit case the optimal $F_3$ stayed unchanged at the value
obtained for the problem with $N_a=N_b=2$.

\subsection{Explicit model for extremal case}

Exemplary optimal settings for violation of local realism
in the experiment with two qutrits and $N_a=N_b=2$ are
\be
&&\vec{\phi_A}^{1}=(0,\frac{\pi}{3},-\frac{\pi}{3})\nonumber\\
&&\vec{\phi_A}^{2}=(0,0,0)\nonumber\\
&&\vec{\phi_B}^{1}=(0,\frac{\pi}{6},-\frac{\pi}{6})\nonumber\\
&&\vec{\phi_B}^{2}=(0,-\frac{\pi}{6},\frac{\pi}{6}).
\label{settings}
\ee

For such settings the quantum matrix 
$\hat{Q}$ (with $F_3=0$) reads\footnote{We present here result of an algebraic
re-calculation of the computer output. Perhaps, the most exciting aspect of
such exercise is the exact value of $F_3=\frac{11-6\sqrt 3}{2}$.}
\be
&&\hat{Q}
=\left(
\begin{array}{cc}
Q_1 & Q_1^* \\
Q_2 & Q_1   \\
\end{array}\right),
\label{Nmod}
\ee
where
$Q_1=\frac{2\sqrt 3 +1}{6}-i\frac{2-\sqrt 3}{6}$ and
$Q_2=-{1\over 3}(1+2i)$. Please, note that the all entries
of \cyt{Nmod} have the same modulus equal $\frac{\sqrt 5}{3}$.

The hidden variables can only reproduce the matrix of 
correlation function given by $(1-F_3)\hat{Q}$ with 
$F_3\leq\frac{11-6\sqrt 3}{2}$.
For the threshold maximal $F_3$ the explicit model is given by
\be
&&\hat{Q}^{F_3}
=p\left(
\begin{array}{cc}
\alpha^3 & \alpha^3\\
\alpha^3 & \alpha^3\\
\end{array}\right)+
p\left(
\begin{array}{cc}
\alpha^2 & \alpha^3\\
\alpha^2 & \alpha^3\\
\end{array}\right)+\nonumber\\
&&p\left(
\begin{array}{cc}
\alpha^3 & \alpha^3\\
\alpha^2 & \alpha^2\\
\end{array}\right)+
p\left(
\begin{array}{cc}
\alpha^3 & \alpha\\
\alpha^2 & \alpha^3\\
\end{array}\right)+\nonumber\\
&&q\left(
\begin{array}{cc}
\alpha^3 & \alpha^2\\
\alpha & \alpha^3\\
\end{array}\right)+
q\left(
\begin{array}{cc}
\alpha & \alpha^3\\
\alpha^2 & \alpha\\
\end{array}\right)=(1-F_3)\hat{Q},
\label{3model}
\ee
where $p=\frac{4-2\sqrt 3}{3}$, 
$q=\frac{8\sqrt 3 -13}{6}$.

This model has to be understood in the following
way. Consider the first term. The rank-one matrix
\be
&&\left(
\begin{array}{cc}
\alpha^3 & \alpha^3\\
\alpha^3 & \alpha^3\\
\end{array}\right)
\ee 
can be factorised into column and row matrices built out of
powers of $\alpha$, in the following three ways:
\be
&&\left(
\begin{array}{cc}
\alpha^3 & \alpha^3\\
\alpha^3 & \alpha^3\\
\end{array}\right)=
\left(\begin{array}{c}
\alpha^3 \\
\alpha^3
\end{array}\right)
\left(\begin{array}{c}
1 \\
1
\end{array}\right)^T=
\left(\begin{array}{c}
\alpha^2\\
\alpha^2
\end{array}\right)
\left(\begin{array}{c}
\alpha \\
\alpha
\end{array}\right)^T=
\left(\begin{array}{c}
\alpha \\
\alpha
\end{array}\right)
\left(\begin{array}{c}
\alpha^2 \\
\alpha^2
\end{array}\right)^T.
\ee
Therefore with probability $p'={1\over 3}p$ each of this
factorisations is present in the model of $\hat{Q}^{F_3}$
(compare \cyt{Nqmodel} and \cyt{Nmodel}).

\section{Conclusions}
It is evident, that indeed two entangled
qu$N$its violate local realism stronger than two entangled qubits, 
and that the violation increases with $N$. It is important to stress that 
the values were obtained using four independently written codes,
one of them employing a different linear optimisation procedure
(from the NAG Library). 

It is interesting to compare our results
with the limit \cite{HOROD99} 
for the non separability of the density matrices \cite{WERNER89}
of the two entangled systems.
The fact that this limit, $\frac{N}{N+1}$, is always higher than 
ours indicates that that requirement of having local 
quantum description of the two subsystems is a much more stringent
condition than our requirement of admitting any possible 
local realistic model.

It will be interesting to consider within our approach 
different families of states, generalisations to more than two qu$N$its, 
extensions of the families of observables,
to see if a wider choice of experiments than can be performed on one
side (i.e., more than two) can lead to even stronger violations
of local realism,
and finally to see experimental realizations of such schemes.

Finally we should mention that this approach can be extended
to the research into the threshold efficiency of detectors
for such experiments. The computer program for imperfect
detectors has been already written.

\appendix

\chapter{Proof of equivalence of the existence of local hidden variables
and a joint probability distribution for incompatible 
measurements}
This appendix is a re-derivation of 
known results (see e.g. \cite{Fine}).

We consider the experiment with two observers
A and B. Observer A can choose to measure $N_A$ observables $A_i$
($i=1,2,\dots,N_A$) whereas observer can choose to measure
$N_B$ observables $B_j$ ($j=1,2,\dots,N_B$). 
In the measurement of the observable $A_i$ observer A
can obtain $N$ results, which we 
denote by $k_i$ ($i=1,\dots,N$). 
Similarly, in the measurement of the observable $B_j$ observer B
can obtain $N$ results, 
which we denote by $l_j$ ($j=1,\dots,N$). In the experiment
the observers can only measure joint probabilities 
$P(k_i,l_j|A_i,B_j)$, i.e., probabilities of obtaining the 
result $k_i$ when measuring the observable $A_i$ and the 
result $l_j$ when measuring the observable $B_j$. 

First, let us show that the existence of stochastic local hidden
variables recovering the marginals $P(k_i,l_j|A_i,B_j)$
implies the existence of a joint probability
distribution 
\be
&&P_{HV}(k_1,\dots,k_{N_A},
l_1,\dots,l_{N_B}|A_1,\dots,A_{N_A},B_1,\dots,B_{N_B}) 
\label{phv}
\ee 
compatible with the given marginals.
The existence of local hidden variable space $\Lambda$
and the "hidden" probabilities $P_i(k_i|\lambda, A_i)$
and $P_j(l_j;\lambda,B_j)$ (see \cyt{stochLHV}) allows one
to define the joint probability distribution 
\cyt{phv} in the following
way
\be
&&P_{HV}(k_1,\dots,k_{N_A},
l_1,\dots,l_{N_B}|A_1,\dots,A_{N_A},B_1,\dots,B_{N_B}) 
\nonumber\\
&&=\int_{\Lambda}d\rho(\lambda)\prod_{i=1}^{N_A}
P_i(k_i|\lambda, A_i)\prod_{j=1}^{N_B}P_j(l_j|\lambda, B_j).
\label{c2}
\ee 
It is evident that \cyt{c2} returns as marginals $P(k_i,l_j|A_i,B_j)$.

Now let us assume that \cyt{phv} exists and is
compatible with the marginals
$P(k_i,l_j|A_i,B_j)$
We can define the local hidden variables as follows.
To each sequence of possible results 
$(\tilde{k}_1,\tilde{k}_2,\dots,\tilde{k}_{N_A},
\tilde{l}_1,\tilde{l}_2,\dots,
\tilde{l}_{N_B})$
we ascribe a hidden variable, which we denote as 
$\lambda(\tilde{k}_1,\tilde{k}_2,\dots,\linebreak\tilde{k}_{N_A},\tilde{l}_1,
\tilde{l}_2,\dots,
\tilde{l}_{N_B})$. Therefore, the local hidden variable space $\Lambda$ consists
of $N^{N_A+N_B}$ different hidden variables. On this space we define 
the discrete probability
distribution of local hidden variables $\rho(\lambda)$
in the following way
\be
&&\rho(\lambda(\tilde{k}_1,\tilde{k}_2,\dots,\tilde{k}_{N_A},\tilde{l}_1,
\tilde{l}_2,\dots,\tilde{l}_{N_B}))\nonumber\\
&&=P_{HV}(\tilde{k}_1,\dots,\tilde{k}_{N_A},\tilde{l}_1,\dots,\tilde{l}_{N_B}|
A_1,\dots,A_{N_A},B_1,\dots,B_{N_B}).
\ee
The hidden probabilities 
$P_i(k_i|\lambda,A_i)$
and 
$P_j(l_j|\lambda,B_j)$
we define as
\be
&&P_i(k_i|\lambda(\tilde{k}_1,\tilde{k}_2,\dots,\tilde{k}_{N_A},\tilde{l}_1,
\tilde{l}_2,\dots,\tilde{l}_{N_B}), A_i)=\delta_{\tilde{k}_i,k_i}\nonumber\\
&&P_j(l_j|\lambda(\tilde{k}_1,\tilde{k}_2,\dots,\tilde{k}_{N_A},\tilde{l}_1,
\tilde{l}_2,\dots,\tilde{l}_{N_B}), B_j)=\delta_{\tilde{l}_j,l_j},
\ee
where $\delta$ denotes the Kronecker delta function. Then the marginals
are recovered in the following way
\be
&&P(k_i,l_j|A_i,B_j)\nonumber\\
&&=\sum_{\lambda}
\rho(\lambda(\tilde{k}_1,\tilde{k}_2,\dots,\tilde{k}_{N_A},\tilde{l}_1,
\tilde{l}_2,\dots,\tilde{l}_{N_B}))
\delta_{\tilde{k}_i,k_i}\delta_{\tilde{l}_j,l_j},
\label{phv3}
\ee
where we sum over all $N^{N_A+N_B}$ hidden variables $\lambda$.
Using the above definitions \cyt{phv3} can be rewritten as
\be
&&P(k_i,l_j|A_i,B_j)=\sum_{\lambda}\rho(\lambda)P_i(k_i|\lambda,A_i)
P_j(l_j|\lambda,B_j),
\ee
i.e. we have a typical structure of a local hidden variable
model (for hidden variables forming a discrete set).


\begin{thebibliography}{99}
\bibitem[Aerts99]{AERTS} S. Aerts, P. Kwiat, J. -\AA. Larsson, and M. \.Zukowski, 
Phys. Rev. Lett. {\bf 83}, 2872 (1999).

\bibitem[Agarwal93]{AGARWAL93} G. S. Agarwal, Phys.Rev. A {\bf 47}, 4608 (1993). 

\bibitem[Ardehali91]{ARDEHALI91} M. Ardehali, Phys. Rev. D {\bf 44}, 3336 (1991). 

\bibitem[Ardehali92]{ARDEHALI92} M. Ardehali, Phys. Rev. A {\bf 46}, 5375 (1992) 


\bibitem[Aspect82]{ASPECT82} A. Aspect, J. Dalibard, and G. Roger, Phys. Rev. Lett. {\bf 49}, 1804 (1982).

\bibitem[Belinfante73]{Belinfante73}F. J. Belinfante, 
{\it Asurvey of Hidden-Variables Theories}, Oxford, Pergamon, 1973.

\bibitem[Bell64]{Bell64} J. Bell, Physics {\bf 1}, 195 (1964).

\bibitem[Bell66]{Bell66} J. S. Bell, Rev. Mod. Phys. {\bf 38}, 447 (1966).

\bibitem[Bell87]{Bell87} J. S. Bell. {\it Speakable and Unspeakable in Quantum Mechanics.} Camridge UP, Cambridge, 1987.

\bibitem[Bennett84]{BB84} C. H. Bennett and G. Brassard, 
in {\it Proceedings of the IEEE International Conference on Computers, 
Systems, and Signal Processing, Bangalore, India} (IEEE, New York, 1984), p. 175.

\bibitem[Bennett93]{TELEPORT95} C. Bennett, G. Brassard, C. Crepeau, R. Jozsa, A. Peres and W. Wooters, 
Phys. Rev. Lett. {\bf 70}, 1895 (1993).

\bibitem[Belinski93]{BIELINSKI93} A. V. Belinskii and D. N. Klyshko, Phys. Usp. {\bf 36}, 653 (1993).

\bibitem[Bohm51]{Bohm-EPR1} D. Bohm. {\it Quantum Theory}, Prentice Hall, Englewood Cliffs, 1951.

\bibitem[Bohm52]{Bohm-EPR2} D. Bohm, Phys. Rev. {\bf 85}, 166 (1952).

\bibitem[Bose98]{BOSE} S. Bose, V. Vedral, P. L. Knight, Phys. Rev. A {\bf 57}, 822 (1998).

\bibitem[Bouwmeester97]{BOUW} D. Bouwmeester, J.-W. Pan, K. Mattle, M. Eible, H. Weinfurter and A. Zeilinger,  
Nature {\bf 390} 575 (1997).

\bibitem[Bouwmeester98]{BZ} D. Bouwmeester, K. Mattle, J.-W. Pan, H. Weinfurter,  A. Zeilinger and M. \.Zukowski,  
Appl. Phys. B, {\bf 67} 749 (1998). 

\bibitem[Bouwmeester99]{INNSBRUCK} D. Bouwmeester, Jian-Wei Pan, M. Daniell, H. Weinfurter 
and A. Zeilinger, Phys. Rev. Lett. {\bf 82}, 1345 (1999). 

\bibitem[Braunstein89]{BR} S. L. Braunstein and C. M. Caves, Ann. Phys. (NY) {\bf 202}, 22 (1990).

\bibitem[Casado97]{CASADO97} A. Casado, A. Fernandez-Rueda, T. W. Marshall, R. Risco-Delgado and E. Santos, Phys. Rev. A {\bf 55}, 3879 (1997).

\bibitem[Clauser69]{CHSH69} J. F. Clauser, M. A. Horne, A. Shimony and R. A. Holt, 
Phys. Rev. Lett. {\bf 23}, 880 (1969).

\bibitem[Clauser74]{CH} J.F Clauser and M. A. Horne, Phys. Rev. D {\bf 10}, 526 (1974).

\bibitem[Clauser78]{CLAUSER78} J. F. Clauser and A. Shimony, Rep. Prog. Phys. {\bf 41}, 1881 (1978).

\bibitem[Deutsch89]{DEUTSCH89} D. Deutsch, Proc. R. Soc. London A {\bf 425}, 73 (1989).

\bibitem[Eckert91]{ECKERT91} A. K. Ekert, Phys. Rev. Lett. {\bf 67}, 661 (1991).

\bibitem[Einstein35]{EPR} A. Einstein, B. Podolsky and N. Rosen, Phys. Rev. {\bf 47}, 777 (1935).

\bibitem[Feynman82]{FEYNMAN82} R. Feynman, Int. J. Theor. Phys. {\bf 21}, 219 (1982).

\bibitem[Fine82]{Fine} A. Fine, Phys. Rev. Lett. {\bf 48}, 291 (1982).

\bibitem[Freedman72]{FREEDMAN72} S. J. Freedman and J. F. Clauser, Phys. Rev. Lett. {\bf 28}, 938 (1972). 

\bibitem[Garg82]{GARG82} A. Garg and N. D. Mermin, Phys. Rev Lett. {\bf 49}, 901 (1982). 

\bibitem[Garg85]{GM} A. Garg and N. D. Mermin, Phys. Rev. D {\bf35}, 3832 (1985).

\bibitem[Garg87]{Garg87} A. Garg, N. D. Mermin, Phys. Rev. D {\bf 35}, 3638 (1987).

\bibitem[Garuccio80]{GARUCCIO80} A. Garuccio and F. Selleri, Found. Phys. {\bf 10}, 209 (1980).

\bibitem[Garuccio94]{GARUCCIO} L. De Caro and A. Garuccio, Phys. Rev. A {\bf 50}, R2803 (1994).   

\bibitem[Gass75]{GASS} S. I. Gass {\it Linear programming} 4th ed. New York: McGraw-Hill 1975. 

\bibitem[Gisin92]{GISIN92} N. Gisin and A. Peres, Phys. Lett. A {\bf 162}, 15 (1992).

\bibitem[Gisin99]{GISIN99} N. Gisin, Phys. Lett. A {\bf260}, 1 (1999). 

\bibitem[Gleason57]{GLEASON57} A. M. Gleason, J. Math. and Mech. {\bf 6}, 885 (1957).

\bibitem[Gondzio95]{GONDZIO} J. Gondzio, European Journal of Operational
Research {\bf 85}, 221 (1995);
J. Gondzio, Computational Optimization and
Applications {\bf 6}, 137 (1996).

\bibitem[Greenberger89]{GHZ89} D. M. Greenberger, M. A. Horne and A. Zeilinger, 
in {\it Bell's Theorem, Quantum Theory, and Conceptions of the Universe,}
edited by Kafatos,\ M. (Kluwer Academics, Dordrecht, The Netherlands,1989), 
p. 73.

\bibitem[Greenberger90]{GHSZ} D. M. Greenberger, M. Horne, A. Shimony 
and A. Zeilinger Am. J. Phys. {\bf 58}, 1131 (1990).

\bibitem[Hagley97]{Hagley97} E. Hagley, X. Maitre, G. Nogues, 
C. Wunderlich, M. Brune, J. M. Raimond, and S. Haroche, Phys. Rev. Lett. {\bf 79}, 1 (1997).

\bibitem[Hong85]{MANDEL2} C. Hong, L. Mandel, Phys. Rev. A {\bf 31}, 2409, (1985).


\bibitem[Hong87]{HOM} C. K. Hong, Z. Y. Ou and L. Mandel, Phys. Rev. Lett. {\bf 59}, 2044 (1987).

\bibitem[Horne85]{HORNE85} M. A. Horne and A. Zeilinger, 
in {\em Symposium on  Foundations of Modern Physics}, eds. P. Lahti and P. Mittelstaedt (World Sc., Singapore, 1985).  

\bibitem[Horne89]{HORNE89} M. A. Horne, A. Shimony and A. Zeilinger, Phys. Rev. Lett. {\bf 62}, 2209 (1989).

\bibitem[Horodecki95]{HOROD95} R. Horodecki, P. Horodecki and M. Horodecki, Phys. Lett. A {\bf 200}, 340 (1995).

\bibitem[Horodecki96]{HOROD96} M. Horodecki, P. Horodecki, and R. Horodecki, Phys. Lett. A 223, 1 (1996). 

\bibitem[Horodecki99]{HOROD99} M. Horodecki and P. Horodecki, Phys. Rev. A {\bf 59}, 4206 (1999).

\bibitem[Ivanovic81]{IVANOVIC81} I. D. Ivanovic, J. Phys. A {\bf 14}, 3241 (1981).

\bibitem[Jex95]{JEX95} I. Jex, S. Stenholm and A. Zeilinger, Opt. Comm. {\bf 117}, 95 (1995).

\bibitem[Jaeger93]{YAGER} G. Jaeger, M. A. Horne and A. Shimony, Phys. Rev. A, {\bf 48}, 1023 (1993).

\bibitem[Klyshko88]{KLYSHKO88} D. N. Klyshko, Phys. Lett. A {\bf 132}, 299 (1988).

\bibitem[Kochen67]{KOCHEN67} S. Kochen and E. Specker, J. Math. Mech. {\bf 17}, 59 (1967).

\bibitem[Kwiat93]{KWIAT93} P. G. Kwiat, A. M. Steinberg and R. Y.  Chiao, Phys. Rev. A {\bf 47}, R2472 (1993)

\bibitem[Kwiat94]{KWIAT1} P. G. Kwiat, P. E. Eberhard, A. M. Steinberg and R. Y. Chiao, Phys. Rev. A {\bf 49}, 3209 (1994).

\bibitem[Kwiat95]{KWIAT2} P. G. Kwiat, Phys. Rev. A {\bf52}, 3380 (1995).

\bibitem[Larsson99]{LARSSON99} J.-\AA. Larsson, Phys. Lett. A {\bf 256}, 245 (1999).  

\bibitem[Mattle95]{MMWZZ95} C.  Mattle, M. Michler,  H. Weinfurter,  A. Zeilinger and 
M. \.Zukowski, Appl. Phys. B {\bf 60}, S111 (1995).

\bibitem[Mattle96]{MATTLE96} K. Mattle, H. Weinfurter, P. G. Kwiat and A. Zeilinger,
Phys. Rev. Lett. {\bf 76}, 4656 (1996).

\bibitem[Mermin80]{MERMIN80} N. D. Mermin, Phys. Rev. D {\bf 22}, 356 (1980). 

\bibitem[Mermin82]{MERMIN82} N. D. Mermin and G. M. Schwarz, Found. Phys. {\bf 12}, 101 (1982).  

\bibitem[Mermin90a]{MERMIN-GHZ} N. D. Mermin Physics Today, p. 9 (June 1990).

\bibitem[Mermin90b]{MERMIN90} N. D. Mermin, Phys. Rev. Lett. {\bf 65}, 1838 (1990). 

\bibitem[Michler96]{MICHLER96} M. Michler, K. Mattle, H. Weinfurter and A. Zeilinger, 
Phys. Rev. A {\bf 53}, R1209 (1996).

\bibitem[Michler00]{MICHLER2000} M. Michler, H. Weinfurter, M. \.Zukowski,
Phys. Rev. Lett. {\bf 84}, 5427 (2000).

\bibitem[Nelder65]{DSM} J. A. Nelder and R. Mead, Computer Journal
{\bf 7}, 308 (1965).

\bibitem[Ou88]{MANDEL88} Z. Y. Ou and L. Mandel, Phys. Rev. Lett. {\bf 61}, 50 (1988).

\bibitem[Pan98]{Pan98} J. W. Pan, D. Bouwmeester, H. Weinfurter, A. Zeilinger, 
Phys. Rev. Lett. {\bf 80}, 3891 (1998).

\bibitem[Peres92]{PERES92} A. Peres, Phys. Rev. A {\bf 46}, 4413 (1992).


\bibitem[Peres96]{PERESSEP96} A. Peres, Phys. Rev. Lett. {\bf 77}, 1413 (1996).

\bibitem[Peres99]{PERESBELL} A. Peres, Found. Phys. {\bf 29}, 589 (1999).

\bibitem[Pittman95]{PITTMAN95} T. B. Pittman, Y. H.  Shih, A. V. Sergienko and M. H. Rubin, Phys. 
Rev. A {\bf 51}, 3495  (1995).

\bibitem[Popescu97]{SHZ} S. Popescu, L. Hardy and M. \.Zukowski, Phys. Rev. A {\bf 56}, R4353 (1997).

\bibitem[Rarity90]{RARITY90} J. G. Rarity and P. R. Tapster, Phys. Rev. Lett. {\bf 64}, 2495 (1990).

\bibitem[Reck94]{RECK94} M. Reck, A. Zeilinger, H. J. Bernstein and P. Bertani, Phys. Rev.  Lett.  
{\bf 73}, 58 (1994).

\bibitem[Reck96]{RECKPHD96} M. Reck, PhD Thesis (supervisor: A. Zeilinger) (University of Innsbruck, 1996, unpublished).

\bibitem[Redhead87]{Redhead87} M. Redhead. {\it Incompletness, Nonlocality, and Realism. A prolegomenon to the philosophy of quantum mechanics.} Clarendon Press, Oxford, 1987.

\bibitem[Roy91]{ROY91} S. M. Roy and V. Singh, Phys. Rev. Lett. {\bf 67}, 2761 (1991).

\bibitem[Santos92]{SANTOS} E. Santos, Phys. Rev. A  {\bf 46}, 3646 (1992).

\bibitem[Schr\"{o}edinger35]{Erwin35} E. Schr\"{o}dinger. The present situation in quantum mechanics. 
In J. Wheeler and W. \.Zurek, editors, {\ it Quantum Theory and Measurement}, 
p. 152. Princeton University Press, 1983.

\bibitem[Schwinger60]{SCHWINGER60} J. Schwinger, Proc. Nat. Acad. Sc. {\bf 46}, 570 (1960).

\bibitem[Selleri88]{Selleri} {\it Quantum MechanicsVersus Local Realism. The Einstein-Podolsky-Rosen Paradox}. 
Edited by F. Selleri. Plenum Press, New York, 1988.

\bibitem[Shih88]{SHIH} C. O. Alley and  Y. H. Shih, Phys. Rev. Lett. {\bf 61}, 2921 (1988).


\bibitem[Stapp77]{Stapp77} H. Stapp, Nuovo Cimento {\bf 40B}, 191 (1977).

\bibitem[Tapster94]{TAPSTER94} P. R.  Tapster, J. G. Rarity, and P. C.  M.  Owens,
Phys. Rev. Lett. {\bf 73}, 1923 (1994).

\bibitem[Tittel98]{TITTEL98} W. Tittel, J. Brendel, H. Zbinden, and N. Gisin, Phys. Rev.  Lett. {\bf 81}, 3563 (1998).

\bibitem[Tsirelson80]{TSIRELSON80} B. S. Cirel'son, Lett. Math. Phys. {\bf 4}, 93 (1980).

\bibitem[W\'odkiewicz91]{WODKIEWICZ91} C. Su and K. W\'odkiewicz, Phys. Rev. A {\bf 44}, 6097 (1991). 

\bibitem[W\'odkiewicz94]{WODKIEWICZ94} K. W\'odkiewicz, Acta Phys. Pol. {\bf 86}, 223 (1994).

\bibitem[W\'odkiewicz95]{WODKIEWICZ95} K. W\'odkiewicz, Phys. Rev {\bf A 51}, 2785 (1995).

\bibitem[Weihs96]{WEIHS96} G. Weihs, M. Reck, H. Weinfurter, A. Zeilinger, Opt. Lett., {\bf 21}, 302 (1996).

\bibitem[Weihs98]{WEIHS98} G. Weihs, T. Jennewein, Ch. Simon, H. Weinfurter and A. Zeilinger, 
Phys. Rev. Lett. {\bf 81}, 5039 (1998). 

\bibitem[Werner89]{WERNER89} R. Werner, Phys. Rev. A {\bf 40}, 4277 (1989).

\bibitem[Wigner70]{Wigner70} E. P. Wigner, Am. J. Phys. {\bf38}, 1005 (1970).

\bibitem[Wooters86]{WOOTERS86} W. K. Wooters, Found. Phys. {\bf 16}, 391 (1986).

\bibitem[Yurke92]{YS} B. Yurke and D. Stoler, Phys. Rev. A {\bf 46} 2229 (1992);
Phys. Rev.  Lett., {\bf 68} 1251 (1992).  

\bibitem[Zeilinger86]{Zeilinger86} A. Zeilinger, Am. J. Phys {\bf 49}, 882 (1981).

\bibitem[Zeilinger93]{ZBGHZ93} A. Zeilinger, H. J. Bernstein,  
D. M. Greenberger, M. A. Horne, and M. \.Zukowski, 
in {\it QuantumControl and Measurement}, eds. H. Ezawa and Y. Murayama (Elsevier, 1993). 

\bibitem[Zeilinger94]{ZZHBG94} A. Zeilinger, M. \.Zukowski, M. A. Horne, H. J. Bernstein and D. M. Greenberger, 
in {\it Quantum Interferometry}, eds. F. DeMartini, A.  Zeilinger, (World  Scientific, Singapore, 1994).

\bibitem[\.Zukowski88]{ZUKOWSKI88} M. \.Zukowski and J. Pykacz, Phys. Lett.  A {\bf 127}, 1 (1988).

\bibitem[\.Zukowski93a]{EVENT} M. \.Zukowski, A. Zeilinger, M. A. Horne and A. K. Ekert, 
Phys. Rev. Lett. {\bf 71}, 4287 (1993).

\bibitem[\.Zukowski93b]{ZUK93} M. \.Zukowski, Phys. Lett. A, {\bf 177}, 290 (1993)

\bibitem[\.Zukowski95]{ZZW} M. \.Zukowski, A. Zeilinger, H. Weinfurter: Ann. N. Y. Acad. Science {\bf 755}, 91 (1995).

\bibitem[\.Zukowski97b]{ZZH} M. \.Zukowski, A. Zeilinger, M. A. Horne, Phys. Rev. A {\bf 55}, 2564 (1997).

\bibitem[\.Zukowski99]{ZZHW} M. \.Zukowski, A. Zeilinger, M. A. Horne and H. Weinfurter, 
Int. J. Theor. Phys. {\bf 38}, 501 (1999). 

\bibitem[\.Zukowski91]{Zukowski91} M. \.Zukowski, Phys. Lett. A {\bf177}, 290 (1991). 


\bibitem[vonNeumann32]{vonNeumann} J. v. Neumann, 
{\it Mathematishe Grundlagender Quantenmechanik.} Springer, Berlin, 1932.

\bibitem{Moja1} M. \.Zukowski, D. Kaszlikowski and E. Santos, 
Phys. Rev. A {\bf 60}, R2614 (1999).

\bibitem{Moja2} M. \.Zukowski and D. Kaszlikowski, 
Acta Phys. Slov. {\bf 49}, 621 (1999).

\bibitem{Moja3} M. \.Zukowski and D. Kaszlikowski, 
Phys. Rev. A, {\bf 56}, R1682 (1997). 

\bibitem{Moja4} D. Kaszlikowski and M. \.Zukowski, 
Phys. Rev. A {\bf 61}, 022114 (2000).

\bibitem{Moja5} M. \.Zukowski and D. Kaszlikowski, 
Phys. Rev. A {\bf 59}, 3200 (1999).

\bibitem{Moja6} M. \.Zukowski and D. Kaszlikowski, 
Vienna Circle Yearbook vol. 7, editors D. Greenberger, 
W. L. Reiter, A. Zeilinger, Kluwer Academic Publishers, 
Dortrecht (1999).

\bibitem{Moja7} M. \.Zukowski, D. Kaszlikowski, A. Baturo
and J. -A. Larsson, quant-ph/9910058

\bibitem{Moja8} D. Kaszlikowski, P. Gnacinski, M. \.Zukowski,
W. Miklaszewski and A. Zeilinger, quant-ph/0005028

\end{thebibliography}
\end{document}